\documentclass[aps,prd,groupedaddress,nofootinbib]{revtex4-2}
\usepackage{graphicx}
\usepackage{slashed}
\usepackage{xcolor}
\usepackage{amsmath,amssymb}
\usepackage{verbatim}
\usepackage{hyperref} 
\hypersetup{
    colorlinks,
    citecolor=black,
    filecolor=black,
    linkcolor=black,
    urlcolor=black
}

\newcommand{\be}{\begin{equation}}
\newcommand{\ee}{\end{equation}}
\newcommand{\ba}{\begin{align}}
\newcommand{\ea}{\end{align}}
\newcommand{\ban}{\begin{eqnarray*}} 
\newcommand{\ean}{\end{eqnarray*}}
\newcommand \nn {\nonumber}
\def\ck{{\check{k}}}
\def\cq{{\check{q}}}
\def\cp{{\check{p}}}

\def\M{{\cal M}}
\def\tr{{\rm Tr}}
\def\sp{\slashed{p}}
\def\sk{\slashed{k}}
\def\sq{\slashed{q}}
\def\A{{\cal A}}

\def\p{{\mathbf p}}
\def\q{{\mathbf q}}
\def\k{{\mathbf k}}
\def\x{{\mathbf x}}  
\def\y{{\mathbf y}}

\def\z{{\mathbf z}}
\def\w{{\mathbf w}}

\def\ux{{\underline{x}}}

\def\uz{{\underline{z}}}
\def\uk{{\underline{k}}}
\def\up{{\underline{p}}}
\def\uq{{\underline{q}}}

\begin{document}

\title{Forward parton-nucleus scattering at next-to-eikonal accuracy in the CGC}
\author{Tolga Altinoluk$^{a}$, Guillaume Beuf$\,^{a}$ and Swaleha Mulani$^{a}$}
\affiliation{ $^{a}$Theoretical Physics Division, National Centre for Nuclear Research,
Pasteura 7, Warsaw 02-093, Poland}

\date{\today}

%%%%%%%%
%%%%%%%%
\begin{abstract}

We derive the full next-to-eikonal (NEik) corrections to the gluon propagator from before to after traversing a highly boosted gluon background field, including corrections both beyond the shockwave limit and beyond the static limit in particular. After summarizing the results of the full NEik corrections to the before-to-after quark propagator computed in our earlier works, we also derive the before-to-inside, inside-to-inside and inside-to-after quark and gluon propagators, which are building blocks to calculate high-energy scattering processes at NEik order. Using these results and also including the NEik corrections that stem from interactions with the target via t-channel quark exchanges, we compute inclusive cross sections for quark and gluon production at forward rapidities in quark-nucleus and gluon-nucleus scatterings at NEik accuracy.   

\end{abstract}
%%%%%%%%
%%%%%%%%

\maketitle

\tableofcontents

%%%%%%%%%%%%%%%%%%%%%%%%%%%%%%%%%%%%%%%%%%%%%
%%%%%%%%%%%%%%%%%%%%%%%%%%%%%%%%%%%%%%%%%%%%%
%%%%%%%%%%%%%%%%%%%%%%%%%%%%%%%%%%%%%%%%%%%%%
%%%%%%%%%%%%%%%%%%%%%%%%%%%%%%%%%%%%%%%%%%%%%
%%%%%%%%%%%%%%%%%%%%%%%%%%%%%%%%%%%%%%%%%%%%%

\section{Introduction}
\label{sec:intro}
%%%%%%%%%%%%%%%%%%%%%%%%%%%%%%%%%%%%%%%%%%%%%
%%%%%%%%%%%%%%%%%%%%%%%%%%%%%%%%%%%%%%%%%%%%%

The Color Glass Condensate (CGC) (see \cite{Gelis:2010nm,Albacete:2014fwa,Blaizot:2016qgz} for recent reviews and references therein) is the effective theory that describes the high energy limit of the hadronic collisions. CGC relies on gluon saturation phenomena that can be reached at sufficiently high scattering energies. The increase in energy is provided by decreasing $x$ where $x$ is the longitudinal momentum fraction carried by the interacting partons. With decreasing $x$ the gluon density of the interacting hadrons increase rapidly and at sufficiently high energies (or sufficiently low values of $x$) this increase is tamed by nonlinear interactions of the emitted gluons and causes the aforementioned gluon saturation phenomena that is characterized by a dynamical scale known as saturation momenta $Q_s$. The evolution in $x$ is governed by the famous Balitsky-Kovchegov/Jalilian-Marian-Iancu-McLerran-Weigert-Leonidov-Kovner (BK-JIMWLK) equation derived in \cite{Balitsky:1995ub,Kovchegov:1999yj,Kovchegov:1999ua,Jalilian-Marian:1996mkd,Jalilian-Marian:1997qno,Jalilian-Marian:1997jhx,Jalilian-Marian:1997ubg,Kovner:2000pt,Weigert:2000gi,Iancu:2000hn,Iancu:2001ad,Ferreiro:2001qy}.    

Over the last three decades a vast amount of effort has been devoted to advance the CGC framework with the motivation of understanding the high energy collision data.  Single inclusive particle/jet production at forward rapidities in proton-nucleus (pA) collisions is one of the observables that is frequently used to test the compatibility of saturation phenomena with the pA collision data from the Relativistic Heavy Ion Collider (RHIC) and the Large Hadron Collider (LHC).  The calculation framework for this observable is known as "hybrid factorization" \cite{Dumitru:2005gt}. This approach allows one to treat the dilute projectile in the spirit of collinear factorization while the scattering of the projectile partons on the dense target is accounted for via eikonal approximation in the CGC framework.  

Even though the CGC framework have shown a great success in phenomenological studies of the high energy collisions, one needs to increase the precision of CGC computations of observables in order to achieve reliable quantitative description of the experimental data. This can be achieved either by computing the next-to-leading order (NLO) corrections in coupling constant $\alpha_s$ to observables or by relaxing the kinematical approximations adopted to compute the leading order (LO) observables. The NLO corrections to single inclusive particle/jet production   at forward rapidity in pA collisions have been computed analytically \cite{Altinoluk:2011qy,Chirilli:2011km,Chirilli:2012jd,Stasto:2014sea,Kang:2014lha,Altinoluk:2014eka,Altinoluk:2015vax,Iancu:2016vyg,Liu:2020mpy,Ducloue:2016shw,Shi:2021hwx,Wang:2022zdu,Altinoluk:2023hfz}.  Numerical studies of these results have been also performed \cite{Stasto:2013cha,Stasto:2014sea,Watanabe:2015tja,Ducloue:2017dit,Ducloue:2017mpb} to test the compatibility with the experimental data. 

As stated earlier, a complementary way to increase the precision of the CGC calculations is to relax the kinematic approximations adopted when computing the LO observables. High energy dilute-dense collisions within the CGC framework relies on two main approximations. The first one is referred to as the "semi-classical approximation". This approximation amounts to representing the dense target in the scattering process by a strong semiclassical background field $\A^{\mu}_a(x)$. The second approximation adopted within in the CGC is the well known eikonal approximation. Eikonal approximation amounts to keeping the leading power terms in energy in the high-energy limit and discarding the finite energy corrections. In the CGC framework, the high-energy limit can be achieved by boosting the target along $x^-$ with a boosting parameter $\gamma_t$. In this limit, the components of the semiclassical background field representing the target are scaled with the boosting parameter $\gamma_t$: 
\begin{align}
\label{eq_-_comp}
& \A^-(x)\propto \gamma_t\gg 1 \\
\label{eq_j_comp}
& \A^j(x)\propto (\gamma_t)^0=1 \\
\label{eq_+_comp}
& \A^+(x)\propto 1/ \gamma_t \ll 1
\end{align}
Moreover, the coordinate dependence of the target fields is also scaled as 
\begin{align}
(x)\to \Big(\gamma_t x^+, \frac{x^-}{\gamma_t},\x\Big)
\end{align}
in the high-energy limit that is achieved by boosting the target in the $x^-$ direction by a large boost parameter $\gamma_t$. In this setup, the eikonal approximation can be understood as the infinite boost of the target field $\A^{\mu}(x)$ that amounts to the following three assumptions:\\
(i) The background field $\A^{\mu}(x)$ is assumed to be independent of $x^-$ coordiante due to the Lorentz time dilation. This is known as the static limit of the target fields and in this case there is no longitudinal momentum $p^+$ transfer from the target to the projectile during the interaction. \\
(ii) The background target fields are localized around $x^+=0$ in the longitudinal direction due to the Lorentz contraction of the background target fields $\A^{\mu}(x)$. This is know as the shockwave limit and in this limit partons from the projectile interact instantly in $x^+$ with the target without having any transverse motion within the target. \\
(iii) Under the boost of the target along $x^-$ direction with the boost parameter $\gamma_t$, the $-$ component of the background field is enhanced, transverse component stays unscaled and $+$ component is suppressed, establishing a strong hierarchy between the components of the target as can be seen from Eqs. \eqref{eq_-_comp},\eqref{eq_j_comp} and \eqref{eq_+_comp} 
\begin{align}
\label{eq:g_field_hierarchy}
\A^-(x)\ll \A^j(x)\ll \A^+(x) 
\end{align}  
in a generic gauge. The Eikonal approximation amounts to accounting for the interaction of the largest component of the background field $\A^-(x)$ and discarding the interactions of the transverse and $+$ components of the background field with the projectile partons. 

All in all, in the eikonal limit the background field takes the following form 
\begin{align}
\A^{\mu}(x^+,x^-,\x)\simeq \delta^{\mu-}\A^-(x^+,\x)\propto \delta(x^+)
\end{align}
and as will be discussed in more detail in the next section, it resums the interactions of $\big(g\A^-(x^+,\x)\big)^n$ to all orders which leads to Wilson lines along $x^+$ direction. 

In order to consider the next-to-eikonal (NEik) corrections one should take into account the corrections that are of the order of $1/\gamma_t$ at the level of the boosted background field. These corrections can arise from relaxing either of the three approximations stated above. 

For the purpose of going beyond the static limit of the target fields (referred to as assumption (i) in the above discussion) one includes the $x^-$ dependence of the background field which goes beyond the infinite Lorentz time dilation. Indeed this correction is treated as a gradient expansion around a common $x^-$ which gives an order $1/\gamma_t$ effect. 
 
Another way to relax the eikonal approximation is to go beyond the so called shockwave limit and consider a finite longitudinal width target along $x^+$ direction (referred to as assumption (ii) in the above discussion). In the rest of the manuscript, in order to make the power counting more intuitive, we consider a finite support from $-L^+/2$ to $L^+/2$ for the target fields with a total longitudinal width $L^+$ of the target. Under the large boost this length $L^+$ is of order $1/\gamma_t$ due to Lorentz contraction. Nevertheless, our results obtained with the assumption of a finite support for the background field remain valid if the background field goes to zero faster than a power for $x^+\rightarrow \pm \infty$. 

Finally, the last source of the NEik corrections is to include the interactions of the transverse components of the background field with the projectile partons. This would go beyond the assumption (iii) listed above. Typically, this corresponds to replacing an enhanced $\A^-(x)$ insertion with an non-enhanced $\A^j(x)$ along the $x^+$ direction which would provide a $1/\gamma_t$ correction compared to the eikonal contribution.   

The above discussion for the sources of the NEik corrections is valid in the presence of a pure gluon background field. There is yet another distinct source of NEik corrections which originates from the interaction of the projectile parton and the target via t-channel quark exchange, which can be accounted for by including a quark background field as well for the target. 
 
Under a boost of the target along $x^-$ direction with parameter $\gamma_t$, a current associated with the target (which can be color, flavor or baryon number for example) scale as  
\begin{align}
\label{eq:current_scaling}
J^-(x)\propto \gamma_t \, , \:\: J^j(x)\propto (\gamma_t)^0\, , \:\: J^+(x)\propto (\gamma_t)^{-1}
\end{align}
In order to understand the scaling behaviour of the quark background field $\Psi(x)$ of the target, it convenient to introduce the projections 
\begin{align}
&\Psi^{(-)}(x)\equiv \frac{\gamma^+\gamma^-}{2}\Psi(x)  \\
&\Psi^{(+)}(x)\equiv \frac{\gamma^-\gamma^+}{2}\Psi(x)
\end{align}
which are known to be good and bad components of $\Psi(x)$ respectively, for a left-moving target. The currents constructed as bilinears of $\Psi(x)$ have components that depends on the good and bad components introduce above which read
\begin{align}
\label{eq:projected_minus}
\overline{\Psi}(x)\gamma^-\Psi(x)&=\overline{\Psi^{(-)}}(x)\gamma^-\Psi^{(-)}(x) 
 \\
\overline{\Psi}(x)\gamma^j\Psi(x)&=\overline{\Psi^{(-)}}(x)\gamma^j\Psi^{(+)}(x) + \overline{\Psi^{(+)}}(x)\gamma^j\Psi^{(-)}(x)
 \\
\overline{\Psi}(x)\gamma^+\Psi(x)&=\overline{\Psi^{(+)}}(x)\gamma^-\Psi^{(+)}(x) 
\end{align} 
Such currents associated with the target have to follow the same scaling behaviour introduced in Eq. \eqref{eq:current_scaling}, so that the components quark background field scale as  
\begin{align}
&\Psi^{(-)}(x)\propto (\gamma_t)^{\frac{1}{2}} \nn\\
& \Psi^{(+)}(x)\propto (\gamma_t)^{-\frac{1}{2}}
\end{align}
under a large boost of the target with $\gamma_t$. The quark background field does not contribute at eikonal order, and thanks to these scaling properties, only the enhanced components $\Psi^{(-)}(x)$ can contribute to NEik corrections, whereas the suppressed components $\Psi^{(+)}(x)$ start to contribute only at  next-to-next-to-eikonal (NNEik) accuracy.

% With this scaling properties only the current  constructed as bilinears of $\Psi(x)$ given in Eq. \eqref{eq:projected_minus} is of ${\cal O}(\gamma_t)$ and contributes to NEik corrections.  The components $\Psi^{(+)}(x)$ would contribute only to the higher order corrections starting at next-to-next-to-eikonal (NNEik) accuracy. 

The studies to include NEik corrections in the CGC calculations have been quite advanced over the last decade. The first studies accounting for the finite longitudinal width of the target in the gluon propagator are performed in \cite{Altinoluk:2014oxa} at NEik accuracy and in \cite{Altinoluk:2015gia} at NNEik accuracy. The results of these studies are used to compute many different observables beyond eikonal approximation. The effects of NEik corrections on particle production and correlations are studied both for dilute-dilute \cite{Altinoluk:2015xuy,Agostini:2019avp,Agostini:2019hkj} and in dilute-dense \cite{Agostini:2022ctk,Agostini:2022oge} collisions. The NEik corrections to the quark propagator that stems from the finite-longitudinal width of target and the interactions with the transverse component of the background field  are studied in \cite{Altinoluk:2020oyd}. A similar study is performed for the scalar propagator in \cite{Agostini:2023cvc}. In \cite{Altinoluk:2021lvu}, NEik corrections that are originating from dynamics of the target by going beyond the static limit of the target fields are studied both for scalar and quark propagators. The results are used to compute the DIS dijet production at full NEik accuracy in a dynamical gluon background both for a dense target in \cite{Altinoluk:2022jkk} and for a dilute target in \cite{Agostini:2024xqs}. On the other hand, the NEik effects stemming from including the interaction of the projectile partons with quark background field via t-channel quark exchanges are studied to compute quark-gluon dijet production in DIS in \cite{Altinoluk:2023qfr}. In this study, the back-to-back limit of the produced jets is considered in order to probe the quark TMDs starting from  CGC calculations with NEik corrections. Last but not least, the back-to-back limit of the DIS dijet production at NEik order is studied in \cite{Altinoluk:2024zom} in order to probe various different gluon TMDs beyond the eikonal and the leading twist approximations.  Apart from the above mentioned works that focus on the derivation of the NEik corrections to parton propagators and their applications to various scattering processes, in \cite{Kovchegov:2015pbl,Kovchegov:2016zex,Kovchegov:2016weo,Kovchegov:2017jxc,Kovchegov:2017lsr,Kovchegov:2018znm,Kovchegov:2018zeq,Kovchegov:2020kxg,Kovchegov:2020hgb,Adamiak:2021ppq,Kovchegov:2021lvz,Kovchegov:2021iyc,Cougoulic:2022gbk,Kovchegov:2022kyy,Borden:2023ugd,Kovchegov:2024aus} quark and gluon helicity evolutions as well as observables such as single and/or double spin asymmetries are computed at NEik accuracy.  In \cite{Cougoulic:2019aja,Cougoulic:2020tbc}, helicity dependent extensions of the CGC have been studied at NEik accuracy. In \cite{Balitsky:2015qba,Balitsky:2016dgz,Balitsky:2017flc}, the rapidity evolution of transverse momentum dependent parton distributions (TMDs) have been computed in a way that would be valid both at moderate and at low values of the momentum fraction $x$. A similar idea is pursued to study the interpolation between the moderate and low values of $x$ both for inclusive DIS \cite{Boussarie:2020fpb,Boussarie:2021wkn} and also for exclusive Compton scattering processes \cite{Boussarie:2023xun}. In \cite{Chirilli:2018kkw,Chirilli:2021lif}, the subeikonal corrections to both quark and gluon propagators are computed in high-energy operator product expansion (OPE) formalism. Subeikonal corrections in the CGC are studied in an effective Hamiltonian approach in \cite{Li:2023tlw,Li:2024fdb,Li:2024xra}. Finally, subeikonal corrections are also investigated in \cite{Jalilian-Marian:2017ttv,Jalilian-Marian:2018iui,Jalilian-Marian:2019kaf} by introducing an approach that allows longitudinal momentum exchange between the projectile and the target during the interaction. Last but not least, the effects of subeikonal corrections are also studied in the context of orbital angular momentum \cite{Hatta:2016aoc,Kovchegov:2019rrz,Boussarie:2019icw, Kovchegov:2023yzd, Kovchegov:2024wjs}.  

In this paper, we present a comprehensive study of both quark and gluon propagators in a dynamical gluon background field at NEik accuracy, and compute various parton production cross sections in parton-nucleus scatterings at forward rapidity at NEik accuracy.  The outline of the paper is as follows. In Section \ref{Sec:before-to-after-gluon_prop}, we derive the before-to-after gluon propagator in the presence of dynamical gluon background field at NEik accuracy. In section \ref{sec:Various_parton_prop}, we present the expressions of various parton propagators with their starting positions either before or inside the medium and final positions either inside or after the medium. Section \ref{sec:gluon_production_in_gA} is devoted to the computation of the forward gluon production cross section in gluon-nucleus scattering at NEik accuracy. Similarly, Section \ref{sec:quark_in_quark-nucleus} is devoted to the computation of the forward quark production cross section in quark-nucleus scattering at NEik accuracy. Sections \ref{sec:g_prod_in_qA} and \ref{sec:q_prod_in_gA} are presenting further applications of the computed propagators. In Section \ref{sec:g_prod_in_qA}, the forward gluon production cross section in quark-nucleus scattering and in Section \ref{sec:q_prod_in_gA} the forward quark production cross section in gluon-nucleus scattering are computed at NEik accuracy. Finally, in Section \ref{sec:outlook} we summarize our results and present an outlook. Moreover, in Appendix \ref{app:derivation_NEik_pure_A-}, we provide the details of the derivation of the NEik corrections beyond the shockwave approximation for the before-to-after gluon propagator. Appendix \ref{App:inside-to-inside-quark} is devoted to the presentation of the derivation of the inside-to-inside quark propagator. Finally, in Appendix \ref{App:inside-to-after-gluon-prop} we present the derivation of the inside-to-after gluon propagator.

%%%%%%%%%%%%%%%%%%%%%%%%%%%%%%%%%%%%%%%%%%%%%
%%%%%%%%%%%%%%%%%%%%%%%%%%%%%%%%%%%%%%%%%%%%%
\section{Gluon Propagator through a shockwave at NEik accuracy}
\label{Sec:before-to-after-gluon_prop}
This section is devoted to the computation of the gluon propagator at NEik accuracy in a highly boosted gluon background field (along the $x^-$ direction), focusing in particular on the case of propagation from a point located before the support of the background field to a point after it, along the $x^+$ direction\footnote{
We use the metric signature $(+,-,-,-)$. We use $x^\mu$ for a Minkowski 4-vector. In a light-cone basis we have
$
x^\mu=(x^+,{\bf x},x^-)
$
where $x^\pm=(x^0\pm x^3)/\sqrt{2}$ and ${\bf x}$ denotes a transverse vector with components $x^i$. %Similarly, for a 4-momentum we use $k^\mu$. 
We will also use the notations 
$
\ux = (x^+,{\bf x})
$
and 
$\uk = (k^+,{\bf k})$.}. 
We start with the derivation of such before-to-after gluon propagator at eikonal order but keeping the $z^-$ dependence of the target fields. Therefore, it goes beyond the static approximation for the target fields and it is referred to as either "Eikonal contribution with $z^-$ dependence" or "Generalized Eikonal contribution" as in our previous works. 
%In our setup, we consider a dense target boosted in the $x^-$ direction. 
The total before-to-after gluon propagator in a dynamic (meaning non-static) gluon background field at NEik accuracy can be schematically written as 
\begin{align}
G^{\mu \nu}_{F}(x,y)\big|_{\rm NEik} =  G^{\mu \nu}_{0,F}(x,y) + \delta G^{\mu \nu}_{F}(x,y)\big|_{\rm NEik}
\end{align} 
with $\delta G^{\mu \nu}_{F}(x,y)\big|_{\rm NEik}$ including both the medium corrections at eikonal order with $z^-$ dependence and the other NEik contributions to medium corrections, while $G^{\mu \nu}_{0,F}(x,y)$ is the vacuum gluon propagator, written in coordinate space. 

The vacuum gluon propagator in light-cone gauge $A^+=0$, when written in momentum space, reads 
\be
 \label{G0}
  \tilde{G}^{\mu\nu}_{0,F}(p) = \frac{i}{p^{2}+i\epsilon}\left[-g^{\mu\nu} + \frac{p^{\mu}\eta^{\nu} + \eta^{\mu}p^{\nu}}{p\cdot \eta}\right]
\ee
with $\eta^{\mu}$ defined as $\eta^{\mu} = g^{\mu+}$, so that  $\eta \cdot p = p^{+}$
and $\eta^{2}=0$. Since we are in the $ \eta\!\cdot\! A \equiv A^+=0$ light cone gauge, one has 
\begin{align}
G^{\mu +}_{0,F}(p) = G^{+\nu}_{0,F}(p)=0
\label{zero_propag_with_plus_index}
\end{align}
In the rest of this section, we compute the medium corrections to the before-to-after gluon propagator in a gluon background field both at Generalized Eikonal order and full NEik order.

%%%%%%%%%%%%%%
 \begin{figure}
    \centering
    \includegraphics[scale=0.5]{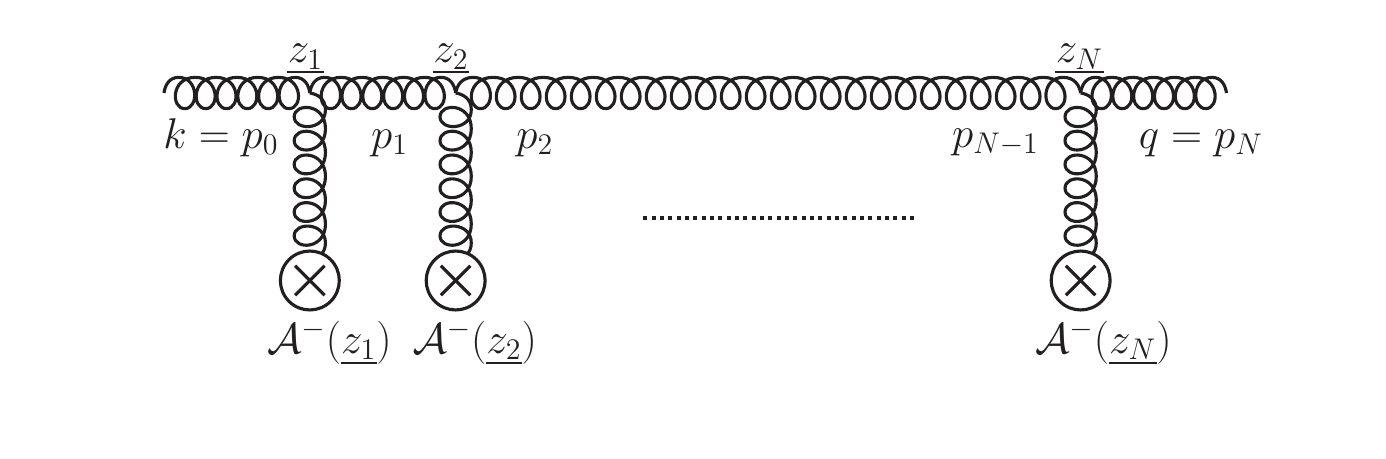}
    \caption{Contribution to the gluon propagator in background field, with insertions of the $\A^{-}$ component via three gluon vertices resummed to all orders.}
    \label{fig: multiple insertions}
\end{figure}
%%%%%%%%%%%%%%%%%%%%%%%%%%%%%%%%%%%%%%%%%%%%%%%%%%%%%%%%%%%%%%%%%%%%%%%%
\subsection{Gluon propagator in a pure ${\cal A}^-$ background beyond the static limit}
%%%%%%%%%%
%%%

A strong gluon background field can be as large as $\A^{\mu}(z)\sim 1/g$, corresponding to the nonlinear regime of QCD.
In order to compute a gluon propagator in gluon background field, valid even in the nonlinear regime, one thus need 
%before-to-after gluon propagator in a dynamical gluon background field, one needs 
to resum multiple interactions of the type $\big(g\A^{\mu}(z)\big)^n$ to all orders in $n$, see Fig. \ref{fig: multiple insertions}. At eikonal order, due to the hierarchy between the components of a highly boosted background field given in Eq. \eqref{eq:g_field_hierarchy}, one actually needs to resum only the $\A^-(z)$ field component insertions, which corresponds to Fig.~\ref{fig: multiple insertions}.\footnote{\label{fnote_4gV}A priori, in addition to the diagrams like Fig.~\ref{fig: multiple insertions}, one could think of diagrams including four gluon vertices as well, corresponding to a local double insertion of the background field on the gluon propagator. However, in the light-cone gauge, such diagrams are found to vanish by numerator algebra for a $\A^-(z)\A^-(z)$ insertion. They vanish even if only one of the two background fields inserted is a $\A^-$ component, the other one being a transverse component.} A similar kind of resummation of multiple interactions with the $\A^-(z)$ fields have been introduced in \cite{Altinoluk:2020oyd} and \cite{Altinoluk:2021lvu} to compute the quark and scalar propagators at eikonal order with a $z^-$ dependent background gluon field. In momentum space, the medium contribution to the gluon propagator with $n$ insertions of the background field $\A^-$ reads,
\begin{align}
\label{ninsertions}
&\delta\tilde{G}_{F}^{\mu\nu}(p_{N},p_{0}) \Big|_{{\rm pure}\, {\cal A}^-, \, z^-} 
= 
\sum_{N=1}^{+\infty} \left( \prod_{n=1}^{N-1} \int \! \frac{d^{4}p_{n}}{(2\pi)^{4}}  \right) \tilde{G}_{0,F}^{\mu\mu_{N}}(p_{N})\; 
 \mathcal{P}_{n} \Bigg\{ \prod_{n=1}^{N} \int \! d^{4}z_{n} \ e^{i(p_{n} - p_{n-1})z_{n}}
 \left[ -ig\A^{-}(z_{n})\cdot T\right] 
\nn\\
&
%\hspace{-2cm}
\times
\bigg[
-(p_{n}^{+} + p_{n-1}^{+}) g_{\mu_{n}\nu_{n}}
+(2{p_{n-1}}_{\mu_{n}}\!-\!{p_{n}}_{\mu_{n}}){g_{\nu_{n}}}^+
+(2{p_{n}}_{\nu_{n}}\!-\!{p_{n-1}}_{\nu_{n}}){g_{\mu_{n}}}^+
\bigg]
\Bigg\}
 \left( \prod_{n=1}^{N-1} \ \tilde{G}_{0,F}^{\nu_{n\!+\!1} \mu_{n}}(p_{n}) \right) \tilde{G}_{0,F}^{\nu_{1}\nu}(p_{0})    
 \, .
\end{align}
Here, ${\cal P}_n$ stands for ordering of the adjoint color generators $T^{a_n}$ from right to left with increasing index n.
% along $z^+$ direction. 
Using the property \eqref{zero_propag_with_plus_index} of the gluon propagator in light-cone gauge, it is clear that among the three terms in the bracket in the second line of Eq.~\eqref{ninsertions} (corresponding to the three gluon vertex), only the first one survives. Moreover, most of the terms dependent on $\eta$ in the gluon propagators in Eq.~\eqref{ninsertions} then drop for similar reasons.
In such a way, the expression \eqref{ninsertions} can be simplified into
\begin{align}
\label{ninsertions_2}
\delta\tilde{G}_{F}^{\mu\nu}(p_{N},p_{0}) \Big|_{{\rm pure}\, {\cal A}^-, \, z^-} 
= &
\sum_{N=1}^{+\infty} \left( \prod_{n=1}^{N-1} \int \! \frac{d^{4}p_{n}}{(2\pi)^{4}}  \right) 
\left( \prod_{n=0}^{N} \frac{i}{(p_n^{2}+i\epsilon)}  \right) 
 \mathcal{P}_{n} \Bigg\{ \prod_{n=1}^{N} \int \! d^{4}z_{n} \ e^{i(p_{n} - p_{n-1})z_{n}}
 \left[ -ig\A^{-}(z_{n})\cdot T\right] 
\Bigg\}
\nn\\
&
%\hspace{-2cm}
\times
\left[-g^{\mu\mu_{N}} + \frac{ \eta^{\mu}p_{N}^{\mu_{N}}}{p_{N}\cdot \eta}\right]
(- g_{\mu_{N}\nu_{1}})
\left[-g^{\nu_{1}\nu} + \frac{p_0^{\nu_{1}}\eta^{\nu} }{p_0\cdot \eta}\right]
\bigg[\prod_{n=1}^{N} \big(p_{n}^{+} + p_{n-1}^{+} \big) \bigg]    
 \, .
\end{align}
For a given momentum 4-vector $k^{\mu}$, let us introduce the notation $\check{k}^{\mu}$ for its on-shell analog. More precisely, it is defined in such a way that their $+$ and transverse components coincide, $\check{k}^{+}={k}^{+}$ and $\check{\k}=\k$, whereas the $-$ component of $\check{k}^{\mu}$ is adjusted to make it on-shell, i.e. $\check{k}^-=\k^2/(2k^+)$ for a gluon. In particular, we thus have 
\begin{align}
k^{\mu} =&\, \check{k}^{\mu} + \frac{k^2}{2k^+} g^{\mu +}=\check{k}^{\mu} + \frac{k^2}{2k^+} \eta^{\mu}
\label{on_vs_off_shell_momentum}
\, .
\end{align}
Using that decomposition for $p_{N}^{\mu_{N}}$ and $p_0^{\nu_{1}}$ in Eq.~\eqref{ninsertions_2}, one finds 
\begin{align}
\left[-g^{\mu\mu_{N}} + \frac{ \eta^{\mu}p_{N}^{\mu_{N}}}{p_{N}\cdot \eta}\right]
(- g_{\mu_{N}\nu_{1}})
\left[-g^{\nu_{1}\nu} + \frac{p_0^{\nu_{1}}\eta^{\nu} }{p_0\cdot \eta}\right]
=&\,
-\left[
%-{g^{\mu}}_{\nu_{1}} + \frac{ \eta^{\mu}}{p_{N}^+} 
%\left(\check{p_{N}}_{\nu_{1}}
%+ \frac{{p_{N}}^2\, \eta_{\nu_{1}}}{2p_{N}^+} \right)
-{g^{\mu}}_{\nu_{1}} + \frac{ \eta^{\mu}}{p_{N}^+} \check{p_{N}}_{\nu_{1}}
+ \frac{ \eta^{\mu}\eta_{\nu_{1}}}{2(p_{N}^+)^2}\, {p_{N}}^2
\right]
\left[
-g^{\nu_{1}\nu} 
+ \check{p_0}^{\nu_{1}}\frac{\eta^{\nu} }{p_0^+}
+{p_0}^2\frac{\eta^{\nu_{1}}\eta^{\nu} }{2(p_0^+)^2}
\right]
\nonumber\\
=&\,
-\left[
-{g^{\mu}}_{\nu_{1}} + \frac{ \eta^{\mu}}{p_{N}^+} \check{p_{N}}_{\nu_{1}}
\right]
\left[
-g^{\nu_{1}\nu} 
+ \check{p_0}^{\nu_{1}}\frac{\eta^{\nu} }{p_0^+}
\right]
\nonumber\\
=&\,
\left[
-{g^{\mu\nu}} + \frac{ \eta^{\mu}}{p_{N}^+} \check{p_{N}}^{\nu}
+ \check{p_0}^{\mu}\frac{\eta^{\nu} }{p_0^+} 
- (\check{p_{N}} \cdot  \check{p_0}) \, \frac{\eta^{\mu}\eta^{\nu} }{p_{N}^+ p_0^+} 
\right]
\, .
\end{align}
Note that this result is expressed in terms of $\check{p_{N}}$ and $ \check{p_0}$, so that it is independent of ${p_{N}^-}$ and ${p_0^-}$.
In order to obtain the medium correction to gluon propagator in position space, we Fourier transform the expression  \eqref{ninsertions_2} as
\begin{align}
\label{Fourier transform G}
\delta G_{F}^{\mu\nu}(x,y) &=  \int \! \frac{d^{4}p_{N}}{(2\pi)^{4}} \int \! \frac{d^{4}p_{0}}{(2\pi)^{4}} e^{-i(x\cdot p_{N} - y\cdot p_{0})}  \ \delta\tilde{G}_{F}^{\mu\nu}(p_{N},p_{0})
\, ,
\end{align}
leading to
\begin{align}
\label{ninsertions_3}
\delta G_{F}^{\mu\nu}(x,y) \Big|_{{\rm pure}\, {\cal A}^-, \, z^-} 
= &
\sum_{N=1}^{+\infty}
{\cal P}_n \Bigg[ \prod_{n=1}^{N} \int \! d^{4}z_{n} \big[-ig\A^{-}(z_{n})\cdot T \big] \Bigg]  
 \left( \prod_{n=0}^{N} \int \! \frac{d^{4}p_{n}}{(2\pi)^{4}}\, \frac{i}{(p_n^{2}+i\epsilon)}\, e^{-i(z_{n+1}- z_{n})\cdot p_{n}}  \right) 
\nn\\
&
%\hspace{-2cm}
\times
\left[
-{g^{\mu\nu}} + \frac{ \eta^{\mu}}{q^+} \check{q}^{\nu}
+ \check{k}^{\mu}\frac{\eta^{\nu} }{k^+} 
- (\check{q} \cdot  \check{k}) \, \frac{\eta^{\mu}\eta^{\nu} }{q^+ k^+} 
\right]
\bigg[\prod_{n=1}^{N} \big(p_{n}^{+} + p_{n-1}^{+} \big) \bigg]    
 \, ,
\end{align}
with the notations $p_{0} \equiv k$, $p_{N} \equiv q$, $z_{0} \equiv y$ and $ z_{N+1} \equiv x$.
At this stage, the integration over each $p_n^-$ can be performed, using the residue theorem, as
\begin{align}
\int \frac{d p_n^-}{2\pi} \frac{i}{p_n^2+i\epsilon}\, e^{-i(z_{n+1}^+-z_n^+)p_n^-}
=&\, 
\frac{1}{2p_n^+}
\Big[ \theta(p_n^+)\theta(z_{n+1}^+-z_n^+)-\theta(-p_n^+)\theta(z_n^+-z_{n+1}^+)\Big] e^{-i(z_{n+1}^+-z_n^+)\cp_n^-}
\, ,
\end{align}
and the result can be written as

\begin{align}
\delta G_{F}^{\mu\nu}(x,y)\Big|_{{\rm pure}\,  \A^-, \, z^-} &= \sum_{N=1}^{+\infty} 
\Bigg[ \prod_{n=1}^{N} \int \! d^{3}\underline {z_{n}} \Bigg]  
\prod_{n=0}^{N}  \int \! \frac{d^2\p_n}{(2 \pi)^2} \ e^{i(\z_{n+1} - \z_n)\p_n} 
\int \! \frac{dq^+}{2\pi}\frac{e^{-i(x^+-z_N^+)\cq^-}}{2q^{+}} 
\int \! \frac{dk^+}{2\pi}\frac{e^{-i(z_1^+-y^+)\ck^-}}{2k^{+}}   
\nn\\ &\times 
\Big[\theta(q^{+}) \theta(x^{+}-z_{N}^{+}) - \theta(-q^{+}) \theta(z_{N}^{+} - x^{+})\Big]
\Big[\theta(k^{+}) \theta(z_{1}^{+}- y^{+}) - \theta(-k^{+}) \theta(y^{+} - z_{1}^{+})\Big]
\nn\\ &\times
\Big[ -g^{\mu \nu} + \frac{\check{k}^{\mu} \eta^{\nu}}{k^{+}} + \frac{\eta^{\mu} \check{q}^{\nu}}{q^{+}} - \frac{\eta^{\mu} \eta^{\nu}}{q^{+} k^{+}} (\check{q} \cdot \check{k})   \Big] \ I_{p^+} 
\, ,
\end{align}
where
\begin{align}
\label{int_p_plus}
I_{p^{+}} &= \int \! \Bigg[ \prod_{n=1}^{N-1}  \frac{dp_{n}^{+}}{2\pi}   \ 
\frac{1}{2p^{+}_{n}} \ e^{\frac{-i(z_{n+1}^{+} -z_{n}^{+})\p_{n}^{2}}{2p_{n}^{+}}}
\Big[ \theta(z_{n+1}^{+} - z_{n}^{+})\ \theta(p_{n}^{+})   -\theta(z_{n}^{+} - z_{n+1}^{+})\ \theta(-p_{n}^{+}) \Big] \Bigg]
%\nn \\ & \times 
\Bigg[\prod_{n=1}^{N} (p_{n}^{+} + p_{n-1}^{+})\Bigg]\ I_{z^{-}}
\end{align}
and
\begin{align}
\label{int_z_minus}
I_{z^{-}} = \int \! \bigg[ \prod_{n=1}^{N} \ dz_{n}^{-}\bigg] \bigg[\mathcal{P}_{n} \prod_{n=1}^{N}(-ig\A^{-}(z_{n})\cdot T )\bigg] e^{-i(x^{-}-z_{N}^{-})q^{+}} \ 
e^{-i(z_{1}^{-}-y^{-})k^{+}} 
\bigg[ \prod_{n=1}^{N-1} e^{-i(z^{-}_{n+1}-z_{n}^{-})p_{n}^{+}} \bigg]
\, .
\end{align}
The integrals that appear in Eqs. \eqref{int_p_plus} and  \eqref{int_z_minus} are the same integrals (Eqs. (D1)-(D3)) that were studied in detail in Appendix D of \cite{Altinoluk:2021lvu}. Therefore, to perform the integrations over $z_n^-$ and $p_n^+$ in Eqs. \eqref{int_p_plus} and  \eqref{int_z_minus}, we follow the same calculation as in \cite{Altinoluk:2021lvu}, performing a gradient expansion of the background fields $\A^{-}(z_{n})$  along the minus direction around a common value $z^-$. As a result, we find
 \begin{align}
 \label{with_cross_terms_z-}
 &
 \delta G_{F}^{\mu\nu}(x,y)\Big|_{{\rm pure}\,  \A^-,\,  z^-} = \sum_{N=1}^{+\infty}   
 \Bigg[ \prod_{n=1}^{N} \int \! d^{3}\underline{z_{n}} \Bigg]  
 \Bigg[ \prod_{n=0}^{N}  \int \! \frac{d^2\p_n}{(2 \pi)^2} \ e^{i(\z_{n+1}  - \z_n)\p_n} \nn \\
 &\times 
 \int \! \frac{dq^+}{2\pi}  \ e^{-i\frac{(x^{+} - z_{N}^{+})\q^2}{2q^{+}}} 
 \Big[\theta(q^{+}) \theta(x^{+}-z_{N}^{+}) - \theta(-q^{+}) \theta(z_{N}^{+} - x^{+})\Big]
 \nn\\
 &\times 
 \int \! \frac{dk^+}{2\pi}  \ e^{-i\frac{(z_{1}^{+} - y^{+})\k^2}{2k^{+}}} 
 \Big[\theta(k^{+}) \theta(z_{1}^{+}- y^{+}) - \theta(-k^{+}) \theta(y^{+} - z_{1}^{+})\Big]\nn \\
 &\times 
 \frac{1}{q^+ + k^+} \Bigg[\prod_{n=1}^{N-1} e^{-i\frac{(z^{+}_{n+1} -z_{n}^{+})\p^2_n}{(q^{+}+ k^{+})}} 
 \Big[\theta(q^{+}+ k^{+}) \theta(z_{n+1}^{+}-z_{n}^{+}) - \theta(-q^{+}-k^{+}) \theta(z_{n}^{+} - z_{n+1}^{+})\Big]\Bigg] \nn \\
&\times
\int\! dz^{-} \ e^{-iq^{+}(x^{-}-z^{-})} \ e^{-ik^{+}(z^{-}-y^{-})}
\Bigg[\mathcal{P}_{n} \prod_{n=1}^{N} \Big(-igA^{-}(\underline{z_{n}},z^-) \cdot T\Big)\Bigg]
\bigg[ -g^{\mu \nu} + \frac{\check{k}^{\mu} \eta^{\nu}}{k^{+}} + \frac{\eta^{\mu} \check{q}^{\nu}}{q^{+}} - \frac{\eta^{\mu} \eta^{\nu}}{q^{+} k^{+}} (\check{q} \cdot \check{k})   \bigg] 
\end{align}
up to next-to-next-to-eikonal corrections. In Eq. \eqref{with_cross_terms_z-}, if one selects $\theta(q^+)$ in the second line and $\theta(k^+)$ in the third line, then only $\theta(q^++k^+)$ term survives in the third line. Similarly, by choosing $\theta(-q^+)$ in the second line and $\theta(-k^+)$ in the third line, then one can show that only $\theta(-q^+-k^+)$ term survives in the third line. As discussed in \cite{Altinoluk:2021lvu}, the cross terms stemming from choosing  $\theta(q^+)$ and  $\theta(-k^+)$ choosing  $\theta(-q^+)$ and  $\theta(-k^+)$ can be shown to correspond to zero modes with $q^+=k^+=0$, which are not relevant for our Eikonal expansion and therefore can be neglected. Thus, one obtains
\begin{align}
\label{Without shockwave approx G}
&
\delta G_{F}^{\mu\nu}(x,y)\Big|_{{\rm pure}\,  \A^-,\,  z^-} = \sum_{N=1}^{+\infty}  
\Bigg[ \prod_{n=1}^{N} \int \! d^{3}\underline{z_{n}} \Bigg]  
\Bigg[ \prod_{n=0}^{N}  \int \! \frac{d^2\p_n}{(2 \pi)^2} \ e^{i(\z_{n+1} - \z_{n})\p_n } \Bigg] \nn \\
&\times 
\int \! \frac{dq^+}{2\pi} \int \! \frac{dk^+}{2\pi}  \ \frac{1}{q^+ + k^+}\ 
\bigg[ -g^{\mu \nu} + \frac{\check{k}^{\mu} \eta^{\nu}}{k^{+}} + \frac{\eta^{\mu} \check{q}^{\nu}}{q^{+}} - \frac{\eta^{\mu} \eta^{\nu}}{q^{+} k^{+}} (\check{q} \cdot \check{k})   \bigg] 
%\nn \\
%& \times
e^{-i\frac{(x^{+} - z_{N}^{+})\q^2}{2q^{+}}}  
\Bigg[\prod_{n=1}^{N-1} e^{-i\frac{(z^{+}_{n+1} -z_{n}^{+})\p_n^2}{(q^+ + k^+)}} \Bigg]  
e^{-i\frac{(z_{1}^{+} - y^{+})\k^2}{2k^{+}}} 
\nn \\
&\times 
\Bigg[\theta(q^{+}) \theta( k^{+}) \bigg[ \prod_{n=0}^{N}\theta(z_{n+1}^{+}-z_{n}^{+}) \bigg] + (-1)^{N+1} \theta(-q^{+}) \theta(-k^{+}) \bigg[ \prod_{n=0}^{N}\theta(z_{n}^{+} - z_{n+1}^{+}) \bigg]\Bigg] \nn \\
&\times 
\int\! dz^{-} \ e^{-iq^{+}(x^{-}-z^{-})} \ e^{-ik^{+}(z^{-}-y^{-})} \bigg[\mathcal{P}_{n} \prod_{n=1}^{N} \Big(-igA^{-}(\underline{z_{n}},z^-) \cdot T\Big)\bigg] + \text{NNEik}
\end{align}
The distance between two successive background field insertions along the $+$ direction, i.e. $z_{n+1}^+-z_n^+$, is suppressed by Lorentz contraction under large boost of the target. 
The shockwave approximation then amounts  to neglecting $z_n^+$ (from $1$ to $N$) in the phase factors in Eq.~\eqref{Without shockwave approx G}. In such a way, one neglects contributions of order NEik, or more suppressed. In that approximation, the integration over the intermediate transverse momenta $\p_n$ can be performed trivially and the result reads 
\begin{align}
\label{With shockwave approx G}
&
\delta G_{F}^{\mu\nu}(x,y)\Big|_{\text{pure}\,  \A^-, z^-} = \sum_{N=1}^{+\infty} \  
\Bigg[ \prod_{n=1}^{N} \int \! d^{3}\underline{z_{n}} \Bigg]  \Bigg[ \prod_{n=1}^{N-1}  \delta^{2}({\z_{n}}- {\z_{n+1}}) \Bigg] \nn \\
&\times 
\int \! \frac{d^{3}\uq}{(2\pi)^{3}} \int \! \frac{d^{3}\uk}{(2\pi)^{3}} 
 e^{i(\x - \z_{N})\cdot \q} 
 \ e^{-i(\y - {\z_{1}})\cdot \k} \ e^{-ix^+\check{q}^{-}} \ e^{i y^{+}\check{k}^{-} } 
  \bigg[ - g^{\mu \nu} + \frac{\check{k}^{\mu} \eta^{\nu}}{k^{+}} + \frac{\eta^{\mu} \check{q}^{\nu}}{q^{+}} - \frac{\eta^{\mu} \eta^{\nu}}{q^{+} k^{+}} (\check{q} \cdot \check{k})   \bigg]
 \nn \\&
\times \, \frac{1}{q^+ +k^+}  \Bigg[ \prod_{n=0}^{N}   \bigg[ \theta(z_{n+1}^{+} - z_{n}^{+})\bigg] \theta(q^{+}) \theta(k^+)
+(-1)^{N+1} \prod_{n=0}^{N} \bigg[ \theta(z_{n}^{+} - z_{n+1}^{+}) \bigg] \theta(-q^{+}) \theta(-k^+) \Bigg] \nn \\
& \times  \int \! dz^{-} \  e^{-iq^{+}(x^{-} -z^{-})} \ e^{-ik^{+}(z^{-} - y^{-})}      
 \bigg[ \mathcal{P}_{n} \prod_{n=1}^{N} \big( -igA^{-}(\underline{z_{n}}, z^{-}) \cdot T \big) \bigg] + \text{NEik}
 \, .
\end{align}
All but one of the integrations over $\z_n$ now can be performed trivially by using the $(N-1)$ delta functions. The integration variable of that leftover transverse integral is then noted $\z$, and the resulting expression can be organized as 
\begin{align}
\label{delta_G_bef_Wilson}
\delta G_{F}^{\mu\nu}(x,y)&\Big|_{\text{pure}\, \A^-,\, z^-} = 
 \int \! \frac{d^{3}\uq}{(2\pi)^{3}} e^{-ix \cdot \check{q}}  
\int \! \frac{d^{3}\uk}{(2\pi)^{3}}  e^{iy \cdot \check{k}}   
\frac{1}{q^{+}+k^{+}} 
%\\&
%\times
 \bigg[- g^{\mu \nu} + \frac{\check{k}^{\mu} \eta^{\nu}}{k^{+}} + \frac{\eta^{\mu} \check{q}^{\nu}}{q^{+}} - \frac{\eta^{\mu} \eta^{\nu}}{q^{+} k^{+}} (\check{q} \cdot \check{k})  \bigg] 
 \nn \\
 & \times
\int \! d^{2}\z \, e^{-i(\q - \k)\cdot\z}  \ 
\int \! dz^{-}   e^{i(q^{+} -k^{+})z^{-}} \   
%\\& \times      
\sum_{N=1}^{+\infty} \     \int \!     \bigg[\prod_{n=1}^{N} dz_{n}^{+} \  \mathcal{P}_{n}  \Big( -igA^{-}(z_{n}^+,\z, z^{-}) \cdot T \Big) \bigg] 
 \nn \\
 &\times  
 \bigg\{ \prod_{n=0}^{N}   \Big[ \theta(z_{n+1}^{+} - z_{n}^{+})\Big] \theta(q^{+}) \theta(k^+)  +( -1)^{N+1} \prod_{n=0}^{N} \Big[ \theta(z_{n}^{+} - z_{n+1}^{+}) \Big] \theta(-q^{+}) \theta(-k^+)  \bigg\} 
 + \text{NEik}
 \end{align}
Using the definition of a Wilson line
\begin{align}
\label{def:Wilson_R}
U_{R}(x^{+}, y^{+}; \z) & = 1 + \sum_{N=1}^{\infty} \frac{1}{N!} \mathcal{P}_{+}\left[-ig \int_{y^{+}}^{x^{+}} \! dz^{+} A^{-}(z^{+},\z) \cdot T_{R}\right]^{N}       
\end{align}
 in a representation $R$, one can rewrite Eq. \eqref{delta_G_bef_Wilson} in terms of an adjoint Wilson line and its conjugate as 
\begin{align}
\label{mediumcorrection}
\delta G_{F}^{\mu\nu}(x,y) \Big|_{\text{pure}\, \A^-,\, z^-} & =   \int \! \frac{d^{3}\uq}{(2\pi)^{3}} \int \! \frac{d^{3}\uk}{(2\pi)^{3}} \   \frac{ e^{-ix \cdot \check{q} } \ e^{iy \cdot \check{k}}}{q^{+}+ k^+} \
\left[ -g^{\mu\nu} + \frac{\check{k}^{\mu} \eta^{\nu} }{k^{+}} + 
   \frac{\eta^{\mu} \check{q}^{\nu}}{q^{+}} - \frac{\eta^{\mu}\eta^{\nu}}{q^{+}k^{+}} (\check{q} \cdot \check{k}) \right]
     \nn \\
 &
% \hspace{-2cm}
 \times  
  \int \! d^{2}\z\, e^{-i(\q - \k)\cdot\z} \int \! dz^{-} e^{i(q^{+} -k^{+})z^{-}}  
  \nn\\
  &\times
\Big\{\theta(x^{+}- y^{+}) \theta(q^+)\theta(k^{+}) \big[  U_{A}(x^{+}, y^{+}; \z, z^-) -1 \big] 
 \nn \\
&
 \hspace{0.8cm}
- \theta(y^{+}- x^{+})\theta(-q^+) \theta(-k^{+}) \big[  U_{A}^{\dagger}(x^{+}, y^{+}; \z, z^-) -1 \big]  \Big\}
 + \text{NEik}
 \, .
 \end{align}
%
%
%{\color{red}minus sign!!!}
%
%
Eq. \eqref{mediumcorrection} is the medium correction to the gluon propagator in a dynamic gluon background field $\A^{-}_{\mu}(z)$, up to NEik corrections beyond the shockwave approximations that have been neglected. In that expression, the dependence of the Wilson lines on $z^-$ is an effect beyond the static approximation, and reflects the slow dependence on a typical $z^-$ for all of the background insertion vertices. Through that  $z^-$ dependence of the Wilson lines, NEik corrections (but not NNEik corrections) beyond the static approximations are accurately included. 

Since the $z^-$ dependence of the Wilson lines is the only difference of the expression \eqref{mediumcorrection} compared with the standard Eikonal result, 
the medium correction  \eqref{mediumcorrection} is referred to as generalized Eikonal as in Ref. \cite{Altinoluk:2021lvu}.
However, for convenience of the notation, in the rest of the manuscript we will also call this contribution "Eikonal with $z^-$ dependence" in the equations. As it will be discussed in more detail later, since it is a slow dependence, one can perform a gradient expansion around $z^-=0$ for the $z^-$ dependent terms. The zeroth order terms in this gradient expansion will be referred to as strict eikonal in the rest of the paper. 

One can further simplify the gluon propagator in a dynamical $\A^-$ background field. For that purpose, it is convenient to further examine the terms in $-1$ (instead of Wilson lines) in  Eq. \eqref{mediumcorrection}, that we can call ''vacuum-like terms'':
% the medium correction given in Eq. \eqref{mediumcorrection} can be written as 
%
%\begin{align}
%\label{medium_corr_with_non_int}
%&
%\delta G_{F}^{\mu\nu}(x,y) \Big|_{\text{pure}\, \A^-,\, z^-}  =   \int \! \frac{d^{3}\uq}{(2\pi)^{3}} \int \! \frac{d^{3}\uk}{(2\pi)^{3}} \   \frac{ e^{-ix \cdot \check{q} } \ e^{iy \cdot \check{k}}}{q^{+}+ k^+} \
%\left[ g^{\mu\nu} - \frac{\check{k}^{\mu} \eta^{\nu} }{k^{+}} - 
%   \frac{\eta^{\mu} \check{q}^{\nu}}{q^{+}} + \frac{\eta^{\mu}\eta^{\nu}}{q^{+}k^{+}} (\check{q} \cdot \check{k}) \right] \int \! d^{2}\z\, e^{-i(\q - \k)\cdot\z} 
%\nn \\
%& \times  
%\int \! dz^{-} e^{i(q^{+} -k^{+})z^{-}}  \Big[-\theta(x^{+}- y^{+}) \theta(q^+)\theta(k^{+})  U_{A}(x^{+}, y^{+}; \z, z^-) 
%+ \theta(y^{+}- x^{+})\theta(-q^+) \theta(-k^{+})  U_{A}^{\dagger}(x^{+}, y^{+}; \z, z^-)   \Big]
%\nn\\
%&+ \delta G_{F}^{\mu\nu}(x,y) \Big|_{\text{pure}\, \A^-,\, z^-}^{\rm non-int.}
%\end{align}
%%
%%
%with the non-interacting part being 
%
%
\begin{align}
\delta G_{F}^{\mu\nu}(x,y) &\Big|_{\text{pure}\, \A^-}^{\rm vacuum-like} = 
\int \! \frac{d^{3}\uq}{(2\pi)^{3}}  \  
\int \! \frac{d^{3}\uk}{(2\pi)^{3}} \ \frac{ e^{-ix \cdot \check{q} } \ e^{iy \cdot \check{k}}}{q^{+}+ k^+} \
\left[ - g^{\mu\nu} + \frac{\check{k}^{\mu} \eta^{\nu} }{k^{+}} + \frac{\eta^{\mu} \check{q}^{\nu}}{q^{+}} - \frac{\eta^{\mu}\eta^{\nu}}{q^{+}k^{+}} (\check{q} \cdot \check{k}) \right] 
\nn \\
&\times\   
   \int \! d^{2}\z\, e^{-i(\q - \k)\cdot\z}  \int \! dz^{-} \  e^{i(q^{+} -k^{+})z^{-}} \  (-1)
   \Big\{\theta(x^{+}- y^{+}) \theta(q^+)\theta(k^{+})
% \nn \\
%&
% \hspace{0.8cm}
- \theta(y^{+}- x^{+})\theta(-q^+) \theta(-k^{+})  \Big\}
\end{align}
Upon integration over $z^-$ and $\z$, one finds 
\begin{align}
\label{non_int_integrated}
\delta G_{F}^{\mu\nu}(x,y) \Big|_{\text{pure}\, \A^-}^{\rm vacuum-like} =& \int \! \frac{d^{3}\uk}{(2\pi)^{3}}  \
e^{-i(x-y) \cdot \check{k}}  \ \frac{1}{2k^{+}} 
 \bigg[- g^{\mu \nu} + \frac{\check{k}^{\mu} \eta^{\nu}}{k^{+}} + \frac{\eta^{\mu} \check{k}^{\nu}}{k^{+}}  \bigg]
 \nn \\
&\times\, 
% \hspace{0.8cm}
(-1)\Big\{\theta(x^{+}- y^{+}) \theta(k^{+})
- \theta(y^{+}- x^{+})\theta(-k^{+})  \Big\}
\, .
\end{align}
On the other hand, the vacuum gluon propagator given in Eq. \eqref{G0} can be written in coordinate space as 
\begin{align}
G_{0,F}^{\mu\nu}(x,y)  &=  \int \! \frac{d^{4}p}{(2\pi)^{4}} \ e^{-i(x-y)\cdot p} \frac{i}{(p^{2}+i\epsilon)}\left(-g^{\mu\nu} + \frac{p^{\mu}\eta^{\nu} + \eta^{\mu}p^{\nu}}{p\cdot \eta}\right) \nn\\
&= \int \! \frac{d^{4}p}{(2\pi)^{4}} \ e^{-i(x-y)\cdot p} 
\bigg[\frac{i}{(p^{2}+i\epsilon)}\left(-g^{\mu\nu} + \frac{\check{p}^{\mu}\eta^{\nu} + \eta^{\mu}\check{p}^{\nu}}{p^{+}}\right)  +i \frac{\eta^{\mu}\eta^{\nu}}{p^{+}p^{+}}\bigg]
\, ,
\end{align}
using the relation \eqref{on_vs_off_shell_momentum}. Upon integration over $p^-$, it reads 
\begin{align}
\label{vacuum contribution}
G_{0,F}^{\mu\nu}(x,y)  &=  \int \! \frac{d^{3}\up}{(2\pi)^{3}}  \frac{e^{-i(x-y)\cdot \check{p}}}{2p^{+}} 
\Big[\theta(p^{+})\theta(x^{+}-y^{+}) - \theta(-p^{+})\theta(y^{+}-x^{+}) \Big] 
\bigg( -g^{\mu\nu} +  \frac{\check{p}^{\mu}  \eta^{\nu} + \eta^{\mu} \check{p}^{\nu}}{p^{+}}\bigg) \nn\\
& + i \delta^{2}(\x - \y) \ \delta(x^{+}-y^{+}) \eta^{\mu} \eta^{\nu}  \int \! \frac{dp^{+}}{2\pi}  \frac{ e^{-i(x^{-}-y^{-})p^{+}}} {p^{+}p^{+}}
\, .
 \end{align}
The first line of Eq.~\eqref{vacuum contribution} identically cancels the vacuum-like contribution \eqref{non_int_integrated}.

The total gluon propagator at Eikonal order with $z^-$ dependence (i.e. Generalized Eikonal order) is defined as 
\begin{align}
\label{gen_eik_imp}
G_{F}^{\mu\nu}(x,y) \Big|_{\rm Eik,\ z^-}  = G_{0,F}^{\mu\nu}(x,y) + \delta G_{F}^{\mu\nu}(x,y)\Big|_{\text{pure}\, \A^-, z^-}
\end{align}
with the vacuum propagator $G_{0,F}^{\mu\nu}(x,y)$ is given in Eq. \eqref{vacuum contribution} and the medium correction in dynamical $\A^-$ background is given in Eq. \eqref{mediumcorrection}. Thanks to the cancellation noted above, the final result reads 
\begin{align}
\label{GEikF}
G_{F}^{\mu\nu}(x,y) \Big|_{\rm Eik.\, z^-} &=  i \delta^{2}(\x - \y) \ \delta(x^{+}-y^{+}) \eta^{\mu} \eta^{\nu} \int \! \frac{dk^{+}}{2\pi}  \frac{ e^{-i(x^{-}-y^{-})k^{+}}} {k^{+}k^{+}} 
\nn\\
&
+  \int \! \frac{d^{3}\uq}{(2\pi)^{3}} \int \! \frac{d^{3}\uk} {(2\pi)^{3}} \   \frac{ e^{-ix \cdot \check{q} } \ e^{iy \cdot \check{k}}}{q^+ + k^{+}} 
\Big[ -g^{\mu\nu} + \frac{\check{k}^{\mu} \eta^{\nu} }{k^{+}} +
\frac{\eta^{\mu} \check{q}^{\nu}}{q^{+}}- \frac{\eta^{\mu}\eta^{\nu}}{q^{+}k^{+}} (\check{q} \cdot \check{k}) \Big] 
\nn\\
&\times
 \int \! d^{2}\z \ e^{-i(\q - \k)\cdot \z}\ \int \! dz^{-} \  e^{i(q^{+} -k^{+})z^{-}}
\nn \\
&\times 
\Big[ \theta(x^{+}- y^{+}) \theta(q^+) \theta(k^{+}) \ U_{A}(x^{+}, y^{+}; \z, z^-)- \theta(y^{+}- x^{+})  \theta(-q^+)\theta(-k^{+}) \  U_{A}^{\dagger}(x^{+}, y^{+}; \z, z^-) \Big]
\end{align}
This is the expression for a gluon propagator in a dynamical $\A^-$ background for any $x^+$ and $y^+$ which can be both inside or outside the medium. 
It contains the NEik corrections beyond the static approximation but no corrections beyond the shockwave limit.
In order to get the gluon propagator at strict Eikonal limit, one can expand the Wilson lines around $z^-=0$ and keep only the zeroth order term in that gradient expansion. In that case, the only $z^-$ dependence appears in the phase which can be integrated trivially and one gets the following strict Eikonal gluon propagator 
\begin{align}
\label{EikF}
G_{F}^{\mu\nu}(x,y) \Big|_{\text{Eik}} &=  i\,  \delta^{2}(\x - \y) \ \delta(x^{+}-y^{+}) \, \eta^{\mu} \eta^{\nu}   \int \! \frac{dk^{+}}{2\pi}  \frac{ e^{-i(x^{-}-y^{-})k^{+}}} {k^{+}k^{+}}  \nn \\
& + \int \! \frac{d^{3}\uq}{(2\pi)^{3}} \int \! \frac{d^{3}\uk}{(2\pi)^{3}} \   
\frac{ e^{-ix \cdot \check{q} } \ e^{iy \cdot \check{k}}}{2k^{+}} \
\Big[2\pi \ \delta(k^{+} - q^{+})\Big] 
\bigg[ -g^{\mu\nu} + \frac{\check{k}^{\mu} \eta^{\nu} }{k^{+}} +\frac{\eta^{\mu} \check{q}^{\nu}}{q^{+}}- \frac{\eta^{\mu}\eta^{\nu}}{q^{+}k^{+}} (\check{q} \cdot \check{k}) \bigg] \nn \\
&\times 
\int \! d^{2}\z \ e^{-i(\q- \k)\cdot\z}
\Big[ \theta(x^{+}- y^{+}) \ \theta(k^{+}) \ U_{A}(x^{+}, y^{+}; \z) 
     - \theta(y^{+}- x^{+}) \ \theta(-k^{+}) \  U_{A}^{\dagger}(x^{+}, y^{+}; \z) \Big]
     \, .
    \end{align}
%
%  
%%%%%%%%%%%%%%%%%%%%%%%%%%%%%%%%%%%%%%%%%%%%%%%%%%%%%%%%%%%%%%%%%%%%%%%%%%%%%%%%%%%%%%%%%%%%%%%%%%%%%%%%%%%%%%%%%%
\subsection{NEik gluon propagator in a gluon background field}

We are now ready to discuss the gluon propagator at full NEik accuracy in a dynamical gluon background field. There are three types of NEik corrections in this context: NEik corrections beyond the static approximation (that are already included in Eq.~\eqref{EikF}), NEik corrections beyond the shockwave approximation (corresponding to the difference between Eqs.~\eqref{Without shockwave approx G} and \eqref{With shockwave approx G}),
and NEik corrections associated with insertions of the transverse components of the background field.

At NEik accuracy, the gluon propagator get contributions from interactions with the "$-$" component of the background field but including the corrections stemming from finite longitudinal extent. Moreover, it also receives contributions from the interactions with the transverse component of the background field. Staying beyond the static limit of the background field,  at NEik accuracy, the medium corrections to the gluon propagator can be written as 
\begin{align}
\label{medium_corrections_NEik}
\delta G_F^{\mu\nu}(x,y)\Big|^{\rm NEik}& =\delta G_F^{\mu\nu}(x,y)\Big|^{\rm Eik}_{\rm pure\, \A^-}
+\delta G_F^{\mu\nu}(x,y)\Big|^{\rm NEik}_{\rm pure\,  \A^-} 
+\delta G_F^{\mu\nu}(x,y)\Big|^{\rm NEik}_{\rm single\, \A_{\perp}}\nn\\
& 
+\delta G_F^{\mu\nu}(x,y)\Big|^{\rm NEik}_{\rm double\, \A_{\perp},\, loc. } 
+\delta G_F^{\mu\nu}(x,y)\Big|^{\rm NEik}_{\rm double\, \A_{\perp},\, non- loc. } 
\end{align}
so that gluon propagator at NEik accuracy reads 
\begin{align}
\label{eq:gluon_prop_NEik_1}
G_F^{\mu\nu}(x,y)\Big|^{\rm NEik}=G_{0,F}^{\mu\nu}(x,y)+ \delta G_F^{\mu\nu}(x,y)\Big|^{\rm NEik}
\end{align}
It is important to note that we keep the $z^-$ dependence in each medium contribution on the right hand side of Eq. \eqref{medium_corrections_NEik} to stay beyond the static limit of the target fields. Moreover, the first term on the right hand side of Eq. \eqref{medium_corrections_NEik} together with the vacuum gluon propagator corresponds to the total gluon propagator at genralized eikonal order given in Eqs. \eqref{gen_eik_imp} and \eqref{GEikF}. 

In the rest of this subsection, we compute and discuss each medium correction to the gluon propagator at NEik accuracy. In our discussions, we will take $x^+> L^+/2$ and $y^+<-L^+/2$, so that the gluon propagator from before to after the medium will be considered. 

%%%%%%%%%%%%%%%%%%%%%%%%%%%%%%%%%%%%%%%%%%%%%%%%%%%%%%%%%%%%%%%%%%%%%%%%%%%%%%%%%%%%%%%%%%%%%%%%%%%%%%%%%%%%%%%%%%%%
\subsubsection{NEik gluon propagator in a pure $\A^-$ background}
%%%%%
%%%%%
The derivation of NEik corrections in pure $\cal A^-$ background have been discussed in detail for gluon propagator in \cite{Altinoluk:2014oxa} and for quark propagator in \cite{Altinoluk:2020oyd, Altinoluk:2021lvu}. Here, we follow the same strategy adopted in \cite{Altinoluk:2020oyd} to compute the medium corrections to a gluon propagator at NEik accuracy in a pure $\cal A^-$ background passing through the whole medium. Since the derivation is almost the same as in the case of  quark propagator, here we only present the result (more details on the derivation can be found in Appendix \ref{app:derivation_NEik_pure_A-}).   

Medium corrections to a before-to-after (traversing the whole medium) gluon propagator (with positive energy) at NEik accuracy in pure $\cal A^-$ background reads 

%%%%%%%%%%%%%%%%%%%%%%%%%%%%%%%%%%%%%%%%%%%%%%%%%%%%%%%%%%%%%%%%%%%%%%%%%%%%%%%%%%%%%%%%%%%%%%%%%%%%%%%%%%%%%%%%%%
%\subsection{NEik gluon propagator in a gluon background field
%
%
\begin{align}
\label{A-NEik}  
&
\delta G_{F}^{\mu\nu}(x,y)\Big|^{\rm Eik}_{\text{pure}\, \A^{-}, z^-} +\delta G_{F}^{\mu\nu}(x,y)\Big|^{\rm NEik}_{\text{pure}\, \A^{-}, z^-} = 
\int \! \frac{d^{3}\underline{q}}{(2\pi)^{3}} \int \! \frac{d^{3}\underline{k}}{(2\pi)^{3}} \frac{\theta(q^+) \theta(k^+)}{q^{+} + k^{+}} \ e^{-ix \cdot \check{q}} \ e^{iy \cdot \check{k}} \nn\\
&\times
\bigg[ -g^{\mu\nu} + \frac{\check{k}^{\mu} \eta^{\nu} }{k^{+}} +
\frac{\eta^{\mu} \check{q}^{\nu}}{q^{+}}
-\frac{\eta^{\mu}\eta^{\nu}}{q^{+}k^{+}} (\check{q} \cdot \check{k}) \bigg]
%\nn \\
%&\times   
\int\! dz^{-}  e^{-iz^{-}(q^{+}-k^{+})} \int \!  d^{2} \z \ e^{-i\z\cdot(\q - \k)} \Bigg\{ \bigg[ U_{A}\bigg(\frac{L^{+}}{2},-\frac{L^{+}}{2}; \z, z^{-}\bigg) -1 \bigg]\nn \\
& \hspace{1.5cm}
 - \frac{(\q^{j}+ \k^{j})}{2(q^{+}+k^{+})} \int_{\frac{-L^{+}}{2}}^{\frac{L^{+}}{2}} \! dz^{+}  \bigg[ U_{A}\bigg({\frac{L^{+}}{2}},z^{+}; \z, z^{-}\bigg)
 \left( \overrightarrow{\frac{d}{d\z^{j}}} - \overleftarrow{\frac{d}{d\z^{j}}}\right) U_{A}\bigg(z^{+},-{\frac{L^{+}}{2}}; \z, z^{-}\bigg) \bigg] \nn\\
 & \hspace{1.5cm}
 -\frac{i}{q^{+} + k^{+}} \int_{-\frac{L^{+}}{2}}^{\frac{L^{+}}{2}} \! dz^{+} \ U_{A}\bigg(\frac{L^{+}}{2},z^{+}; \z, z^{-}\bigg) \overleftarrow{\frac{d}{d\z^{j}}}\ \overrightarrow{\frac{d}{d\z^{j}}} \ U_{A}\bigg(z^{+},-\frac{L^{+}}{2}; \z, z^{-}\bigg)
 \Bigg\} + \text{NNEik}
\end{align}
It is worth to emphasize that in Eq. \eqref{A-NEik} derivatives act only on the Wilson lines but not on the phase factors.  
%
%
%%%%%%%%%%%%%%%%%%%%%%%%%%%%%%%%%%%%%%%%%%%%%%%%%%%%%%%%%%
%%%%%%%%%%%%%%%%%%%%%%%%%%%%%%%%%%%%%%%%%%%%%%%%%%%%%%%%%%
\subsubsection{NEik contribution to the gluon propagator from single $\A_{\perp}$ insertion}
\begin{figure}
\centering
\includegraphics[scale=0.5]{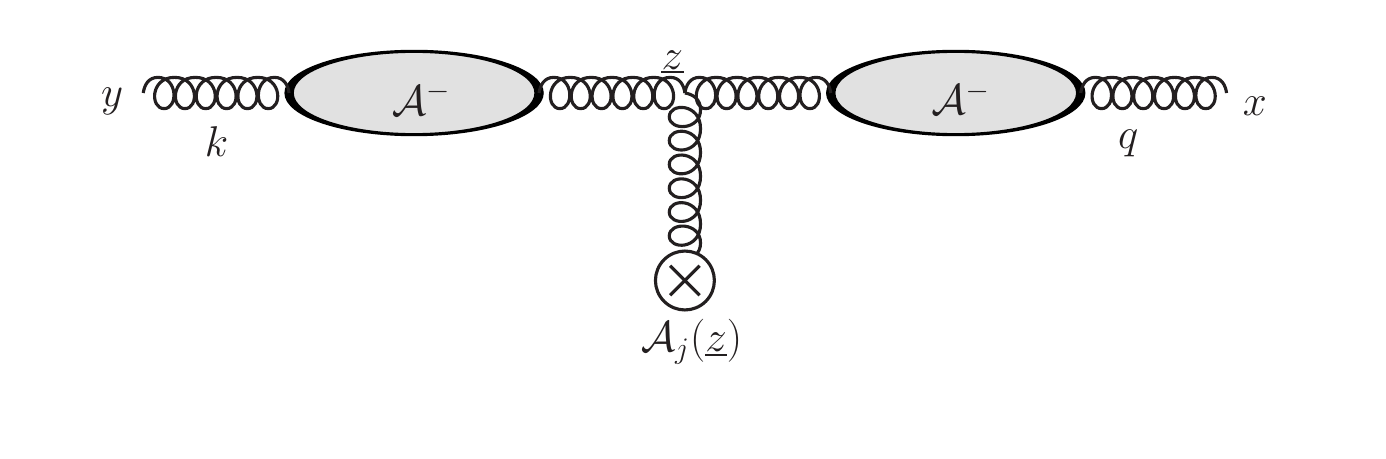}
\caption{Contribution to the gluon propagator in background field, with a single transverse component of the background field inserted via a three gluon vertex, and with interactions with the $\A^{-}$ component resummed to all orders.}
 \label{fig: Three gluon vertex in medium}
\end{figure}

%%%%%%%%%%%%%%%%%%%%%%%%%%%%%%%%%%%%%%%%%%%%%%%%%%%%%%%%%%
%%%%%%%%%%%%%%%%%%%%%%%%%%%%%%%%%%%%%%%%%%%%%%%%%%%%%%%%%%
Now that the gluon propagator in a pure $\A^-$ background at NEik accuracy is known, one can compute the interactions with the transverse component of the background field in coordinate space using perturbation theory. The first such contribution is obtained by replacing one of the interactions with the background field $\A^-$ with an interaction with $\A_{\perp}$ (see Fig. \ref{fig: Three gluon vertex in medium}) which reads 
\begin{align}
\label{Single A perp general}
    \delta {G_{F}^{\mu\nu}}_{ab}(x,y) \Big|_{\text{single}\,  \A_{\perp}}^{\rm NEik} = 
        \int d^{4}z \  \bigg[ G_{F}^{\mu\mu'}(x,z) \Big|_{ \text{Eik}} \bigg]_{aa'} \ \bigg[{\rm X}^{3g}_{\mu'\nu'}(z) \bigg]^{a'b'} \ \bigg[ G_{F}^{\nu'\nu}(z,y) \Big|_{\text{Eik}} \bigg]_{b'b}
\end{align}
%
%
%%%%%%%%%%%
%%%%%%%%%%%
%%%%%%%%%%%
%%%%%%%%%%%
Here $a,b,a'$ and $b'$ represent color indices, ${\rm X}^{3g}_{\mu'\nu'}(z)$ is the effective single $\cal A_{\perp}$ insertion factor that is computed from triple gluon vertex using the Feynman rules and it reads  
\begin{align}
\label{three gluon vertex insertion}
%\operatorname{X}
\Big[{\rm X}^{3g}_{\mu'\nu'}(z)\Big]^{ab} &= -gf^{abc}  
\Bigg[ ig_{\mu'\nu'}  \overleftarrow{\frac{d}{d\z^{j}}}\A_{c}^{j}(z) 
         -ig_{\mu'\nu'}  \A_{c}^{j}(z)  \overrightarrow{\frac{d}{d\z^{j}}} 
         - 2i g^{j}_{\; \nu'} \A_{c}^{j}(z)  \overrightarrow{\frac{d}{dz^{\mu'}}} 
         \nn \\
&\hspace{1.5cm}
 - i g^{j}_{\;\nu'}   \overleftarrow{\frac{d}{dz^{\mu'}}} \A_{c}^{j}(z) 
+ 2i g^{j}_{\;\mu'}  \overleftarrow{\frac{d}{dz^{\nu'}}} \A_{c}^{j}(z)  + i g^{j}_{\;\mu'}  \A_{c}^{j}(z) \overrightarrow{\frac{d}{dz^{\nu'}}} \,   \Bigg]
\end{align}
It is worth mentioning that in Eq. \eqref{Single A perp general}, the integration over $z^+$ amounts to a factor of $L^+$ in the limit $L^+\to 0$, since the background field is assumed to vanish outside of a range of width $L^+$. However, the field components $\cal A_{\perp}$ themselves are of order $(L^+)^0$. Thus,  the contribution of a single $\cal A_{\perp}$ insertion via a three gluon vertex is in general of order $L^+$ overall, and thus NEik.
% and is a NEik contribution for $x^+>L^+/2$ and $y^+<-L^+/2$. 
The only exception in the following. If one takes into account the instantaneous contribution to the gluon propagator at eikonal order \eqref{EikF}, either after or before the $\cal A_{\perp}$ insertion in Eq.~\eqref{Single A perp general}, the Dirac delta along the longitudinal direction makes the $z^+$ integral trivial, and the transverse component is inserted at an endpoint of the propagator, either at $z^+=x^+$ or $z^+=y^+$. This is only possible if  $x^+$ or $y^+$ belongs to the support $[L^+/2,-L^+/2]$. In that case, the gluon does not propagate through the whole medium. These cases are referred to as before-to-inside or inside-to-after gluon propagators (or inside-inside) and they will be considered in the next section. 

For the case $x^+>L^+/2$ and $y^+<-L^+/2$, the medium correction to the before-to-after gluon propagator at NEik accuracy stemming from a single $\A_{\perp}$ insertion can be computed using Eq. \eqref{three gluon vertex insertion} and it reads
\begin{align}
\label{eq:single_A_perp_ins_fin}
\delta {G_{F}^{\mu\nu}}_{ab}(x,y) \Big|_{\text{single}\,  \A_{\perp}}^{\rm NEik} 
& = g\ \int d^{3}\uz   \int \! \frac{d^{3}\uq}{(2\pi)^{3}}    \frac{ e^{-ix \cdot \check{q} } }{2q^{+}} %\theta(x^{+}- z^{+}) 
\theta(q^{+}) 
\int \! \frac{d^{3}\uk}{(2\pi)^{3}} \   \frac{ e^{iy \cdot \check{k}}}{2k^{+}}  \theta(k^{+})  \int\! dz^{-} e^{-iz^{-}(q^{+}-k^{+})}
\nn  \\
& \times \Big[U_{A}(x^{+}, z^{+}; \z, z^{-}) \Big]_{aa'}  \ e^{-i\q\cdot \z} \  
\Bigg\{ 2\bigg[ \bigg( g^{\mu j} g^{\nu i}- \frac{\eta^{\mu}g^{\nu i} \q^{j}}{q^{+}}- \frac{g^{\mu j} \k^{i}\eta^{\nu}}{k^{+}} + \frac{\eta^{\mu}\eta^{\nu} \k^{i}\q^{j}}{q^{+}k^{+}}\bigg) 
\nn \\
& - \bigg(  g^{\mu i} g^{ j \nu} - \frac{\eta^{\mu} \q^{i}g^{ j \nu}}{q^{+}} - \frac{ g^{\mu i}\k^{j} \eta^{\nu}}{k^{+}}+ \frac{\eta^{\mu} \eta^{\nu} \q^{i}\k^{j} }{q^{+}k^{+}}\bigg)  \bigg] 
\bigg[ \overleftarrow{\frac{d}{d\z^{i}}} \left(T \cdot \A^{j}(z) \right)  + \left(T \cdot \A^{j}(z) \right) \overrightarrow{\frac{d}{d\z^{i}}}  \bigg] 
\nn \\  
& + \bigg( -g^{\mu\nu} + \frac{\eta^{\mu } \check{q}^{\nu}}{q^{+}} + \frac{\check{k}^{\mu}\eta^{\nu}}{k^{+}} - \frac{\eta^{\mu}\eta^{\nu}}{q^{+}k^{+}} (\check{q} \cdot \check{k} )  \bigg)
\bigg[   -\overleftarrow{\frac{d}{d\z^{j}}}\big( T \cdot \A^{j}(z) \big)+ \big(T \cdot \A^{j}(z) \big)  \overrightarrow{\frac{d}{d\z^{j}}} \bigg]\Bigg\}
\nn \\
& \times e^{i\k\cdot\z} \ \Big[ U_{A}(z^{+}, y^{+}; \z, z^{-}) \Big]_{b'b} %\ \theta(z^{+}- y^{+}) 
+ \text{NNEik}  
    \end{align}
Note that in Eq.~\eqref{eq:single_A_perp_ins_fin} the transverse derivative act both on the Wilson lines and on the phase factors, by contrast to the convention in Eq.~\eqref{A-NEik}. Hence, in the first part of the expression, the derivative over $\z^i$ acts on everything apart from the transverse background field. Integrating by parts then makes the derivative over $\z^i$ act only on the transverse background field. In the second part of the expression \eqref{eq:single_A_perp_ins_fin} instead, let us perform the derivatives over the phase factors explicitly, so that one arrives at
\begin{align}
\label{eq:single_A_perp_ins_fin_2}
\delta {G_{F}^{\mu\nu}}_{ab}(x,y)& \Big|_{\text{single}\,  \A_{\perp}}^{\rm NEik} 
 = g\ \int d^{3}\uz    \int \! \frac{d^{3}\uq}{(2\pi)^{3}}    \frac{ e^{-ix \cdot \check{q} } }{2q^{+}} %\theta(x^{+}- z^{+}) 
\theta(q^{+}) 
\int \! \frac{d^{3}\uk}{(2\pi)^{3}} \   \frac{ e^{iy \cdot \check{k}}}{2k^{+}}  \theta(k^{+})  \int\! dz^{-} e^{-iz^{-}(q^{+}-k^{+})}\ e^{-i(\q-\k)\cdot \z} \  
\nn  \\
& \times \Big[U_{A}(x^{+}, z^{+}; \z, z^{-}) \Big]_{aa'}  
\Bigg\{ -2\bigg[ \bigg( g^{\mu j} g^{\nu i}- \frac{\eta^{\mu}g^{\nu i} \q^{j}}{q^{+}}- \frac{g^{\mu j} \k^{i}\eta^{\nu}}{k^{+}} + \frac{\eta^{\mu}\eta^{\nu} \k^{i}\q^{j}}{q^{+}k^{+}}\bigg) 
\nn \\
& - \bigg(  g^{\mu i} g^{ j \nu} - \frac{\eta^{\mu} \q^{i}g^{ j \nu}}{q^{+}} - \frac{ g^{\mu i}\k^{j} \eta^{\nu}}{k^{+}}+ \frac{\eta^{\mu} \eta^{\nu} \q^{i}\k^{j} }{q^{+}k^{+}}\bigg)  \bigg] 
  \left(T \cdot \partial_{\z^i}\A^{j}(z) \right)  
\nn \\  
& + \bigg( -g^{\mu\nu} + \frac{\eta^{\mu } \check{q}^{\nu}}{q^{+}} + \frac{\check{k}^{\mu}\eta^{\nu}}{k^{+}} - \frac{\eta^{\mu}\eta^{\nu}}{q^{+}k^{+}} (\check{q} \cdot \check{k} )  \bigg)
\bigg[   -\overleftarrow{\frac{d}{d\z^{j}}}\big( T \cdot \A^{j}(z) \big)+ \big(T \cdot \A^{j}(z) \big)  \overrightarrow{\frac{d}{d\z^{j}}} +i(\q^i\!+\!\k^i) T \cdot \A^{j}(z)\bigg]\Bigg\}
\nn \\
& \times \ \Big[ U_{A}(z^{+}, y^{+}; \z, z^{-}) \Big]_{b'b} %\ \theta(z^{+}- y^{+}) 
+ \text{NNEik}  
 \, ,
\end{align}
where the derivatives in $\z^i$ now act only on the Wilson lines, not on the phase factor. Finally, by relabeling $i$ and $j$ into each other in some of the terms, one obtains
\begin{align}
\label{eq:single_A_perp_ins_fin_2}
\delta {G_{F}^{\mu\nu}}_{ab}&(x,y) \Big|_{\text{single}\,  \A_{\perp}}^{\rm NEik} 
 = g\ \int d^{3}\uz    \int \! \frac{d^{3}\uq}{(2\pi)^{3}}    \frac{ e^{-ix \cdot \check{q} } }{2q^{+}} %\theta(x^{+}- z^{+}) 
\theta(q^{+}) 
\int \! \frac{d^{3}\uk}{(2\pi)^{3}} \   \frac{ e^{iy \cdot \check{k}}}{2k^{+}}  \theta(k^{+})  \int\! dz^{-} e^{-iz^{-}(q^{+}-k^{+})}\ e^{-i(\q-\k)\cdot \z} \  
\nn  \\
& \times \Big[U_{A}(x^{+}, z^{+}; \z, z^{-}) \Big]_{aa'}  
\Bigg\{ -2\bigg(  g^{\mu i} g^{ j \nu} - \frac{\eta^{\mu} \q^{i}g^{ j \nu}}{q^{+}} - \frac{ g^{\mu i}\k^{j} \eta^{\nu}}{k^{+}}+ \frac{\eta^{\mu} \eta^{\nu} \q^{i}\k^{j} }{q^{+}k^{+}}\bigg)
\bigg[ \left(T \cdot \partial_{\z^i}\A_{j}(z) \right) -  \left(T \cdot \partial_{\z^j}\A_{i}(z) \right) \bigg] 
\nn \\  
& + \bigg( -g^{\mu\nu} + \frac{\eta^{\mu } \check{q}^{\nu}}{q^{+}} + \frac{\check{k}^{\mu}\eta^{\nu}}{k^{+}} - \frac{\eta^{\mu}\eta^{\nu}}{q^{+}k^{+}} (\check{q} \cdot \check{k} )  \bigg)
\bigg[   \overleftarrow{\frac{d}{d\z^{j}}}\big( T \cdot \A_{j}(z) \big)- \big(T \cdot \A_{j}(z) \big)  \overrightarrow{\frac{d}{d\z^{j}}} -i(\q^i\!+\!\k^i) T \cdot \A_{j}(z)\bigg]\Bigg\}
\nn \\
& \times \ \Big[ U_{A}(z^{+}, y^{+}; \z, z^{-}) \Big]_{b'b} %\ \theta(z^{+}- y^{+}) 
+ \text{NNEik}  
 \, .
\end{align}

%%%%%%%%%%%%%%%%%%%%%%%%%%%%%%%%%%%%%%%%%%%%%%%%%%%%%%%%%%%%
%%%%%%%%%%%%%%%%%%%%%%%%%%%%%%%%%%%%%%%%%%%%%%%%%%%%%%%%%%%%
\subsubsection{NEik contribution to the gluon propagator from local double $\A_{\perp}$ insertion}
\begin{figure}
\centering
\includegraphics[scale=0.5]{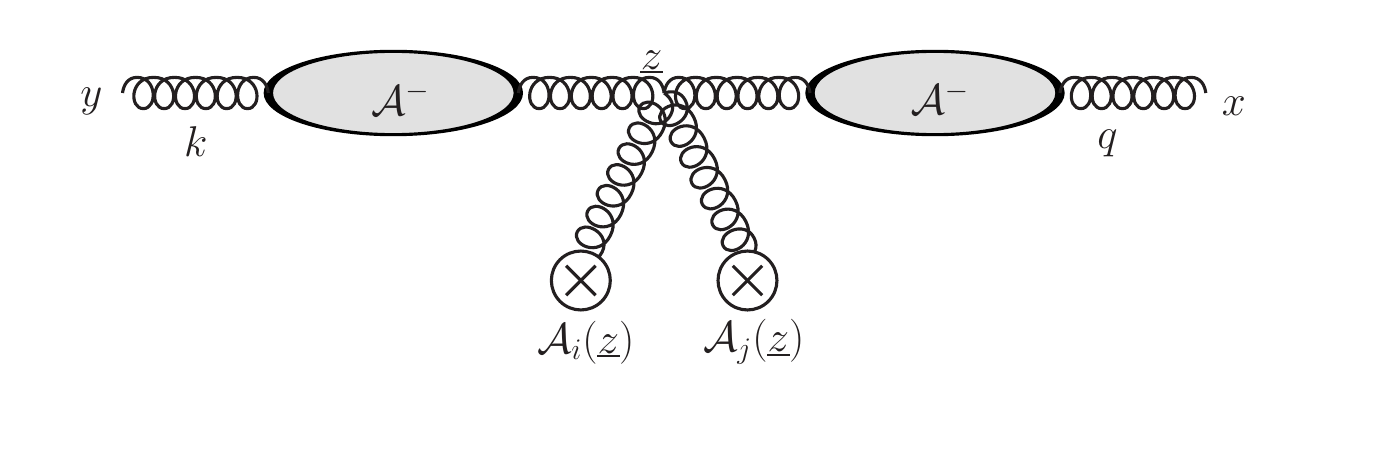}
\caption{Contribution to the gluon propagator in background field, with a local insertion of two transverse components of the background field via a four gluon vertex, and with interactions with the $\A^{-}$ component resummed to all orders.}
\label{fig: Four gluon vertex in medium}
\end{figure}

%%%%%%%%%%%%%%%%%%%%%%%%%%%%%%%%%%%%%%%%%%%%%%%%%%%%%%%%%%%%
%%%%%%%%%%%%%%%%%%%%%%%%%%%%%%%%%%%%%%%%%%%%%%%%%%%%%%%%%%%%
Another medium correction to the gluon propagator at NEik accuracy is stemming from the insertion of two background fields at the same points thanks to a four-gluon vertex. As we have explained in the footnote~\ref{fnote_4gV}, such diagrams vanish in the light-cone gauge $A^+=0$ if at least one of the two background fields inserted at that vertex is a $\A^-$ component. This leaves only the possibility of inserting two transverse components of the background field. Again, both field components $\cal A_{\perp}$  are of order $(L^+)^0$, whereas the integration over the coordinate $z^+$ of the four gluon vertex, restricted to the support of the background field, brings a suppression by a factor of $L^+$. Therefore, such contribution, drawn on Fig.~\ref{fig: Four gluon vertex in medium}, is indeed of NEik order. 
% $\A_{\perp}$ fields along the $+$ longitudinal direction. Naively, one can argue that this contribution would only start at NNEik order, since one should integrate over two different longitudinal positions $z^+$ and $z^{\prime +}$ where the two $\A_{\perp}$ fields are inserted, and each integration brings factor of $L^+$. This argument fails in two cases: (i) if the fields are inserted at the same longitudinal position which we refer to as local double $\A_{\perp}$ contribution (see Fig. \ref{fig: Four gluon vertex in medium}) or (ii) if the fields are inserted at two different longitudinal positions $z^+$ and $z^{\prime +}$ but these insertions are connected with the instantaneous part of the gluon propagator which we refer to as non-local double $\A_{\perp}$ contribution (see Fig. \ref{fig: Inst four gluon vertex in medium}). We will discuss the latter one in the next subsection, for now let us focus on the local double $\A_{\perp}$ contribution. For this case, since the two $\A_{\perp}$ fields are inserted at the same longitudinal position $z^+$, the integration over it amounts to a factor of $L^+$, and thus that contribution starts at NEik order. This contribution originates from the four gluon vertex (see Fig. \ref{fig: Four gluon vertex in medium}) and i
It can be written as 
%
%%%%%%%%%%%%%%
%%%%%%%%%%%%%%
%%%%%%%%%%%%%%
%%%%%%%%%%%%%%
%
\begin{align}
\delta G_{F}^{\mu\nu}(x,y) \Big|_{\rm double\, \A_{\perp}, \, loc.}^{\rm NEik}
= 
 \int d^{4}z \  \bigg[ G_{F}^{\mu\mu'}(x,z) \Big|_{\rm Eik} \bigg]_{aa'} 
\Big[ {\rm X}^{4g}_{\mu'\nu'}(z) \Big]^{a'b'}\ 
\bigg[  G_{F}^{\nu'\nu}(z,y) \Big|_{\rm Eik} \bigg]_{b'b}
\end{align}
 ${\rm X}^{4g}_{\mu'\nu'}(z)$ is the effective double $\A_{\perp}$ insertion factor which is obtained from the four gluon vertex using the Feynman rules and it reads 
\begin{align}
\Big[ {\rm X}^{4g}_{\mu'\nu'}(\underline{z})\Big]^{a'b'} &=  -ig^{2}  
\Big[ f^{ea'b'}f^{edc} \big(g_{\nu' i}g_{\mu' j} - g_{\nu' j}g_{\mu' i}\big)
+ f^{eb'd}f^{ea'c} \big(g_{\nu'\mu'}g_{ij} - g_{\nu'j}g_{\mu'i}\big) 
\nn\\
& \hspace{1.3cm}
+\,  f^{eb'c}f^{ea'd} \big(g_{\nu'\mu'}g_{ij} - g_{\nu' i}g_{\mu' j}\big) \Big]\,  \frac{1}{2}  \A_{c}^{i}(z)  \A_{d}^{j}(z)\ 
\end{align}
where the factor of one half comes from the Bose symmetry of exchange of the two background fields inserted.
Using this effective double $\A_{\perp}$ insertion factor, the medium correction to the gluon propagator at NEik accuracy stemming from local double $\A_{\perp}$ contribution can be computed as 
\begin{align}
\label{eq:loc_double_A-perp_1}
\delta G_{F}^{\mu\nu}(x,y) \Big|_{\rm double \, \A_{\perp},\, loc.}^{\rm NEik} &=
\frac{1}{2}\int d^{4}z\,  \bigg\{ 
\int \! \frac{d^{3}\uq}{(2\pi)^{3}} \int \! \frac{d^{3}\uk_{1}}{(2\pi)^{3}} \   \frac{ e^{-ix \cdot \check{q} } \ e^{iz \cdot \check{k}_{1}}}{2k_{1}^{+}} \Big[2\pi \ \delta(k_{1}^{+} - q^{+})\Big]   
% \theta(x^{+}- z^{+}) 
 \theta(k_{1}^{+})
\nn \\
& \times  
\bigg[ -g^{\mu\mu'} + \frac{\check{k}_{1}^{\mu} \eta^{\mu'} }{k_{1}^{+}} + \frac{\eta^{\mu} \check{q}^{\mu'}}{q^{+}}
     - \frac{\eta^{\mu}\eta^{\mu'}}{q^{+}k_{1}^{+}} (\check{q} \cdot \check{k}_{1}) \bigg]
\int \! d^{2}\z_{1} \ e^{-i(\q - \k_{1})\cdot \z_{1}} 
\ U_{A}(x^{+},z^{+}; \z_{1},z^{-})  \bigg\}_{aa'} 
\nn \\
&
%\hspace{-2.5cm}
\times 
\bigg\{\!  -ig^{2} 
\bigg[ \big[ T \cdot \A^{i}(z), T \cdot \A^{j}(z)\big]  \big(g_{\nu' i}g_{\mu' j} - g_{\nu' j}g_{\mu' i}\big)
\nn\\
& 
\hspace{1.6cm}
         +\big[T \cdot \A^{i}(z)\big] \big[T \cdot \A^{j}(z)\big]  \big(g_{\nu'\mu'}g_{ij} - g_{\nu'j}g_{\mu'i}\big)
 \nn\\
 & \hspace{1.6cm}
 + \big[T \cdot \A^{j}(z)\big] \big[T \cdot \A^{i}(z)\big] \big(g_{\nu'\mu'}g_{ij} - g_{\nu' i}g_{\mu' j}\big) \bigg]\bigg\}_{a'b'}
 \nn\\
&\times  
   \bigg\{ \int \! \frac{d^{3}\uq_{1}}{(2\pi)^{3}} \int \! \frac{d^{3}\uk}{(2\pi)^{3}}    \frac{ e^{-iz \cdot \check{q}_{1} }  e^{iy \cdot \check{k}}}{2k^{+}} 
  \Big[2\pi \ \delta(q_{1}^{+} - k^{+})\Big]  %\theta(z^{+}- y^{+}) 
  \theta(k^{+})
  \nn \\
&
\times
 \bigg[ -g^{\nu' \nu} + \frac{\check{k}^{\nu'} \eta^{\nu} }{k^{+}} +
   \frac{\eta^{\nu'} \check{q}_{1}^{\nu}}{q_{1}^{+}} - \frac{\eta^{\nu'}\eta^{\nu}}{q_{1}^{+}k^{+}} (\check{q_{1}} \cdot \check{k}) \bigg]
    \int \! d^{2}\z_{2 } \ e^{-i(\q_{1} - \k)\cdot \z_{2}} 
U_{A}(z^{+},y^{+}; \z_{2},z^{-})  \bigg\}_{b'b} 
    \end{align}
Using the simple relation 
\begin{align}
\big[ T\cdot \A_{j}(z) \big]\big[T\cdot \A_{i}(z)\big] &= \big[T\cdot \A_{j}(z), T\cdot \A_{i}(z)\big] + \big[T\cdot \A_{i}(z)\big] \big[T\cdot \A_{j}(z)\big]
\nn \\
&=  - \big[T\cdot \A_{i}(z), T\cdot \A_{j}(z)\big] + \big[T\cdot \A_{i}(z)\big] \big[T\cdot \A_{j}(z)\big]    
 \end{align}   
Eq. \eqref{eq:loc_double_A-perp_1} can be further simplified and the medium correction to the gluon propagator at NEik accuracy stemming from local double $\A_{\perp}$ insertion can be written as 
\begin{align}
\label{eq:double_A_perp_loc_fin}
\delta {G_{F}^{\mu\nu}}_{ab}(x,y) &\Big|_{\rm double\, \A_{\perp}, \, loc.}^{\rm NEik}  = 
\int \! \frac{d^{3}\uq}{(2\pi)^{3}} \   \frac{ e^{-ix \cdot \check{q} } }{2q^{+}} %\theta(x^{+}- z^{+})
 \theta(q^{+}) \
\int \! \frac{d^{3}\uk}{(2\pi)^{3}} \   \frac{ e^{iy \cdot \check{k}}}{2k^{+}} \ \theta(k^{+})  
\int\! dz^{-} e^{iz^{-}(q^{+} - k^{+})}\int\! d^{2}\z \, e^{-i(\q-\k)\cdot \z} 
\nn \\
& \hspace{-1.2cm}
\times 
 \int\! dz^{+} \big( ig^{2} \big) \  \Big[U_{A}(x^{+}, z^{+}; \z,z^{-})\Big]_{aa'}  
%\nn \\
%& \hspace{-1.2cm}
%\times
\bigg\{  g_{ij}\big[ T \cdot \A^{i}(z)\big] \big[ T \cdot \A^{j}(z)\big] 
\Big[ 
- g^{\mu \nu} + \frac{\check{k}^{\mu} \eta^{\nu} }{k^{+}} + \frac{\eta^{\mu}\check{q}^{\nu} }{q^{+}} \ - \frac{\eta^{\mu} \eta^{\nu} }{k^{+}q^{+}} \bigl( \check{k} \cdot \check{q} \bigr) 
\Big]
\nn \\
&\hspace{-1.2cm}
+
\Big[ -2\big[T\cdot \A_{i}(z),T\cdot \A_{j}(z)\big] + \big[T\cdot \A_{i}(z) \big]\big[T\cdot \A_{j}(z)\big] \Big]
\Big[ g^{\mu i} g^{j\nu}   - \frac{\k^{j} g^{\mu i} \eta^{\nu}}{k^{+}}  - \frac{\eta^{\mu} g^{j\nu} \q^{i}}{q^{+}}+ \frac{\eta^{\mu} \eta^{\nu} \k^{j} \q^{i}}{q^{+}q^{+}}  \Big] \bigg\}_{a'b'}
\nn \\ 
& \hspace{-1.2cm}
\times 
\Big[ U_{A}(z^{+}, y^{+}; \z,z^{-}) \Big]_{b'b} %\ \theta(z^{+}- y^{+})  
+ \text{NNEik}
\, .
\end{align}
%
%

%%%%%%%%%%%%%%%%%%%%%%%%%%%%%%%%%%%%%%%%%%%%%%%%%%%%%%%%%%%%
%%%%%%%%%%%%%%%%%%%%%%%%%%%%%%%%%%%%%%%%%%%%%%%%%%%%%%%%%%%%
\subsubsection{NEik contribution to the gluon propagator from non-local double $\A_{\perp}$ insertion}
%%%%%%%%%%%%%%%%%%%%%%%%%%%%%%%%%%%%%%%%%%%%%%%%%%%%%%%%%%%%
%%%%%%%%%%%%%%%%%%%%%%%%%%%%%%%%%%%%%%%%%%%%%%%%%%%%%%%%%%%%

\begin{figure}
    \centering
    \includegraphics[scale=0.5]{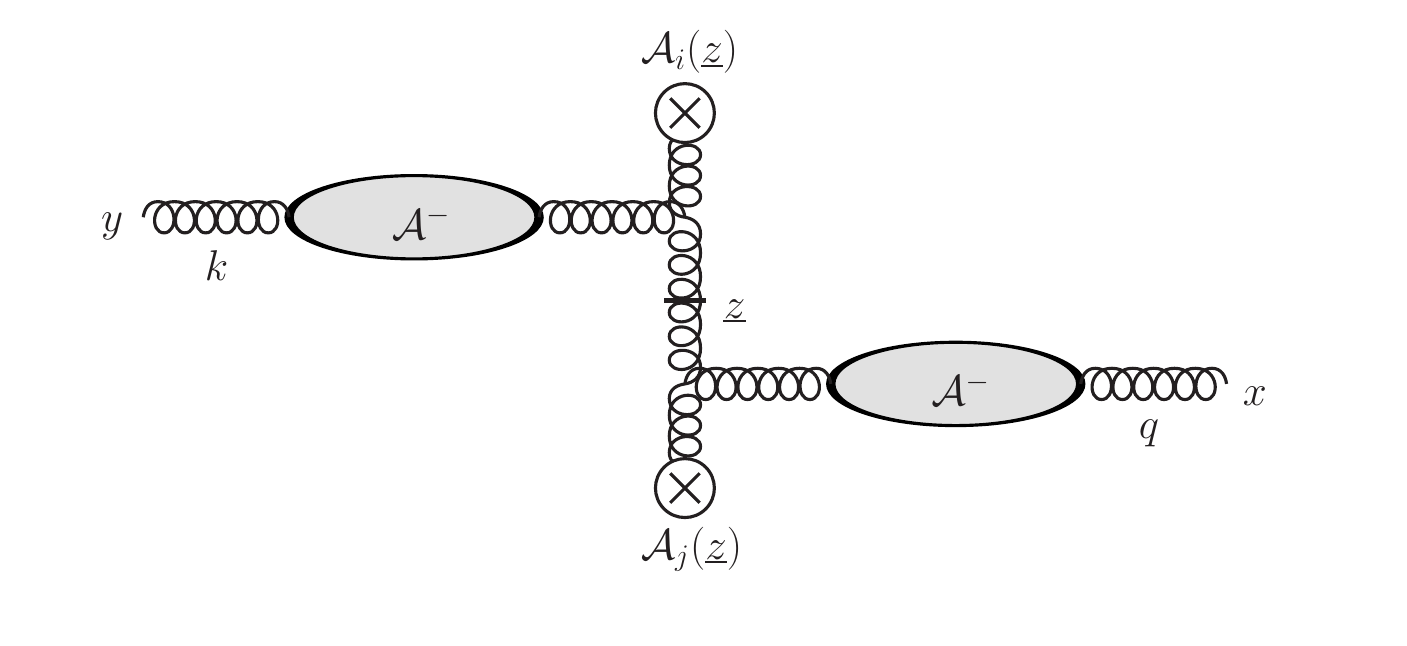}
    \caption{Contribution to the gluon propagator in background field, with an instantaneous non-local insertion of two transverse components of the background field, via two three-gluon vertices, and with interactions with the $\A^{-}$ component resummed to all orders. For the gluon vacuum propagator between the two transverse field insertions, only the instantaneous term is kept.}
    \label{fig: Inst four gluon vertex in medium}
\end{figure}

Another medium correction to the gluon propagator at NEik accuracy is stemming from the insertion of two $\A_{\perp}$ fields from two different three gluon vertices. Naively, one can argue that this contribution would only start at NNEik order, since one should integrate over two different longitudinal positions $z^+$ and $z^{\prime +}$ of the vertices where the two $\A_{\perp}$ fields are inserted, and each integration brings factor of $L^+$. This is usually true, except  in the following case: if the two vertices  where the background fields $\A_{\perp}$ are inserted are connected with the instantaneous part of the gluon propagator, which contain a Dirac delta enforcing $z^+=z^{\prime +}$. Hence, in that case, there is only one non-trivial longitudinal integration, over $z^+$, and thus only one suppression factor $L^+$ instead of two, so that a NEik contribution is obtained, instead of a NNEik one.
We refer to that as the non-local double $\A_{\perp}$ contribution (see Fig. \ref{fig: Inst four gluon vertex in medium}), because the two insertions do not happen at the same $z^-$ and ${z'}^-$  a priori, until the static approximation is taken. 
That contribution to the medium correction to the gluon propagator at NEik accuracy can be written as   
\begin{align}
\label{eq: double_A_perp_non_local_1}
\delta G^{\mu \nu }_{ab}(x,y) \Big|_{\rm double\, \A_{\perp},\, non-loc.}^{\rm NEik}= & 
\int \! d^{4}z' \int \! d^{4}z'' \ \bigl[G^{\mu\mu'}_{F}(x,z')\bigr]_{aa'} \bigl[{\rm X}^{3g}_{\mu'\mu''}(z')\bigr]^{a'a''} 
\nn \\ & \times
\bigl[ G^{\mu'' \nu''}_{F} (z',z'')\bigr]_{a''b''} \bigl[{\rm X}^{3g}_{\nu''\nu'}(z'')\bigr]^{b''b'} \bigl[G^{\nu'\nu}_{F}(z'',y)\bigr]_{b'b}
    \end{align}
%
%
%%%%%%%%%%%%%%%%%%%%%%%%%%%%%%%%%%%%%%%%%%%%%%%%%%%%%%%%%%%%%%%%%%%%%%%%%%%%%%%%%%%%%%%%%%%%%%%%%%%%%%%%%
%%%%%%%%%%%%%%%%%%%%%%%%%%%%%%%%%%%%%%%%%%%%%%%%%%%%%%%%%%%%%%%%%%%%%%%%%%%%%%%%%%%%%%%%%%%%%%%%%%%%%%%%%%%%%
where the effective single $\A_{\perp}$ insertion factor is given in Eq. \eqref{three gluon vertex insertion}. Substituting corresponding expressions to Eq. \eqref{eq: double_A_perp_non_local_1}, and keeping only the instantaneous part of the gluon propapagtor between the $z'$ and $z''$, one gets
\begin{align}
\label{eq:double_A_perp_non_loc_fin}
\delta G^{\mu \nu }_{ab}(x,y) \Big|_{\rm double\, \A_{\perp}, \, non-loc.}^{\rm NEik}= &\,   
g^{2} \int \! d^{3}\uz \int \! \frac{d^{3}\uq}{(2\pi)^{3}}    \frac{ e^{-ix \cdot \check{q} }}{2q^{+}} \ 
%\theta(x^{+}- z^{+}) \ 
\theta(q^{+})
\int \! \frac{d^{3}\uk}{(2\pi)^{3}} \   \frac{ \ e^{iy \cdot \check{k}}}{2k^{+}} \theta(k^{+}) \int\! dz^{-} \ e^{iz^{-}(q^{+}-k^{+})} 
e^{-i \z\cdot(\q-\k)}
\nn\\
& \times U_{A}\big(x^{+}, z^{+}; \z, z^{-}\big)_{aa'}\  (-i)        
\Big[\big(T \cdot \A_{i}(z) \big)\big(T \cdot \A_{j}(z) \big)\Big]_{a'b'}  %\theta(z^{+}- y^{+}) \ 
\
\nn \\
&\times
\Big[ g^{\mu i} g^{j \nu } - \frac{g^{\mu i} \k^{j} \eta^{\nu}}{k^{+}}- \frac{\eta^{\mu} \q^{i}g^{j \nu }}{q^{+}} + \frac{\eta^{\mu}\eta^{\nu} \q^{i} \k^{j}}{q^{+}k^{+}} \Big]      
 \ U_{A}\big(z^{+}, y^{+}; \z, z^{-}\big)_{b'b} + \text{NNEik}
\, .
 \end{align}
%
%
%%%%%%%%%%%%%%%%%%%%%%%%%%%%%%%%%%%%%%%%%%%%%%%%%%%%%%%%%%%%%%%%%%%%%%%%%%%%%%%%%%%%%%%%%%%%%%%%%%%%%%%%%
\subsection{Full NEik Gluon Propagator Through the Medium}
%%%%%%%%%%%%%%%%%%%%%%%%%%%%%%%%%%%%%%%%%%%%%%%%%%%%%%%%%%%%%%%%%%%%%%%%%%%%%%%%%%%%%%%%%%%%%%%%%%%%%%%%%
So far, we have obtained all the medium corrections to the gluon propagator passing through the whole medium at NEik accuracy in a dynamical gluon background field. Their expressions are given in Eqs.~\eqref{A-NEik}, \eqref{eq:single_A_perp_ins_fin_2}, \eqref{eq:double_A_perp_loc_fin} and \eqref{eq:double_A_perp_non_loc_fin}.
 %We can now combine these expressions and write down the before-to-after gluon propagator at NEik accuracy. The result can simply be written as in Eq. \eqref{eq:gluon_prop_NEik_1} with the medium corrections at NEik accuracy to the gluon propagator given Eq. \eqref{medium_corrections_NEik} with each different correction given in Eqs. \eqref{A-NEik}, \eqref{eq:single_A_perp_ins_fin}, \eqref{eq:double_A_perp_loc_fin} and \eqref{eq:double_A_perp_non_loc_fin}.  
 Using these explicit expressions and focusing only on the case where $x^+>y^+$ (with  $x^+>L^+/2$ and $y^+<-L^+/2$), the total before-to-after gluon propagator at NEik accuracy can be organized as 
\begin{align}
\label{general NEik G}
G^{\mu \nu}_{F}(x,y) =G^{\mu \nu}_{F}(x,y)\Big|_{\text{Eik}, \, z^-} +  \delta G^{\mu \nu}_{F}(x,y)\Big|_{\rm NEik \, (1)} + \delta G^{\mu \nu}_{F}(x,y)\Big|_{\rm NEik \, (2)}
+ \text{NNEik} 
\end{align}
where the first contribution is the $x^+>y^+$ contribution in Eq. \eqref{GEikF} 
%(only the $\theta (x^+-y^+) term would contribiute $) with the generalized eikonal contribution  
%% 
%%
%\begin{align}
%\label{GEik}
%\delta G_{F}^{\mu\nu}(x,y) \Big|_{{\rm Eik},\, z^{-}} &= 
%\int \! \frac{d^{3}\uq}{(2\pi)^{3}} \ e^{-ix \cdot \check{q}}  \ \theta(q^{+}) \
%\int \! \frac{d^{3}\uk}{(2\pi)^{3}}  \ e^{iy \cdot \check{k}}  \ \theta(k^{+}) \ \frac{1}{q^{+}+k^{+}} 
%\Big[- g^{\mu \nu} + \frac{\check{k}^{\mu} \eta^{\nu}}{k^{+}} + \frac{\eta^{\mu} \check{q}^{\nu}}{q^{+}} - \frac{\eta^{\mu} \eta^{\nu}}{q^{+} k^{+}} (\check{q} \cdot \check{k}) \Big]
%\nn \\
%& \times
%\int \! d^{2}\z e^{-i(\q - \k)\cdot \z} 
%\int \! dz^{-} \  e^{i(q^{+} -k^{+})z^{-}} \     U_{A}\big(x^{+},y^{+},\z,z^{-}\big)
%\end{align}
%%
%%
and the two NEik corrections are given as  
\begin{align}
\label{GNEik1}
\delta G^{\mu \nu}_{F}(x,y)\Big|_{\rm NEik \, (1)} &= 
\int \! \frac{d^{3}\underline{q}}{(2\pi)^{3}} \  \frac{e^{-ix \cdot \check{q}}}{2q^{+}} \ \theta(q^{+})\ \int \! \frac{d^{3}\underline{k}}{(2\pi)^{3}} \  \frac{e^{iy \cdot \check{k}}}{2k^{+}}\ \theta(k^{+})
\Big[ -g^{\mu\nu} + \frac{\check{k}^{\mu} \eta^{\nu} }{k^{+}} + \frac{\eta^{\mu} \check{q}^{\nu}}{q^{+}} -\frac{\eta^{\mu}\eta^{\nu}}{q^{+}k^{+}} (\check{q} \cdot \check{k}) \Big]
\nn \\
&\times  \int \! dz^{-} \ e^{iz^{-}(q^{+} - k^{+})} \ \int \!  d^{2} \z \ e^{-i\z\cdot(\q-\k)} 
\nn \\
& \times
\bigg\{ - \frac{(\q^{j}+ \k^{j})}{2} \int_{\frac{-L^{+}}{2}}^{\frac{L^{+}}{2}} \! dz^{+}
\Big[ 
U_{A}\Big({\frac{L^{+}}{2}},z^{+}; \z, z^{-}\Big)  
\Big(\overrightarrow{D}_{\z^{j}}^{A}-  \overleftarrow{D}_{\z^{j}}^{A}\Big)  
U_{A}\Big(z^{+},-{\frac{L^{+}}{2}}; \z, z^{-}\Big) \Big] 
\nn \\
& \hspace{0.7cm} 
-i \int_{-\frac{L^{+}}{2}}^{\frac{L^{+}}{2}} \! dz^{+} \Big[ U_{A}\Big(\frac{L^{+}}{2},z^{+}; \z, z^{-}\Big) 
\Big( \overleftarrow{D}_{\z^{j}}^{A} \overrightarrow{D}_{\z^{j}}^{A} \Big) \
U_{A}\Big(z^{+},-\frac{L^{+}}{2}; \z, z^{-}\Big)\Big]
 \bigg\} 
 %+ \text{NNEik}
 \end{align}
%
%
%%%%%%%%%%%%%%%%%%%%%%%%%%%%%%%%%%%%%%%%%%%%%%%%%%%%%%%%
%%%%%%%%%%%%%%%%%%%%%%%%%%%%%%%%%%%%%%%%%%%%%%%%%%%%%%%
%
%
\begin{align}
\label{GNEik2}
\delta {G_{F}^{\mu\nu}}(x,y) \Big|_{\rm NEik \, (2)} &=  -2
\int \! \frac{d^{3}\uq}{(2\pi)^{3}} \   \frac{ e^{-ix \cdot \check{q} } }{2q^{+}} \ \theta(q^{+})
\int \! \frac{d^{3}\uk}{(2\pi)^{3}} \   \frac{ e^{iy \cdot \check{k}}}{2k^{+}}  \theta(k^{+}) \ 
\Big[ g^{\mu i}  - \frac{\eta^{\mu} \q^{i}}{q^{+}}  \Big] \Big[  g^{j \nu } - \frac{ \k^{j} \eta^{\nu}}{k^{+}}  \Big] 
%\Big[ g^{\mu j} g^{\nu i}- \frac{\eta^{\mu}g^{\nu i} \q^{j}}{q^{+}}- \frac{g^{\mu j} \k^{i}\eta^{\nu}}{k^{+}} + \frac{\eta^{\mu}\eta^{\nu} \k^{i}\q^{j}}{q^{+}k^{+}}\Big] 
\nn \\ 
& \hspace{-3cm}
\times          
\int\! dz^{-}\ e^{iz^{-}(q^{+}-k^{+})}
\int \! d^{2}\z e^{-i(\q-\k)\cdot\z} \int dz^{+}  
U_{A}\Big(\frac{L^{+}}{2}, z^{+}; \z, z^{-}\Big) \big[ gT \cdot {\cal F}_{ij}(z)\big] \ U_{A}\Big(z^{+}, -\frac{L^{+}}{2}; \z, z^{-}\Big) 
%+ \text{NNEik} 
\end{align}
Here, we have used 
\begin{align}
{\overrightarrow D}_{\z^i}^R&=\partial_{\z^j}+ig\ T_R\cdot  {\A_j} \\
{\overleftarrow D}_{\z^i}^R&=\partial_{\z^j}-ig\ T_R\cdot  {\A_j} \\
 \overleftarrow{D}^R_{\z^j} \overrightarrow{D}^R_{\z^j} &= 
 \overleftarrow{\partial}_{\z^j} \overrightarrow{\partial}_{\z^j} - ig \bigl( T_R \cdot \A_{j}\bigr) \overrightarrow{\partial}_{\z^j} + ig \overleftarrow{\partial}_{\z^j} \bigl(T_R \cdot \A_{j}\bigr) 
 + g^{2} \bigl( T_R \cdot \A_{j}\bigr)\bigl( T_R \cdot \A_{j}\bigr) \\
  T_R \cdot {\cal F}_{ij} &=  T_R \cdot \partial_{\z^i}\A_j - T_R \cdot \partial_{\z^j}\A_i+ig \big[ T_R \cdot  \A_i, T_R \cdot \A_j\big] 
 %= \partial_{i} \big(igT \cdot A_{j}\big) - \partial_{j} \big( igT \cdot A_{i}\big) + \big[igT \cdot A_{i} , igT \cdot A_{j}\big]$
\end{align}
where $R$ is the representation (adjoint or fundamental), $T_R$ is color generator of the representation $R$  and ${\cal F}_{ij}$ is $ij$ component of the background field strength tensor.  

The NEik corrections given in Eq. \eqref{GNEik1} and \eqref{GNEik2} can be written in a more compact form. For that purpose, let us first remind a  property of  a generic Wilson line which reads
\begin{align}
\partial_{z^\mu}U_{R}(x^{+},y^{+};\z,z^{-})&= 
-ig \int_{y^{+}}^{x^{+}} dz^{+}
\ U_{R}(x^{+},y^{+};\z,z^{-})\  T_{R}\cdot \mathcal{F}^{\:-}_{\mu}(z) \ U_{R}(x^{+},y^{+};\z,z^{-})  
\nn \\
&
-ig\ T_{R}\cdot \A_{\mu}(x^{+},\z,z^{-}) \ U_{R}(x^{+},y^{+};\z,z^{-})
+ ig\ U_{R}(x^{+},y^{+};\z,z^{-})\ T_{R}\cdot \A_{\mu}(y^{+},\z,z^{-})
 \end{align}
 where the component $\mu$ can be the transverse ($j$) or minus ($-$) component. Note that we are using a gauge in which the gauge fields vanish outside of the target, so that we have 
 \begin{align}
  \A_{\mu}\Big(\frac{L^+}{2},\z,z^{-}\Big)= \A_{\mu}\Big(\frac{-L^+}{2},\z,z^{-}\Big)=0
  \, .
 \end{align}
 Therefore, decorated Wilson lines that appear at NEik order corrections to the gluon propagator given in Eqs. \eqref{GNEik1} and \eqref{GNEik2} can be written in the following compact forms: 
\begin{align}
\label{eq:def_Dec_Wil_1}
U_{R;j}^{(1)}(\z,z^{-}) &= 
\int_{-\frac{L^+}{2}}^{\frac{L^+}{2}}\! dz^{+}\ U_{R}\Big(\frac{L^+}{2},z^{+},\z,z^{-}\Big)
\Big(\overrightarrow{D}_{\z^{j}}- \overleftarrow{D}_{\z^{j}}\Big) 
\ U_{R}\Big(z^{+},-\frac{-L^+}{2},\z,z^{-})
\nn \\
&= -2\int_{-\frac{L^+}{2}}^{\frac{L^+}{2}}\! dz^{+} \ z^{+} \ 
U_{R}\Big(\frac{L^+}{2},z^{+},\z,z^{-}\Big) \ 
\Big[-igT_{R}\cdot\mathcal{F}^{\:-}_{j}(z^{+},\z, z^{-})\Big]
\ U_{R}\Big(z'^{+},\frac{-L^+}{2},\z,z^{-}\Big)
\end{align}
\begin{align}
\label{eq:def_Dec_Wil_2}
U_{R}^{(2)}(\z,z^{-}) &= 
\int_{-\frac{L^+}{2}}^{\frac{L^+}{2}}\! dz^{+}
\ U_{R}\Big(\frac{L^+}{2},z^{+},\z,z^{-}\Big) 
\overleftarrow{D}_{\z^{j}} \overrightarrow{D}_{\z^{j}} 
\ U_{R}\Big(z^{+},\frac{-L^+}{2},\z,z^{-}\Big)
\nn \\
&
= \int_{-\frac{L^+}{2}}^{\frac{L^+}{2}}\! dz^{+}
\int_{-\frac{L^+}{2}}^{\frac{L^+}{2}}\! dz'^{+} 
\ \theta(z^{+}-z'^{+}) \ (z^{+}-z'^{+}) 
\ U_{R}\Big(\frac{L^+}{2},z^{+},\z,z^{-}\Big) 
\big[-igT_{R}\cdot\mathcal{F}^{-}_{j}(z^{+}\z,z^{-})\big] 
\nn \\
&
\times
U_{R}\big(z^{+},z'^{+},\z,z^{-}\big) \ 
\big[-igT_{R}\cdot\mathcal{F}_{j}^{-}(z'^{+},\z,z^{-})\big] 
\ U_{R}\Big(z'^{+},\frac{-L^+}{2},\z,z^{-}\Big)
\end{align}
and 
\begin{align}
\label{eq:def_Dec_Wil_3}
U_{R;ij}^{(3)}(\z,z^{-}) &= 
\int_{-\frac{L^+}{2}}^{\frac{L^+}{2}}\! dz^{+}
\ U_{R}\Big(\frac{L^+}{2},z^{+},\z,z^{-}\Big) \ gT_{R}\cdot\mathcal{F}_{ij}(z) 
\ U_{R}\Big(\frac{L^+}{2},z^{+},\z,z^{-}\Big)
\end{align}
As a remark, the target width $L^+$ was introduced as a tool to clarify the power counting at high energy. In the expressions \eqref{eq:def_Dec_Wil_1}, \eqref{eq:def_Dec_Wil_2} and \eqref{eq:def_Dec_Wil_3} in terms of background field strength insertions, it is safe to replace $L^+$ by $+\infty$ both in the integrations bounds and in the endpoints of Wilson lines. Indeed, our power counting is ensured by the faster than power decay of the background field strength  components (and gauge field, in light cone gauge) for $x^+\rightarrow \pm \infty$, making $L^+$ now redundant. In particular, in the following, when we omit the initial and final $z^+$ coordinates of a Wilson line, we mean a Wilson line through the whole target, which can be equivalently considered to go from $-L^+/2$ to $L^+/2$, or from $-\infty$ to $+\infty$, as
\begin{align}
U_{R}(\z,z^{-}) &\equiv  U_{R}\Big(\frac{L^+}{2},-\frac{L^+}{2},\z,z^{-}\Big) = U_{R}(+\infty,-\infty,\z,z^{-})
\, .
\end{align}

Using the compact expressions for the decorated Wilson lines introduced in Eqs. \eqref{eq:def_Dec_Wil_1}, \eqref{eq:def_Dec_Wil_2} and \eqref{eq:def_Dec_Wil_3} to rewrite the NEik corrections to the gluon propagator given in Eqs. \eqref{GNEik1} and \eqref{GNEik2}, one obtains 
\begin{align}
\label{eq:NEik_g_Comp_1}
\delta G_{1,F}^{\mu\nu}(x,y)  &= 
\int \! \frac{d^{3}\underline{q}}{(2\pi)^{3}} \  \frac{\theta(q^{+})}{2q^{+}} \ e^{-ix \cdot \check{q}}  \int \! \frac{d^{3}\underline{k}}{(2\pi)^{3}} \  \frac{\theta(k^{+})}{2k^{+}} \ e^{iy \cdot \check{k}}
\Big[ -g^{\mu\nu} + \frac{\check{k}^{\mu} \eta^{\nu} }{k^{+}} +
\frac{\eta^{\mu} \check{q}^{\nu}}{q^{+}}
-\frac{\eta^{\mu}\eta^{\nu}}{q^{+}k^{+}} (\check{q} \cdot \check{k}) \Big]
\nn \\
&
\times  
\int \! dz^{-} \ e^{iz^{-}(q^{+} - k^{+})}
\int \!  d^{2} \z \ e^{-i\z\cdot(\q-\k)} 
\Big[ - \frac{(\q^{j}+ \k^{j})}{2} U_{A,j}^{(1)}(\z,z^{-}) -i U_{A}^{(2)}(\z,z^{-})\Big]
%+ \text{NNEik}
 \end{align}
and 
\begin{align}
\label{eq:NEik_g_Comp_2}
\delta {G_{2,F}^{\mu\nu}}(x,y)  & = 
-2
\int \! \frac{d^{3}\uq}{(2\pi)^{3}}  \frac{ \theta(q^{+}) }{2q^{+}}  e^{-ix \cdot \check{q} } 
\int \! \frac{d^{3}\uk}{(2\pi)^{3}}  \frac{ \theta(k^{+}) }{2k^{+}}  e^{iy \cdot \check{k}}    
\Big[ g^{\mu i}  - \frac{\eta^{\mu} \q^{i}}{q^{+}}  \Big] \Big[  g^{j \nu } - \frac{ \k^{j} \eta^{\nu}}{k^{+}}  \Big] 
\nn \\
&
\times          
\int\! dz^{-} e^{iz^{-}(q^{+}-k^{+})}
\int \! d^{2}\z \ e^{-i\z\cdot(\q-\k)}  \ U_{A;ij}^{(3)}(\z,z^{-}) 
%+ \text{NNEik} 
\, .
\end{align}
The final expression for the total before-to-after gluon propagator at NEik accuracy is given by Eq. \eqref{general NEik G} where the eikonal propagator (with $z^-$ dependence) is given by Eq. \eqref{GEikF} and the NEik corrections are given by Eqs. \eqref{GNEik1} and \eqref{GNEik2}. As discussed above, the total gluon propagator can also be written in a more compact form by using the Eqs. \eqref{eq:NEik_g_Comp_1} and \eqref{eq:NEik_g_Comp_2} in Eq. \eqref{general NEik G} for the NEik corrections. 

As a reminder, the first term, given in Eq. \eqref{GEikF}, contain both the strict Eikonal contribution, and the NEik corrections beyond the static limit. The second term, given in 
Eq. \eqref{eq:NEik_g_Comp_1}, contains the NEik corrections beyond the shockwave limit, written in terms of a gauge covariant operator by including some of the $\A_{\perp}$ insertions as well. These two terms are direct analogs to the ones found for the scalar and quark propagators at NEik accuracy \cite{Altinoluk:2020oyd,Altinoluk:2021lvu}. By contrast, there is no further term in the scalar propagator. In the quark propagator, there is an analog to the last term of our results \eqref{GNEik2}, which was interpreted as a coupling term between the light-front helicity of the propagating parton with the longitudinal chromomagnetic field of the target $\mathcal{F}_{12}$.

\section{Various parton propagators in a gluon background field}
\label{sec:Various_parton_prop}
%%%%%%%%%%%%%%%%%%%%%%%%%%%%%%%%%%%%%%%%%%%%%%%%%%%%%%%%%%%%%%%%%%%%%%%%%%%%%%%%%%%%%%%%%%%%%%%%%%%%%%%%%%%%%%%%%%%%%%%%%%%%%%%%%%%%%%%%%%%%%%%%%%
%%%%%%%%%%%%%%%%%%%%%%%%%%%%%%%%%%%%%%%%%%%%%%%%%%%%%%%%%%%%%%%%%%%%%%%%%
This section is devoted to provide the expressions for both quark and gluon propagators with either their starting position or final position inside the medium. In the calculation of scattering processes at NEik accuracy, the expression of the propagators in these special configurations are typically needed only at eikonal (or generalized eikonal) order, and we will thus restrict ourselves to that precision in this section. Some of those propagators have been computed in our earlier works \cite{Altinoluk:2022jkk,Altinoluk:2023qfr}. 
The details of the derivation of the new propagators are provided in Appendices \ref{App:inside-to-inside-quark} and \ref{App:inside-to-after-gluon-prop} while their results are given in this section.  

%%%%%%%%%%%%%%%%%%%%%%%%%%%%%%%%%%%%%%%%%%%%%%%%%%%%%%%%%%%%%%%%%%%%%%%%%
\subsection{Inside-to-after quark propagator} 
%%%%%%%%%%%%%%%%%%%%%%%%%%%%%%%%%%%%%%%%%%%%%%%%%%%%%%%%%%%%%%%%%%%%%%%%%
The quark propagator from inside to after the medium, corresponding to $x^+>L^+/2$ and $-L^+/2<y^+<L^+/2$,  is computed at generalized eikonal order in Ref.~\cite{Altinoluk:2022jkk} and it reads 
\begin{align}
\label{q IA}
S_{F}(x,y)%_{\beta\alpha}
\big|^{\rm IA,q}_{\rm Gen. Eik} &= \int\! \frac{d^{3}\uq}{(2\pi)^{3}} \frac{\theta(q^{+})}{2q^{+}}\ 
e^{-ix\check{q}} \ (\check{\sq}+m) \ U_{F}(x^{+},y^{+};\y, y^{-})%_{\beta\alpha}\ 
\bigg[1-\frac{\gamma^{+}\gamma^{i}}{2q^{+}}i\overleftarrow{D}^{F}_{\y^i}\bigg] \ e^{iy^{-}q^{+}} \ e^{-i\y\cdot\q}
\end{align}
which in the strict eikonal limit can be written as 
\begin{align}
\label{q_IA_Eik}
S_{F}(x,y)%_{\beta\alpha}
\big|^{\rm IA,q}_{\rm Eik} &= \int\! \frac{d^{3}\underline{q}}{(2\pi)^{3}} \frac{\theta(q^{+})}{2q^{+}}\ e^{-ix\check{q}} \ (\check{\sq}+m) \ U_{F}(x^{+},y^{+};\y)%_{\beta\alpha}\ 
\bigg[1-\frac{\gamma^{+}\gamma^{i}}{2q^{+}}i\overleftarrow{D}^{F}_{\y^i}\bigg] \ e^{iy^{-}q^{+}} \ e^{-i\y\cdot\q}
\, .
\end{align}
We also need to consider the case in which the points $x$ and $y$ are interchanged, meaning that $y^+>L^+/2$ and $-L^+/2<x^+<L^+/2$, which corresponds to what we call 
the inside-to-after antiquark propagator. It
 is also computed in \cite{Altinoluk:2022jkk} and at strict eikonal accuracy it reads 
\begin{align}
S_{F}(x,y)%_{\beta\alpha}
\big|^{\rm IA,\bar{q}}_{\rm Eik} &= \int\! \frac{d^{3}\underline{q}}{(2\pi)^{3}} (-1)\frac{\theta(-q^{+})}{2q^{+}}\ e^{iy\check{q}} \
e^{-ix^{-}q^{+}} \ e^{i\x\cdot\q}\  
\bigg[1-\frac{\gamma^{+}\gamma^{i}}{2q^{+}}i\overrightarrow{D}^{F}_{\x^i}\bigg] \ 
      \ U_{F}^{\dagger}(y^{+},x^{+},\x)%_{\beta\alpha}\ 
       (\check{\sq}+m)
\end{align}
%
%
%%%%%%%%%%%%%%%%%%%%%%%%%%%%%%%%%%%%%%%%%%%%%%%%%%%%%%%%%%%%%%%%%%%%%%%%%
\subsection{Before-to-inside quark propagator} 
%%%%%%%%%%%%%%%%%%%%%%%%%%%%%%%%%%%%%%%%%%%%%%%%%%%%%%%%%%%%%%%%%%%%%%%%%
The quark propagator from before to inside the medium, with $-L^+/2<x^+<L^+/2$ and $y^+<-L^+/2$, can be computed in a similar way and at generalized eikonal order it reads 
\begin{align}
\label{BI quark}
S_{F}(x,y)%_{\beta\alpha} 
\big|_{\rm Gen. Eik}^{\rm BI,q} &= 
 \int \! \frac{d^{3}\uk}{(2 \pi)^3} \frac{\theta(k^+)}{2k^+} \ e^{iy \cdot \check{k}}\ e^{-ix^-k^+} e^{i\x\cdot \k} \ 
 \bigg[1 - \frac{i\gamma^+ \gamma^i}{2k^+}\overrightarrow{D}_{x^i}^{F}\bigg] (\check{\sk} + m) 
 \ U_{F}(x^+,y^+;\x, x^{-})%_{\beta\alpha}
 \end{align}
which in the strict eikonal limit can be written as 
\begin{align}
S_{F}(x,y)%_{\beta\alpha} 
\big|_{\rm Eik}^{\rm BI,q} &= 
 \int \! \frac{d^{3}\underline{k}}{(2 \pi)^3} \frac{\theta(k^+)}{2k^+} 
 \ e^{iy \cdot \check{k}} \ e^{-ix^-k^+} \ e^{i\x\cdot \k} \ 
 \bigg[1 - \frac{i\gamma^+ \gamma^i}{2k^+}\overrightarrow{D}_{\x^i}^{F}\bigg] (\check{\sk} + m) 
 \ U_{F}(x^+,y^+,\x)%_{\beta \alpha}
\end{align}
The before-to-inside antiquark propagator, corresponding to  $x^+<-L^+/2$ and $-L^+/2<y^+<L^+/2$, can be computed in a similar manner and at strict eikonal accuracy it reads  
\begin{align}
S_{F}(x,y)%_{\beta\alpha}
 \big|_{\rm Eik}^{\rm BI,\bar q} &= 
\int \! \frac{d^{3}\underline{k}}{(2 \pi)^3} (-1)\frac{\theta(-k^+)}{2k^+}  e^{-ix \cdot \check{k}} (\check{\sk} + m) 
\ U_{F}^{\dagger}(y^+,x^+,\y)%_{\beta \alpha}\  \
\bigg[1 - \frac{i\gamma^+ \gamma^i}{2k^+}\overleftarrow{D}_{\y^i}^{F}\bigg] 
e^{iy^-k^+} \ e^{-i\y\cdot \k} 
\, .
\end{align}
%
%
%%%%%%%%%%%%%%%%%%%%%%%%%%%%%%%%%%%%%%%%%%%%%%%%%%%%%%%%%%%%%%%%%%%%%%%%%
\subsection{Inside-to-inside quark propagator} 
%%%%%%%%%%%%%%%%%%%%%%%%%%%%%%%%%%%%%%%%%%%%%%%%%%%%%%%%%%%%%%%%%%%%%%%%%
The quark propagator from inside-to-inside the medium with $-L^+/2<x^+<L^+/2$ and $-L^+/2<y^+<L^+/2$ have not been computed previously. Here, we provide the final result of the inside-to-inside quark propagator for $x^+>y^+$ at generalized eikonal order while a detailed derivation of this propagator can be found in Appendix \ref{App:inside-to-inside-quark}. The final result reads 
\begin{align}
 \label{eq:inside-inside_quark}
 S_{F}(x,y)%_{\beta \alpha} 
 \big|_{\rm Gen. Eik}^{\rm II,q} &=  \int \! \frac{dk^+}{2\pi}   \frac{\theta(k^+)}{2k^+} \ e^{-ik^+(x^- - y^-)}  \bigg\{   
 \bigg[ k^+ \gamma^- + m + i \gamma^i \overrightarrow{D}_{\x^i}^{F} \bigg] \, \frac{\gamma^+}{2k^+}
 \nn \\ 
 &\times 
 \bigg[ \int \! d^{2}\z \
\delta^2(\x - \z) \ \delta^2(\z- \y) \ U_{F}(x^+,y^+;\z, x^{-})%_{\beta \alpha}
\bigg]  
\ \bigg[k^+ \gamma^- + m -i \gamma^j \overleftarrow{D}_{\y^j}^{F}\bigg] \bigg\}
\end{align}
A similar derivation can be performed for the inside-to-inside antiquark propagator with $-L^+/2<x^+<L^+/2$ and $-L^+/2<y^+<L^+/2$, and for $y^+>x^+$ the final result at generalized eikonal accuracy is given as 
\begin{align}
\label{eq:inside-inside-antiquark}
S_{F}(x,y)%_{\beta \alpha}
 \big|_{\rm Gen. Eik}^{\rm II, \bar q} &=  \int \! \frac{dk^+}{2\pi}   \frac{\theta(-k^+)}{2k^+} \ e^{-i(x^- - y^-)k^+}  
\bigg\{ \bigg[ k^+ \gamma^- + m + i \gamma^i \overrightarrow{D}_{\x^i}^{F} \bigg]  
\frac{\gamma^+}{2k^+} \nn 
\\ & \times 
\bigg[ \int \! d^{2}\z \, \delta^2(\y - \z) \ \delta^2(\z- \x) \
(-1)\ U_{F}^{\dagger}(y^+,x^+,\z, y^{-})%_{\beta \alpha}
\bigg]  
\ \bigg[k^+ \gamma^- + m -i \gamma^j \overleftarrow{D}_{\y^j}^{F}\bigg] \bigg\}
\, .
\end{align}
In the most general inside-inside configuration, where both points $x$ and $y$ are inside the medium, but no ordering between them is assumed, both contributions \eqref{eq:inside-inside_quark} and \eqref{eq:inside-inside-antiquark} should be summed, as well as the instantaneous term from the vacuum quark propagator. Hence, assuming only $-L^+/2<x^+<L^+/2$ and $-L^+/2<y^+<L^+/2$, the full inside-inside  quark propagator is found to be
\begin{align}
 \label{eq:inside-inside_quark_full}
 S_{F}(x,y)%_{\beta \alpha} 
 \big|_{\rm Gen. Eik}^{\rm II, \textrm{ full}} = &\, \int \! \frac{dk^+}{2\pi}   \frac{1}{2k^+} \ e^{-ik^+(x^- - y^-)}  \bigg\{  
i\gamma^+ \delta(x^{+}\!-\!y^{+}) \delta^{2}(\x \!-\! \y)
 \nn \\ 
 &
 +
 \bigg[ k^+ \gamma^- + m + i \gamma^i \overrightarrow{D}_{\x^i}^{F} \bigg] \, \frac{\gamma^+}{2k^+}
 %\bigg[ 
 \int \! d^{2}\z \
\delta^2(\x - \z) \ \delta^2(\z- \y) \ 
 \nn \\ 
 &\times 
\bigg[\theta(x^+\!-\!y^+)\theta(k^+) U_{F}(x^+,y^+;\z, x^{-})%_{\beta \alpha}
-\theta(y^+\!-\!x^+)\theta(-k^+)
U_{F}^{\dagger}(y^+,x^+,\z, y^{-})%_{\beta \alpha}
\bigg]  
\nn \\ 
 &\times 
\ \bigg[k^+ \gamma^- + m -i \gamma^j \overleftarrow{D}_{\y^j}^{F}\bigg] \bigg\}
\end{align}
\, .
%
%

%%%%%%%%%%%%%%%%%%%%%%%%%%%%%%%%%%%%%%%%%%%%%%%%%%%%%%%%%%%%%%%%%%%%%%%%%
\subsection{Inside-to-after gluon propagator} 
%%%%%%%%%%%%%%%%%%%%%%%%%%%%%%%%%%%%%%%%%%%%%%%%%%%%%%%%%%%%%%%%%%%%%%%%%
The derivation of the inside-to-after gluon propagator, with $x^+>L^+/2$ and $-L^+/2<y^+<L^+/2$, %, and for $x^+>y^+$ 
 is presented in detail in Appendix \ref{App:inside-to-after-gluon-prop}. At generalized eikonal accuracy the final result can be written as
\begin{align}
\label{eq:inside-after_gluon}
 G_{F}^{\mu\nu}(x,y) \big|_{\rm Gen. Eik}^{\text{IA}} 
 =&\,   \int \! \frac{d^{3}\uq}{(2\pi)^{3}} 
\ \frac{\theta(q^{+})}{2q^+} \ e^{-ix \cdot \check{q}} \   
U_{A}(x^{+},y^{+},\y,y^{-}) 
%\nn \\
%& \times  
\bigg[  g^{\mu j} - \frac{\eta^{\mu}  \q^{j}}{q^{+}}
  \bigg] 
 \bigg[ g^{j \nu} -\frac{  \eta^{\nu}}{q^{+}}
 \Big(i\overleftarrow{D}_{\y^j}^{A} +\q^{j} \Big)  \bigg] 
 \ e^{-i\q\cdot \y} \ e^{iy^-q^+}
 \, .
\end{align}
In order to get the strict eikonal limit of the inside-to-after gluon propagator, one can perform a gradient expansion in $y^-$ of the Wilson line in Eq. \eqref{eq:inside-after_gluon}  and keep only the zeroth order term in that gradient expansion, meaning that the dependence on $y^-$ of the Wilson line is neglected.  Thus, it reads
\begin{align}
\label{IA G_Eik}
 G_{F}^{\mu\nu}(x,y) \big|_{\rm Eik}^{\text{IA}} 
 =&\,   \int \! \frac{d^{3}\uq}{(2\pi)^{3}} 
\ \frac{\theta(q^{+})}{2q^+} \ e^{-ix \cdot \check{q}} \   
U_{A}(x^{+},y^{+},\y)
% \nn \\
%& \times  
\bigg[  g^{\mu j} - \frac{\eta^{\mu}  \q^{j}}{q^{+}}
  \bigg] 
 \bigg[ g^{j \nu} -\frac{  \eta^{\nu}}{q^{+}}
 \Big(i\overleftarrow{D}_{\y^j}^{A} +\q^{j} \Big)  \bigg] 
 \ e^{-i\q\cdot \y} \ e^{iy^-q^+}
 \, .
\end{align}
%
%
%%%%%%%%%%%%%%%%%%%%%%%%%%%%%%%%%%%%%%%%%%%%%%%%%%%%%%%%%%%%%%%%%%%%%%%%%
\subsection{Before-to-inside gluon propagator} 
%%%%%%%%%%%%%%%%%%%%%%%%%%%%%%%%%%%%%%%%%%%%%%%%%%%%%%%%%%%%%%%%%%%%%%%%%
The before-to-inside gluon propagator, with $-L^+/2<x^+<L^+/2$ and $y^+<-L^+$/2
%, and for $x^+>y^+$ 
can be computed in a similar way to the inside-to-after gluon propagator. The final result at generalized eikonal order can be organized as 
\begin{align}
\label{eq:before-inside_gluon}
G^{\mu \nu}_{F}(x,y)\big|^{\rm BI}_{\rm Gen. Eik} 
&= \int \! \frac{d^{3}\uk}{(2\pi)^{3}}\ \frac{\theta(k^{+})}{2k^{+}} 
\ e^{iy\check{k}} \ e^{-ix^{-}k^{+}}\ e^{i\x\cdot\k} 
%\nn \\
%&\times
\bigg[g^{\mu i}
+\frac{\eta^{\mu}}{k^{+}}
\Big(i\overrightarrow{D}_{\x^i}^{A} -\k^{i}\Big) \bigg] \ U_{A}(x^{+},y^{+},\x,x^{-})
\bigg[g^{i\nu} - \frac{\eta^{\nu}\k^{i}}{k^{+}} 
 \bigg]
\end{align}
which in the strict eikonal limit can be written as 
\begin{align}
G^{\mu \nu}_{F}(x,y)\big|^{\rm BI}_{\rm Eik} 
&= \int \! \frac{d^{3}\uk}{(2\pi)^{3}}\ \frac{\theta(k^{+})}{2k^{+}} 
\ e^{iy\check{k}} \ e^{-ix^{-}k^{+}}\ e^{i\x\cdot\k} 
%\nn \\
%&\times
\bigg[g^{\mu i}
+\frac{\eta^{\mu}}{k^{+}}
\Big(i\overrightarrow{D}_{\x^i}^{A} -\k^{i}\Big) \bigg] \ U_{A}(x^{+},y^{+},\x)
\bigg[g^{i\nu} - \frac{\eta^{\nu}\k^{i}}{k^{+}} 
 \bigg]
 \, .
\end{align}
%
%
%%%%%%%%%%%%%%%%%%%%%%%%%%%%%%%%%%%%%%%%%%%%%%%%%%%%%%%%%%%%%%%%%%%%%%%%%
\subsection{Inside-to-inside gluon propagator} 
%%%%%%%%%%%%%%%%%%%%%%%%%%%%%%%%%%%%%%%%%%%%%%%%%%%%%%%%%%%%%%%%%%%%%%%%%
The derivation of the inside-to-inside gluon propagator can be performed by following the same steps that are introduced in Appendix \ref{App:inside-to-inside-quark} for the derivation of the inside-to-inside quark propagator. The final result of the inside-to-inside gluon propagator with $-L^+/2<x^+<L^+/2$ and $-L^+/2<y^+<L^+/2$ for $x^+>y^+$ at generalized eikonal order can be written as 
\begin{align}
 \label{eq:inside-inside_gluon}
G^{\mu \nu}_{F}(x,y)\big|^{\rm II,\, x^+>y^+}_{\rm Gen. Eik} &= \int\! \frac{dk^+}{2\pi} \frac{\theta(k^+)}{2k^{+}} \ e^{-ik^{+}(x^{-}-y^{-})} 
\bigg[g^{\mu}_{i} - \frac{i\eta^{\mu}}{k^{+}} \overrightarrow{D}_{\x^i}^{A} \bigg] \nn \\
& \times
\bigg[\int\! d^{2}\z \ \delta^{2}(\x-\z) \ \delta^{2}(\z-\y) \ U_{A}(x^{+},y^{+},\z, x^{-})\bigg] 
\bigg[g^{\nu}_{i} + \frac{i\eta^{\nu}}{k^+}\overleftarrow{D}_{\y^i}^{A}\bigg]
\, .
\end{align}    
Like in the quark case, if one assumes only $x$ and $y$ to be inside, meaning $-L^+/2<x^+<L^+/2$ and $-L^+/2<y^+<L^+/2$, but no specific ordering between $x^+$ and $y^+$, one needs to include not only the contribution \eqref{eq:inside-inside_gluon}, but also the symmetric one with  $y^+>x^+$, as well as the instantaneous contribution from the vacuum gluon propagator. In such a way, one finds the total inside-to-inside gluon propagator
\begin{align}
 \label{eq:inside-inside_gluon}
G^{\mu \nu}_{F}(x,y)\big|^{\rm II,\, full}_{\rm Gen. Eik} &= \int\! \frac{dk^+}{2\pi} \frac{1}{2k^{+}} \ e^{-ik^{+}(x^{-}-y^{-})} 
\Bigg\{\frac{2i}{k^+}\, \eta^{\mu}\eta^{\nu}\,   \delta(x^{+}\!-\!y^{+}) \delta^{2}(\x \!-\! \y)
 \nn \\
& \times
\bigg[g^{\mu}_{i} - \frac{i\eta^{\mu}}{k^{+}} \overrightarrow{D}_{\x^i}^{A} \bigg]\int\! d^{2}\z \ \delta^{2}(\x-\z) \ \delta^{2}(\z-\y) \  
 \nn \\
&
+\bigg[\theta(x^+\!-\!y^+)\theta(k^+) U_{A}(x^{+},y^{+},\z, x^{-})
-\theta(y^+\!-\!x^+)\theta(-k^+)
U_{A}^{\dagger}(y^+,x^+,\z, y^{-})
\bigg] 
\bigg[g^{\nu}_{i} + \frac{i\eta^{\nu}}{k^+}\overleftarrow{D}_{\y^i}^{A}\bigg]
\Bigg\}
\, .
\end{align}    
%
%

%%%%%%%%%%%%%%%%%%%%%%%%%%%%%%%%%%%%%%%%%%%%%%%%%%%%%%%%%%%%%%%%%%%%%%%%%%%%%%%%%%%%%%%%%%%%%%%%%%%%%%%%%%%%%%%%%%%%%%%%%%%%%%%%%%%%%%%%%%%%%%%%%%%%%%%%%%%%%%%%%%%%%%%%
%%%%%%%%%%%%%%%%%%%%%%%%%%%%%%%%%%%%%%%%%%%%%%%%%%%%%%%%%%%%%%%%%%%%%%%%%%%%%%%%%%%%%%%%%%%%%%%%%%%%%%%%%%%%%%%%%%%%%%%%%%%%%%%%%%%%%%%%%%%%%%%%%%%%%%%%%%%%%%%%%%%%%%%%
%%%%%%%%%%%%%%%%%%%%%%%%%%%%%%%%%%%%%%%%%%%%%%%%%%%%%%%%%%%%%%%%%%%%%%%%%%%%%%%%%%%%%%%%%%%%%%%%%%%%%%%%%%%%%%%%%%%%%%%%%%%%%%%%%%%%%%%%%%%%%%%%%%%%%%%%%%%%%%%%%%%%%%%%
%%%%%%%%%%%%%%%%%%%%%%%%%%%%%%%%%%%%%%%%%%%%%%%%%%%%%%%%%%%%%%%%%%%%%%%%%%%%%%%%%%%%%%%%%%%%%%%%%%%%%%%%%%%%%%%%%%%%%%%%%%%%%%%%%%%%%%%%%%%%%%%%%%%%%%%%%%%%%%%%%%%%%%%%
%%%%%%%%%%%%%%%%%%%%%%%%%%%%%%%%%%%%%%%%%%%%%%%%%%%%%%%%%%%%%%%%%%%%%%%%%%%%%%%%%%%%%%%%%%%%%%%%%%%%%%%%%%%%%%%%%%%%%%%%%%%%%%%%%%%%%%%%%%%%%%%%%%%%%%%%%%%%%%%%%%%%%%%%
  \section{Forward gluon production in gluon-nucleus scattering at NEik accuracy}
  \label{sec:gluon_production_in_gA}
%%%%%%%%%%%%%%%%%%%%%%%%%%%%%%%%%%%%%%%%%%%%%%%%%%%%%%%%%%%%%%%%%%%%%%%%%%%%%%%%%%%%%%%%%%%%%%%%%%%%%%%%%%%%%%%%%%%%
So far we have computed various parton propagators relevant for scattering processes at NEik accuracy in Sec. \ref{Sec:before-to-after-gluon_prop} and \ref{sec:Various_parton_prop}. This section is devoted to an application of those results to forward single inclusive gluon production in gluon-nucleus collisions at NEik accuracy. The cross sections that we consider in the rest of this manuscript are partonic cross sections. In general, in order to be able to compare with the experimental data one should consider the hadronic cross sections. This would require convolution of the computed partonic cross section with the relevant parton distribution functions (PDFs) and fragmentation functions (FFs). Our goal in this manuscript is to provide the partonic cross sections at full NEik accuracy for the first time and since we do not intend to provide numerical comparison of our results with the experimental data, we only consider the partonic cross sections. Moreover, the analysis and discussion of the NEik corrections to cross sections are independent of the convolution of the partonic results with PDFs and FFs. This step, together with the numerical comparison of our results with the experimental data, is left for future studies since models to describe the operators that are obtained at NEik order are not available yet. 

At partonic level, forward gluon production in gluon-nucleus scattering gets two separate contributions at NEik accuracy. 
\begin{align}
\label{eq:gAtogX_gen}
\frac{d \sigma^{gA\rightarrow g+X}}{d\rm P.S.} = \frac{d \sigma^{gA\rightarrow g+X}}{d\rm P.S.}\bigg|_{g \, \rm  backg.} 
+ \frac{d \sigma^{gA\rightarrow g+X}}{d\rm P.S.}\bigg|_{q \, \rm backg.} 
\end{align}
As discussed in detail in Sec. \ref{Sec:before-to-after-gluon_prop}, one contribution originates from relaxing all three assumptions that are required by the eikonal approximation in a pure gluon background (first term in Eq. \eqref{eq:gAtogX_gen}). For the second contribution  in Eq. \eqref{eq:gAtogX_gen}, a quark background field for the target is turned on, on top of the gluon background field, and interactions of the propagating gluon with that quark background field are calculated at NEik order, accounting for scattering via t-channel quark exchange with the target.
%  and another contribution originates from the scattering of the incoming gluon with the target via a t-channel quark exchange, therefore refereed to as quark background contribution (second term in Eq. \eqref{eq:gAtogX_gen}). Note however that the gluon background field is still  
In the rest of this section, we compute each term separately and finally give the total result.  
%%%%%%%%%%%%%%%%%%%%%%%%%%%%%%%%%%%%%%%%%%%%%%%%%%%%%%%%%%%%%%%%%%%%%%%%%%%%%%%%%%%%%%%%%%%%%%%%%%%%%%
\subsection{Forward gluon production cross section in gluon-nucleus scattering at NEik accuracy: pure gluon background contribution}   
%%%%%%%%%%%%%%%%%%%%%%%%%%%%%%%%%%%%%%%%%%%%%%%%%%%%%%%%%%%%%%%%%%%%%%%%%%%%%%%%%%%%%%%%%%%%%%%%%%%%%%
First we focus on the gluon background contribution to the forward gluon production cross section in gluon-nucleus scattering. For that purpose, consider an incoming gluon with four momenta $p_1$, polarization $\lambda_1$  and color $a_1$ in the initial state and the outgoing gluon with four momenta $p_2$, polarization $\lambda_2$  and color $a_2$ in the final state. Then, the S-matrix element for that process can be obtained thanks to the following LSZ-type reduction formula
\begin{align}
\label{S_matrix}
S_{g_2 \leftarrow \, g_1} &=
\lim_{x^{+}\to +\infty} ( -1)(2p_2^{+})\int d^{2}\x \int dx^{-}\, e^{+ix\cdot \check{p}_{2}}\, \epsilon^{\lambda_{2}}_{\mu}(p_2)^* 
\nn \\& \times\
\lim_{y^{+}\to -\infty} ( -1)(2p_{1}^{+})\int d^{2}\y \int dy^{-}\, e^{-iy\cdot \check{p}_{1}}\, \epsilon^{\lambda_{1}}_{\nu}(p_{1})\; 
G_{F}^{\mu\nu}(x,y)_{a_2\, a_1} \: .
\end{align}
Substituting the total before-to-after gluon propagator at NEik accuracy given in Eq. \eqref{general NEik G} together with Eqs. \eqref{GEikF}, \eqref{GNEik1} and \eqref{GNEik2} into the above LSZ-type reduction formula one can perform the integrations over $x$ and $y$, which provide Dirac deltas enforcing $\uk = \underline{p_1}$ and $\uq = \underline{p_2}$, and thus removing the integrations over $\uk$ and $\uq$ present in the NEik gluon propagator. For simplicity, we keep the result written in terms of $\uk$ and $\uq$ which are thus now the momenta of the incoming and of the outgoing gluon respectively. In that way, one obtains
\begin{align}
S_{g_2 \leftarrow \, g_1} &= 
\int \! d^{2}\z \ e^{-i\z\cdot(\q -\k)} \int\! dz^{-}\ e^{i(q^{+} - k^{+})z^{-}} 
% \nn \\
%&
%\times 
\Bigg\{ 
 {\epsilon_{\lambda_{2}}^{i}}^{*} \ \epsilon_{\lambda_{1}}^{i}
 \bigg[\frac{(2q^{+})(2k^{+})}{(q^{+}+k^{+})} 
\ U_{A}(\z,z^{-}) 
%\nn \\
%& \hspace{0.5cm}
-   \frac{(\q^{j}+ \k^{j})}{2} U_{A;j}^{(1)}(\z,z^{-})
% \int_{\frac{-L^{+}}{2}}^{\frac{L^{+}}{2}} 
% \! dz^{+} \ U_{A}\Big({\frac{L^{+}}{2}},z^{+}; \z, z^{-}\Big)
%  \Big( \overrightarrow{D}_{\z^{j}}^{A}-  \overleftarrow{D}_{\z^{j}}^{A}\Big) \
% U_{A}\Big(z^{+},-{\frac{L^{+}}{2}}; \z,z^{-}\Big) 
% \nn \\
% & \hspace{0.5cm}
 -i \  U_{A}^{(2)}(\z,z^{-}) \bigg]
%\int_{-\frac{L^{+}}{2}}^{\frac{L^{+}}{2}} \! dz^{+} 
%U_{A}\Big(\frac{L^{+}}{2},z^{+}; \z,z^{-}\Big) 
%\Big( \overleftarrow{D}_{\z^{j}}^{A} \overrightarrow{D}_{\z^{j}}^{A} \Big) \
%U_{A}\Big(z^{+},-\frac{L^{+}}{2}; \z,z^{-}\Big) 
 \nn \\
&  \hspace{0.5cm}
-2 {\epsilon_{\lambda_{2}}^{i}}^{*} \ \epsilon_{\lambda_{1}}^{j}
\ U_{R;ij}^{(3)}(\z,z^{-}) 
%\int_{\frac{-L^{+}}{2}}^{\frac{L^{+}}{2}}  \! dz^{+} 
%\ U_{A}\Big(\frac{L^{+}}{2}, z^{+}; \z, z^{-}\Big) \ gT \cdot F_{ij}(z) \ U_{A}\Big(z^{+}, -\frac{L^{+}}{2}; \z,z^{-}\Big)  
\Bigg\} 
+ \text{NNEik}  
\; .
\end{align}
Our aim in this subsection is to compute the gluon background contribution to the gluon production cross section beyond the static limit of the target, i.e. we would like to keep the $z^-$ dependence of the gluon background field. The computation of a cross section beyond the static limit of the target fields is discussed in detail in \cite{Altinoluk:2022jkk}. Here, we follow the same method. For that purpose, we first get the $z^-$ dependent scattering amplitude from the S-matrix element, which reads 
%
%Now, we want to compute a cross-section for gluon scattering in $z^{-}$ dependent gluon background field. For that we will use a similar method used in \cite{Altinoluk:2022jkk} to compute $z^{-}$ dependent cross-section.\\
%For this first, we have to extract the scattering amplitude from the above-obtained s-matrix. Therefore from equation (B11) and (B19) of \cite{Altinoluk:2022jkk}, we get,
%
%
\begin{align}
\label{Scattering amplitude gluon on gluon}
 i {\cal M}_{ab}^{\lambda_{1} \lambda_{2}}(z^{-}, \underline{k}, \underline{q}) =&      
 \int \! d^{2}\z\  e^{-iz\cdot(\q -\k)} 
 \Bigg\{ 
 {\epsilon_{\lambda_{2}}^{i}}^{*} \ \epsilon_{\lambda_{1}}^{i}
 \bigg[\frac{(2q^{+})}{(q^{+}+k^{+})} 
\ U_{A}(\z,z^{-}) 
-   \frac{(\q^{j}+ \k^{j})}{4k^{+}} U_{A;j}^{(1)}(\z,z^{-})
 -\frac{i}{2k^{+}} \  U_{A}^{(2)}(\z,z^{-}) \bigg]
 \nn \\
&  \hspace{0.5cm}
-\frac{2}{2k^{+}} {\epsilon_{\lambda_{2}}^{i}}^{*} \ \epsilon_{\lambda_{1}}^{j}
\ U_{R;ij}^{(3)}(\z,z^{-}) 
\Bigg\} 
+ \text{NNEik}  
%  \nn \\
%& 
% \Big\{ {\epsilon_{\lambda_{2}}^{i}}^{*} \ \epsilon_{\lambda_{1}}^{i}
% \ U_{A} \Big(\frac{L^{+}}{2},\frac{-L^{+}}{2};\z, z^{-} \Big) 
% \nn \\ 
% &-  
% \frac{ {\epsilon_{\lambda_{2}}^i}^{*} \ \epsilon_{\lambda_{1}}^{i}}{2k^{+}}
% \frac{\q^{j}+ \k^{j}}{2} \int_{\frac{-L^{+}}{2}}^{\frac{L^{+}}{2}} \! dz^{+}
%\  U_{A}\Big({\frac{L^{+}}{2}},z^{+}; \z, z^{-}\Big)
% \Big(    \overrightarrow{D}_{\z^{j}}^{A}-  \overleftarrow{D}_{\z^{j}}^{A}\Big) \
% U_{A}\Big(z^{+},-{\frac{L^{+}}{2}}; \z, z^{-}\Big) 
% \nn \\
% &
% -i  \frac{ {\epsilon_{\lambda_{2}}^i}^{*} \ \epsilon_{\lambda_{1}}^{i}}{2k^{+}} \ \int_{-\frac{L^{+}}{2}}^{\frac{L^{+}}{2}} \! dz^{+} 
% \ U_{A}\Big(\frac{L^{+}}{2},z^{+}; \z, z^{-}\Big) \Big( \overleftarrow{D}_{\z^{j}}^{A} \overrightarrow{D}_{\z^{j}}^{A} \Big) \
%   U_{A}\Big(z^{+},-\frac{L^{+}}{2}; \z, z^{-}\Big)    
%\nn \\
%&  
%+ \frac{{\epsilon_{\lambda_{2}}^{j}}^{*} \ \epsilon_{\lambda_{1}}^{i}}{2k^{+}}  
%    \ \int_{\frac{-L^{+}}{2}}^{\frac{L^{+}}{2}}  \! dz^{+} \ U_{A}\Big(\frac{L^{+}}{2}, z^{+}; \z, z^{-}\Big) \ gT \cdot F_{ij}(z) \ U_{A}\Big(z^{+}, -\frac{L^{+}}{2}; \z, z^{-}\Big)  \Big\}_{ab}
    \end{align}
From the $z^-$ dependent scattering amplitude, one can compute the cross section as\footnote{For interested reader, we refer to Appendix B of \cite{Altinoluk:2022jkk} for a detailed discussion of how to get the cross section from the scattering amplitude in the case of $z^-$ dependent gluon background field.}
\begin{align}
\label{Definition of cross section z- dependent for gluon}
{\frac{d \sigma^{gA\rightarrow g+X}}{d\rm P.S.}} &= 2k^{+} \int\! dr^{-} \ e^{ir^{-}(q^{+}-k^{+})}  \frac{1}{2(N_c^2 -1)}
    \sum_{\lambda_1, \lambda_2} \sum_{a,b} 
\Big\langle {\mathcal{M}_{ab}^{\lambda_1 \lambda_2}\Big(\frac{-r^{-}}{2},  \underline{k}, \underline{q}\Big)^{\dagger}\ 
\mathcal{M}_{ab}^{\lambda_1 \lambda_2}\Big(\frac{r^{-}}{2},  \underline{k}, \underline{q}\Big)}\Big\rangle
\end{align}
where the one-particle phase space is defined as 
\begin{align}
d{\rm P.S}. =  \frac{1}{2q^+}\frac{dq^{+}}{(2\pi)} \frac{d^{2} \q}{(2\pi)^2} \:  . 
\end{align}
A few comments are in order for Eq. \eqref{Definition of cross section z- dependent for gluon}. While the scattering amplitude depends on $z^-$, its complex conjugate is given with $z^{\prime -}$ dependence. The $r^-$ coordinate in Eq. \eqref{Definition of cross section z- dependent for gluon} is simply defined as the difference between the $z^-$
coordinate in the amplitude and $z^{\prime -}$ in the complex conjugate amplitude, i.e. $r^{-} = z^{-} - z'^{-}$. In the case of slow $z^-$ dependence of the background field, one can gradient expand  the scattering amplitudes in Eq. \eqref{Definition of cross section z- dependent for gluon} around $r^-=0$ and perform the integration over $r^-$ explicitly. The zeroth order term in this expansion gives the strict Eikonal contribution while the the first term gives an explicit NEik contribution. On the other hand, one performs the summation over the gluon polarizations $\lambda_i$ and the gluon colors. The normalization factor $2$ in the denominator comes from averaging over the initial gluon polarizations and the normalization factor $(N_c^2-1)$ comes from averaging over the colors. Finally, $\langle \cdots\rangle$ corresponds to averaging over the target fields which is standard in CGC. 
%
%
%Where,
%The above expression is only valid at the level of the target-average cross-section, and not for each configuration of the background field. Also, here $r^{-} = z^{-} - z'^{-}$ is a coordinate we get after transformation.\\
%
%In the above expression, factor two in the denominator originates from averaging over the initial state gluon polarization vector $\lambda_{1}$ and $\lambda_{2}$ and $N_{c}^{2}-1$ comes from averaging over the color while summing over the final state polarization vector and color.\\
%Let, $U_{A}(z_{\perp}, \frac{r^{-}}{2}) =  U_{A} \bigg(\frac{L^{+}}{2},\frac{-L^{+}}{2},z_{\perp}, \frac{r^{-}}{2} \bigg) $, and $U^{\dagger}_{A}(z'_{\perp}, \frac{-r^{-}}{2}) =  U^{\dagger}_{A} \bigg(\frac{L^{+}}{2},\frac{-L^{+}}{2},z'_{\perp}, \frac{-r^{-}}{2} \bigg) $\\
%Also, note that there will be the following kind of terms contributing to the NEik accuracy cross-section:
%    1. Square of extended Eik with $z^{-}$ dependence( right now, here, we are only considering square of Eik).\\
%    2. Mix of Eik (strict Eik term) and NEik term.\\
%    3. Square of NEik will be NNEik term, so we will be neglecting them.\\
%By, using the explicit expression for the forward gluon scattering amplitude given in the equation \eqref{Scattering amplitude gluon on gluon}, the partonic level production cross section at NEik accuracy can be written as,\\
%\\
%\textcolor{red}{In below equation for first term we are assuming that q++k+ = 2k+ = 2q+, should e do that pr just keep sum of q+ and k+ as it is??!!!} \\
%

After squaring the amplitude, we get the gluon production cross section at forward rapidity in gluon-nucleus scattering in a pure gluon background at NEik accuracy as 
\begin{align}
\label{eq:g_backg_X_sec_1}
&
\frac{  d\sigma^{gA\rightarrow g+ X}}{d\rm P.S.}\bigg|_{g \, \rm backg.} = 
 \frac{1}{2(N_{c}^{2}-1)} \ 2k^{+} \int\! dr^{-} \ e^{ir^{-}(q^{+}-k^{+})} 
\int \! d^{2}\z' \  \int \! d^{2}\z\ e^{-i(\q -\k)\cdot(\z- \z')} \nn \\
& 
 \times 
 \Bigg\{ \sum_{\lambda_{1}, \lambda_{2}} \ {\epsilon_{\lambda_{2}}^{k}} \ {\epsilon_{\lambda_{1}}^{k}}^{*}
 {\epsilon_{\lambda_{2}}^{i}}^{*} \ \epsilon_{\lambda_{1}}^{i}
 \Bigg[ \frac{(2q^{+})^2}{(q^{+}+k^{+})^2}
\Big\langle \tr \Big\{U_{A}^{\dagger}\Big(\z', \frac{-r^-}{2}\Big) U_{A}\Big(\z, \frac{r^-}{2}\Big) \Big\} \Big\rangle \nn \\
& \hspace{2cm}
 %\pmb
 {+}\frac{(2q^{+})}{2k^{+}(q^{+}+k^{+})} 
 %\int_{\frac{-L^{+}}{2}}^{\frac{L^{+}}{2}} \! dz^{+}  
\bigg\langle \tr \bigg\{ 
 U_{A}^{\dagger}\Big(\z', \frac{-r^-}{2}\Big) \ 
 %U_{A}\Big({\frac{L^{+}}{2}},z^{+}; \z, \frac{r^-}{2}\Big)
%  \nn \\
%& \hspace{5cm}
%\times 
\bigg[
-   \frac{(\q^{j}+ \k^{j})}{2} U_{A;j}^{(1)}\Big(\z,\frac{r^-}{2}\Big) -i \  U_{A}^{(2)}\Big(\z,\frac{r^-}{2}\Big) \bigg]
% \bigg[ - \frac{\q^{j}+ \k^{j}}{2}   \Big(\overrightarrow{D}_{\z^{j}}^{A}-  \overleftarrow{D}_{\z^{j}}^{A} \Big) 
%-i\Big( \overleftarrow{D}_{\z^{j}}^{A} \overrightarrow{D}_{\z^{j}}^{A} \Big) \bigg] \
 %U_{A}\Big(z^{+},-{\frac{L^{+}}{2}}; \z, \frac{r^-}{2}\Big)
  \bigg\} \bigg\rangle  \nn \\
 &  \hspace{2cm}
  +  \frac{(2q^{+})}{2k^{+}(q^{+}+k^{+})} 
\bigg\langle \tr \bigg\{ 
\bigg[
-   \frac{(\q^{j}+ \k^{j})}{2} U_{A;j}^{(1) \dagger}\Big(\z',-\frac{r^-}{2}\Big) +i \  U_{A}^{(2) \dagger}\Big(\z',-\frac{r^-}{2}\Big) \bigg]
 U_{A}\Big(\z, \frac{r^-}{2}\Big) \ 
  \bigg\} \bigg\rangle
  \Bigg]   \nn \\
 & \hspace{0.5cm}
 +  \sum_{\lambda_{1} \lambda_{2}} \ {\epsilon_{\lambda_{2}}^{k}} \ {\epsilon_{\lambda_{1}}^{k}}^{*} {\epsilon_{\lambda_{2}}^{i}}^{*} \ \epsilon_{\lambda_{1}}^{j}\, (-2) \frac{(2q^{+})}{2k^{+}(q^{+}+k^{+})} 
% \int_{\frac{-L^{+}}{2}}^{\frac{L^{+}}{2}}  \! dz^{+} \ 
\bigg\langle \tr \bigg\{ U_{A}^{\dagger}\Big(\z, \frac{-r^-}{2}\Big)\  
U_{R;ij}^{(3)}\Big(\z,\frac{r^-}{2}\Big) 
%U_{A}\Big(\frac{L^{+}}{2}, z^{+}; \z, \frac{r^-}{2}\Big) \nn \\
% & \hspace{7cm}
% \times g\ T\cdot F_{ij}\Big(\uz, \frac{r^-}{2}\Big)\  U_{A}\Big(z^{+}, -\frac{L^{+}}{2}; \z, \frac{r^-}{2}\Big) 
 \bigg\} \bigg\rangle
  \nn \\
% &  \hspace{0.5cm}
% + \sum_{\lambda_{1}, \lambda_{2}} \ {\epsilon_{\lambda_{2}}^{k}} \ {\epsilon_{\lambda_{1}}^{k}}^{*}
% {\epsilon_{\lambda_{2}}^{i}}^{*} \ \epsilon_{\lambda_{1}}^{i}  \  \frac{1}{2k^{+}}   
% \int_{\frac{-L^{+}}{2}}^{\frac{L^{+}}{2}} \! dz'^{+}  
% \bigg\langle \tr\bigg\{  U_{A}^{\dagger}\Big(z'^{+},-{\frac{L^{+}}{2}}; \z', \frac{-r^-}{2}\Big)  \nn \\
% & \hspace{5cm}
% \times 
% \bigg[  \frac{\q^{l}+ \k^{l}}{2}   \Big(\overrightarrow{D}_{\z'^{l}}^{A}-  \overleftarrow{D}_{\z'^{l}}^{A} \Big) 
% +i\Big( \overleftarrow{D}_{\z'^{l}}^{A} \overrightarrow{D}_{\z'^{l}}^{A} \Big) \bigg] 
% \ U_{A}^{\dagger}\Big({\frac{L^{+}}{2}},z'^{+}; \z', \frac{-r^-}{2}\Big) U_{A}\Big(\z, \frac{r^-}{2}\Big) \ \bigg\} \bigg\rangle \nn \\
 & \hspace{0.5cm}
 +   \sum_{\lambda_{1} \lambda_{2}} \ {\epsilon_{\lambda_{2}}^{l}} \ {\epsilon_{\lambda_{1}}^{k}}^{*} {\epsilon_{\lambda_{2}}^{i}}^{*} \ \epsilon_{\lambda_{1}}^{i} \,  
 (-2) \frac{(2q^{+})}{2k^{+}(q^{+}+k^{+})} 
\bigg\langle \tr \bigg\{U_{R;lk}^{(3) \dagger}\Big(\z, \frac{-r^-}{2}\Big)\  
U_{A}\Big(\z,\frac{r^-}{2}\Big) 
 \bigg\} \bigg\rangle
% \frac{1}{2k^{+}}
%\int_{\frac{-L^{+}}{2}}^{\frac{L^{+}}{2}}  \! dz'^{+} \  
%\bigg\langle \tr\bigg\{ U_{A}^{\dagger}\Big(z'^{+}, -\frac{L^{+}}{2}; \z', \frac{-r^-}{2}\Big)  \nn   \\
%&  \hspace{7cm}
%\times 
%gT \cdot F_{kl}\Big(\uz, \frac{-r^-}{2}\Big)
% U_{A}^{\dagger}\Big(\frac{L^{+}}{2}, z'^{+}; \z', \frac{-r^-}{2}\Big) U_{A}\Big(\z, \frac{r^-}{2}\Big) \  \bigg\} \bigg\rangle
 \Bigg\}  + \text{NNEik}
 \, .  
 \end{align}
%
%
%where we have introduced a short hand notation for $ U_{A} \Big(\frac{L^{+}}{2},\frac{-L^{+}}{2},\z, \frac{r^{-}}{2} \Big) \equiv U_{A}\Big(\z, \frac{r^{-}}{2}\Big)$. 
Summation over the polarizations can be performed trivially and one gets
\begin{align}
\sum_{\lambda_{1}, \lambda_{2}} \ {\epsilon_{\lambda_{2}}^{k}} \ {\epsilon_{\lambda_{1}}^{k}}^{*}
 {\epsilon_{\lambda_{2}}^{i}}^{*} \ \epsilon_{\lambda_{1}}^{i} & = \sum_{\lambda_{1}, \lambda_{2}} \ {\epsilon_{\lambda_{2}}^{k}} \ {\epsilon_{\lambda_{1}}^{k}}^{*}
 {\epsilon_{\lambda_{2}}^{i}}^{*} \ \epsilon_{\lambda_{1}}^{i} = \delta^{ki} \ \delta^{ik} =2  \\ 
  \sum_{\lambda_{1} \lambda_{2}} \ {\epsilon_{\lambda_{2}}^{k}} \ {\epsilon_{\lambda_{1}}^{k}}^{*} {\epsilon_{\lambda_{2}}^{j}}^{*} \ \epsilon_{\lambda_{1}}^{i}  &   = \delta^{kj} \delta^{ik} = \delta^{ij} \\
   \sum_{\lambda_{1} \lambda_{2}} \ {\epsilon_{\lambda_{2}}^{l}} \ {\epsilon_{\lambda_{1}}^{k}}^{*} {\epsilon_{\lambda_{2}}^{i}}^{*} \ \epsilon_{\lambda_{1}}^{i}  & = \delta^{li} \delta^{ik}= \delta^{lk} 
\end{align}
Note that $\delta^{ij}$ is the structure that multiplies the field strength tensor $F_{ij}$. Since the delta function structure is symmetric under the exchange of $i \leftrightarrow j$ while $F_{ij}$ is antisymmetric under the same exchange, that contribution vanishes, and similarly for the term with $\delta^{lk}$ and $F_{lk}$. %Similar argument is true for the delta function structure and the fields strength tensor  $F_{lk}$ under the exchange of $l \leftrightarrow k$. 
%Therefore, these two terms in Eq. \eqref {eq:g_backg_X_sec_1} vanish yielding the following result:   
Noting as well that 
\begin{align}
\frac{(2k^{+})(2q^{+})}{(q^{+}+k^{+})^2} 
=&\, 1 - \left(\frac{q^{+}-k^{+}}{q^{+}+k^{+}}\right)^2 =1 + \text{NNEik}\, ,
\end{align}
since $q^{+}\!-\!k^{+}$ is the conjugate to the $r^-$ variable, one arrives at
\begin{align}
\label{g_background_NEik_fin}
\frac{  d\sigma^{gA\rightarrow g+ X}}{d\rm P.S.} \bigg|_{g\, \rm backg. } &= 
 \frac{1}{(N_{c}^{2}-1)} \ \int\! dr^{-} \ e^{ir^{-}(q^{+}-k^{+})} 
\int \! d^{2}\z' \  \int \! d^{2}\z\ e^{-i(\q -\k)\cdot(\z- \z')} \nn \\
 & \times
 \Bigg\{ (2q^+)\, 
 \Big\langle \tr\Big[ U_{A}^{\dagger}\Big(\z', \frac{-r^-}{2}\Big) U_{A}\Big(\z, \frac{r^-}{2}\Big)\Big] \Big\rangle
  \nn \\
 &+
\bigg\langle
 \tr \ \bigg\{  U_{A}^{\dagger}\Big(\z', \frac{-r^-}{2}\Big) \ 
 \bigg[  - \frac{(\q^{j}+ \k^{j})}{2}  U_{A;j}^{(1)}\Big(\z,\frac{r^{-}}{2}\Big) -i U_{A}^{(2)}\Big(\z,\frac{r^{-}}{2}\Big)\bigg] \bigg\}
 \bigg\rangle
 \nn \\
 &+ 
 \bigg\langle
 \tr \ \bigg\{ \bigg[  -\frac{(\q^{j}+ \k^{j})}{2}  U_{A;j}^{(1) \dagger}\Big(\z',\frac{-r^{-}}{2}\Big) 
   +i  U_{A}^{(2)\dagger}\Big(\z',\frac{-r^{-}}{2}\Big) \bigg] U_{A}\Big(\z, \frac{r^-}{2}\Big) \bigg\}
   \bigg\rangle
    \Bigg\}
 \end{align}
This is the final result for the gluon background contribution to the forward gluon production cross section in gluon-nucleus scattering at NEik accuracy. \\

%%%%%%%%%%%%%%%%%%%%%%%%%%%%%%%%%%%%%%%%%%%%%%%%%%%%%%%%%%%%%%%%%%%%%%%%%%%%%%%%%%%%%%%%%%%%%%%%%%%%%%%%%%%%%%%%%%%%%%%%%%%%     
\subsection{Forward gluon production cross section in gluon-nucleus scattering at NEik accuracy: quark background contribution}
At NEik accuracy, the other contribution to the forward gluon production cross section in gluon-nucleus scattering originates from the interaction with the quark background field. This corresponds to the second term in Eq. \eqref{eq:gAtogX_gen}. This section is devoted to the computation of this contribution. 

\begin{figure}
\centering 
\includegraphics[width=0.9\linewidth]{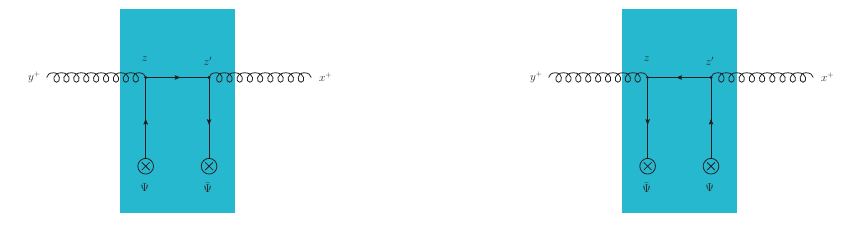}  
\caption{The two diagrams with double quark background field insertion contributing to gluon production in gluon-target scattering. They corresponds to the temporary conversion of the propagating gluon into a quark for the diagram on the left, and to an antiquark for the diagram on the right.}
\label{fig:gluon in q BF case 1 q}
\end{figure}

The quark background contribution to the gluon production in gluon-nucleus scattering can be considered as two separate sub contributions which can be written as 
\begin{align}
\label{eq:q_backg_total_gen}
\frac{d \sigma^{gA\rightarrow g+X}}{d\rm P.S.}\bigg|_{q\, \rm backg.}= \frac{d \sigma^{gA\rightarrow g+X}}{d\rm P.S.}\bigg|_{gqg}+\frac{d \sigma^{gA\rightarrow g+X}}{d\rm P.S.}\bigg|_{g{\bar q}g} \: .
\end{align}
The first term in Eq. \eqref{eq:q_backg_total_gen} corresponds to the case where the incoming gluon interacts with a quark background field converting the gluon into a quark which later interacts with antiquark field converting the quark back into a gluon before it goes out of the medium (see the left panel of Fig. \ref{fig:gluon in q BF case 1 q}). On the other hand, the second term in Eq. \eqref{eq:q_backg_total_gen} corresponds to the case where the incoming gluon interacts with an antiquark background field converting the gluon into an antiquark that later interacts with a quark background field converting it back into a gluon and finally gluon exceeds the medium (see the right panel of Fig. \ref{fig:gluon in q BF case 1 q})\footnote{In principle, there is a third contribution, with the instantaneous term from the quark propagator between the two quark background field insertions. However, in that contribution, the enhanced components of the quark background field are projected out, and only the suppressed component survive, giving a contribution to the process only at NNEik order.}  

Let us start our analysis by considering the first case where the propagator can be written as 
\begin{align}
M^{\mu \nu}_{F}(x,y)\big|_{gqg} &= \int \! d^{4} z \int \! d^{4} z' 
\ G^{\mu \mu'}_{F}(x,z')_{aa'}|^{\rm IA}_{\rm Eik} \ \overline{\Psi}^{-}_{\beta}(z')
\ \Big[ \big(-igt^{a'} \gamma_{\mu'}\big) \ S_{F}(z',z)|^{\rm II,q}_{\rm Eik} \ \big(-igt^{b'}\gamma_{\nu'}\big) \Big]_{\beta \alpha} \nn \\
&\times  
\Psi^{-}_{\alpha}(z) \ G^{\nu' \nu}_{F}(z,y)_{b'b}|^{\rm BI}_{\rm Eik}
    \end{align}
After substituting the explicit expressions for the before-to-inside gluon propagator given in Eq. \eqref{eq:before-inside_gluon}, inside-to-inside quark propagator given in Eq. \eqref{eq:inside-inside_quark} and inside-to-after gluon propagator given in Eq. \eqref{eq:inside-after_gluon}, performing the integrations over $p^+,\z', z^{\prime -}$ and simplifying the gamma matrix structure, one gets      
\begin{align}
\label{eq:prop_gqg}
M^{\mu \nu}_{F}(x,y)\big|_{gqg} &= 
\int\! \frac{d^{3}\uq}{(2\pi)^{3}} \frac{\theta(q^{+})}{2q^{+}} \ e^{-ix\check{q}}
\int\! \frac{d^{3}\uk}{(2\pi)^{3}} \frac{\theta(k^{+})}{2k^{+}} \ e^{iy\check{k}} 
\int\! d^{2}\z\ e^{-i\z\cdot(\q-\k)} \int\! dz^{-} e^{i(q^{+}-k^{+})z^{-}} \nn \\
&
\times
\int\! dz^{+}\int\! dz'^{+} \  \theta(z'^{+}-z^{+})
 \ U_{A}\big(x^{+},z'^{+};\z,z^{-}\big)_{aa'}\,  
 \overline{\Psi}_{\beta}(z'^{+},\z,z^{-})
\Big[g^{\mu i} - \frac{\eta^{\mu}\q^{i}}{q^{+}}\Big]  
\Big[ \frac{\gamma^{i}\gamma^{-}\gamma^{j}}{2}\Big] 
\nn \\
&
\times
\Big[ g^{j\nu}- \frac{\eta^{\nu}\k^{j}}{k^{+}}\Big] 
\Big[ \big(-igt^{a'}\big) \ U_{F}\big(z'^{+},z^{+};\z,z^{-}\big) \ \big(-igt^{b'}\big) \Big]_{\beta\alpha}
\Psi_{\alpha}(z) \ U_{A}\big(z^{+},y^{+}; \z,z^{-}\big)_{b'b}
    \end{align}
As usual, we can use the obtained propagator in Eq. \eqref{eq:prop_gqg} to compute the S-matrix element for the scattering process with the LSZ-type reduction formula which is given as  
\begin{align}
S_{g_{2}\leftarrow g_{1}}\big|_{gqg} &= \lim_{x^{+}\rightarrow +\infty} (-1) (2p_{2}^{+}) \int\! d^{2}\x \int\! dx^{-}\  e^{ix\check{p}_{2}} \ \epsilon_{\mu}^{\lambda_{2}} (p_{2})^{*} \nn \\
&
\times
\lim_{y^{+}\rightarrow -\infty}(-1) (2p_{1}^{+})\int\! d^{2}\y \int\! dy^{-} \ e^{-iy\check{p}_{1}} \epsilon_{\nu}^{\lambda_{1}}(p_{1}) 
\ M^{\mu \nu}_{F}(x,y)\big|_{gqg}
 \end{align}
Substituting the explicit expression of the propagator Eq. \eqref{eq:prop_gqg} inside the LSZ-type formula and simplifying the Dirac matrix structure, we get  
%
%
%\begin{align}
%S_{g_{1}(\check{k},a,\lambda_1)\rightarrow g_{2}(\check{q},b,\lambda_2)} &= 
%\int\! d^{2}\z \ e^{-i\z\cdot(\q-\k)} \int\! dz^{-} e^{i(q^{+}-k^{+})z^{-}}\ {\epsilon^{\lambda_2}_{\mu}(q)}^{*}
%\int\! dz^{+}\int\! dz'^{+}\ \theta(z'^{+}-z^{+})\nn \\
%& 
%\times
% \bigg\{ 
% U_{A}\Big(\frac{L^{+}}{2},z'^{+},\z,z^{-}\Big)_{aa'} 
% \overline{\Psi}_{\beta}\big(z'^{+},\z,z^{-}\big)
%\Big[g^{\mu i} - \frac{\eta^{\mu}\q^{i}}{q^{+}}\Big]
%\Big[\frac{\gamma^{i}\gamma^{-}\gamma^{j}}{2}\Big]
%\Big[g^{j\nu}- \frac{\eta^{\nu}\k^{j}}{k^{+}}\Big] \nn \\
%&
%\times
% \Big[\big(-igt^{a'}\big) U_{F}\big(z'^{+},z^{+},\z,z^{-}\big) \big(-igt^{b'}\big)\Big]_{\beta\alpha}
% \ \Psi_{\alpha}(z) \ U_{A}\Big(z^{+},\frac{-L^{+}}{2},\z,z^{-}\Big)_{b'b} \bigg\}\, 
% \epsilon^{\lambda_1}_{\nu}(k)
%\end{align}
%%
%%
%Now, using the definition of polarization vector from \eqref{pol_vect}, we can contract the indices and simplify the Dirac matrix structure, we get,
%
%
\begin{align}
S_{g_{2}(\check{q},b,\lambda_2)\leftarrow g_{1}(\check{k},a,\lambda_1)}\big|_{gqg}& =  
\int\! d^{2}\z\ e^{-i\z\cdot(\q-\k)} \int\! dz^{-} e^{i(q^{+}-k^{+})z^{-}}\ 
\int\! dz^{+}\int\! dz'^{+} \ \theta(z'^{+}-z^{+}) \nn \\
& 
\times
U_{A}\Big(\frac{L^{+}}{2},z'^{+}; \z,z^{-}\Big)_{aa'} 
\overline{\Psi}_{\beta}\big(z'^{+},\z,z^{-}\big)
\ {\epsilon_{\lambda_2}^{i}}^{*} \ \epsilon_{\lambda_{1}}^{j}
\Big[\frac{\gamma^{i}\gamma^{-}\gamma^{j}}{2}\Big] \nn  \\
&
\times
\Big[\big(-igt^{a'}\big) \  U_{F}\big(z'^{+},z^{+};\z,z^{-}\big) \ \big(-igt^{b'}\big) \Big]_{\beta\alpha}
\ \Psi_{\alpha}(z) 
\ U_{A}\Big(z^{+},\frac{-L^{+}}{2};\z,z^{-}\Big)_{b'b}
\end{align}
One can easily extract the $z^-$ dependent scattering amplitude from the S-matrix element and it reads  
\begin{align}
\label{scattering amplitude gqg}
i\M^{\lambda_1 \lambda_2}_{ab}(z^{-}, \underline{k}, \underline{q})\big|_{gqg} &= 
\frac{1}{2k^{+}}\int\! d^{2}\z \ e^{-i\z\cdot(\q-\k)} 
\int\! dz^{+}\int\! dz'^{+} \ \theta(z'^{+}-z^{+})
\ U_{A}\Big(\frac{L^{+}}{2},z'^{+};\z,z^{-}\Big)_{aa'} 
\overline{\Psi}_{\beta}\big(z'^{+},\z,z^{-}\big) 
\nn \\
&
\times
{\epsilon_{\lambda_2}^{i}}^{*}\ \epsilon_{\lambda_{1}}^{j}
\ \Big[\frac{\gamma^{i}\gamma^{-}\gamma^{j}}{2}\Big]
\Big [\big(-igt^{a'}\big) U_{F}\Big(z'^{+},z^{+};\z,z^{-}\Big)\big(-igt^{b'}\big)\Big]_{\beta\alpha}
\Psi_{\alpha}(z) 
\ U_{A}\Big(z^{+},\frac{-L^{+}}{2};\z,z^{-}\Big)_{b'b}
\end{align}
We are now ready to compute the cross section for this sub contribution. Since the amplitude computed for this process given in Eq. \eqref{scattering amplitude gqg} is NEik itself, when computing the cross section it has to be multiplied by the eikonal contribution of gluon scattering amplitude on the complex conjugate amplitude. Again using the cross section beyond the static target limit, as discussed in previous section, the first sub contribution to quark background contribution to the forward gluon production in gluon-nucleus scattering can be written generically as 
%(z^{-}, \underline{k}, \underline{q})
%
\begin{align}
\label{cross section in qg BF formula}
&
\frac{d \sigma^{gA\rightarrow g+X}}{d\rm P.S.}\bigg|_{gqg} = 2k^{+} \int\! dr^{-} \ e^{ir^{-}(q^{+}-k^{+})} \frac{1}{2(N_{c}^{2}-1)}
% \sum_{\lambda_1 \lambda_2}\sum_{a,b} 
\nn \\
&\times 
\sum_{\lambda_1,\lambda_2} \ \sum_{a,b}
\bigg[
\Big\langle
\M\Big(\frac{-r^-}{2}, \underline{k}, \underline{q}\Big)^{\dagger}\Big|_{\rm Eik} \
\M\Big(\frac{r^-}{2}, \underline{k}, \underline{q}\Big)\Big|_{\rm NEik} 
\Big\rangle +
%\nn \\
%& +
\Big\langle
\M\Big(\frac{-r^-}{2}, \underline{k}, \underline{q}\Big)^{\dagger}\Big|_{\rm NEik} \
\M\Big(\frac{r^-}{2}, \underline{k}, \underline{q}\Big)\Big|_{\rm Eik} 
\Big\rangle 
\bigg]
\end{align}
The cross section in Eq. \eqref{cross section in qg BF formula} can be computed explicitly by using Eq. \eqref{scattering amplitude gqg} for the NEik amplitude and the eikonal part of gluon scattering amplitude (first term in Eq. \eqref{Scattering amplitude gluon on gluon}. It is more convenient to present the result as two separate contributions where the first one corresponds to the first term in the square brackets in Eq. \eqref{cross section in qg BF formula} and the second contribution corresponds to the second term in  Eq.  \eqref{cross section in qg BF formula}. When written explicitly, the first term reads 
\begin{align}
\label{eq:gqg_1_1}
&
\frac{d \sigma^{gA\rightarrow g+X}}{d\rm P.S.}\bigg|_{gqg, \, (1)} =  2k^{+} \int\! dr^{-} \ e^{ir^{-}(q^{+}-k^{+})} \frac{1}{2(N_{c}^{2}-1)}
 %   \sum_{\lambda_1 \lambda_2}\sum_{a,b} 
 \sum
 \bigg\{\frac{(2q^{+})}{2k^{+}(q^{+}+k^{+})}
 \int\!d^{2}\z'\int\! d^{2}\z\ e^{-i(\q-\k)\cdot(\z-\z')} \nn \\
 & \hspace{0.5cm}
 \times
\int\! dz^{+}\int\! dz_{1}^{+} \ \theta(z_{1}^{+}-z^{+}) 
\  {\epsilon_{\lambda_{1}}^{k}}^{*}\epsilon_{\lambda_2}^{k}
\Big\langle  U_{A}^{\dagger}\Big(\z',\frac{-r^{-}}{2}\Big)_{ab}
\ U_{A}\Big(\frac{L^{+}}{2},z_{1}^{+};\z,\frac{r^{-}}{2}\Big)_{aa'} 
\overline{\Psi}_{\beta}\Big(z_{1}^{+},\z,\frac{r^{-}}{2}\Big) 
\nn \\
& \hspace{0.5cm}
\times
{\epsilon_{\lambda_2}^{i}}^{*} \epsilon_{\lambda_{1}}^{j}
\Big[ \frac{\gamma^{i}\gamma^{-}\gamma^{j}}{2}\Big]
\Big[  \big(-igt^{a'}\big) \ U_{F}\Big(z_{1}^{+},z^{+};\z,\frac{r^{-}}{2}\Big) \ \big(-igt^{b'}\big)\Big]_{\beta\alpha} 
\Psi_{\alpha}\Big(\underline{z}, \frac{r^-}{2}\Big) 
U_{A}\Big(z^{+},\frac{-L^{+}}{2};\z,\frac{r^{-}}{2}\Big)_{b'b} \Big\rangle\bigg\}
\end{align}
The summation over the gluon polarizations can be performed easily as  
\begin{align}
N_{1}&=\sum_{\lambda_{1}\lambda_{2}}{\epsilon_{\lambda_{1}}^{k}}^{*} \epsilon_{\lambda_2}^{k} {\epsilon_{\lambda_2}^{i}}^{*}\epsilon_{\lambda_{1}}^{j}\
\overline{\Psi}_{\beta}\Big(z_{1}^{+},\z,\frac{r^{-}}{2}\Big) 
\big[ \gamma^{i}\gamma^{-}\gamma^{j} \big]
\Psi_{\alpha}\Big(\underline{z},\frac{r^-}{2}\Big)
= 
-\delta^{ki}\delta^{kj} 
\ \overline{\Psi}_{\beta}\Big(z_{1}^{+},\z,\frac{r^{-}}{2}\Big) 
\big[\gamma^{i}\gamma^{j}\gamma^{-}\big]
\ \Psi_{\alpha}\Big(\underline{z},\frac{r^-}{2}\Big)
\nn \\ 
&
=
-\delta^{ij}\ 
\overline{\Psi}_{\beta}\Big(z_{1}^{+},\z,\frac{r^{-}}{2}\Big) 
\big[ \gamma^{i}\gamma^{j}\gamma^{-}\big] 
\Psi_{\alpha}\Big(\underline{z},\frac{r^-}{2}\Big)
= -\delta^{ij} 
\ \overline{\Psi}_{\beta}\Big(z_{1}^{+},\z,\frac{r^{-}}{2}\Big)
\ \frac{1}{2} \big[ \{\gamma^{i},\gamma^{j}\}\gamma^{-} \big]
\ \Psi_{\alpha}\Big(\underline{z},\frac{r^-}{2}\Big)
\nn \\&
=-\delta^{ij} g^{ij} 
\ \overline{\Psi}_{\beta}\Big(z_{1}^{+},\z,\frac{r^{-}}{2}\Big) 
\ \gamma^{-} \ \Psi_{\alpha}\Big(\underline{z},\frac{r^-}{2}\Big)
%=\delta^{ij}\delta^{ij} \overline{\Psi}_{\beta}(z_{1}^{+},z_{\perp},\frac{r^{-}}{2}) \gamma^{-}\Psi_{\alpha}(\underline{z},\frac{r^-}{2})\\
%&
=2 \ \overline{\Psi}_{\beta}\Big(z_{1}^{+},\z,\frac{r^{-}}{2}\Big) \ \gamma^{-} \ \Psi_{\alpha}\Big(\underline{z},\frac{r^-}{2}\Big)
\end{align}
Substituting this back in the cross section in Eq. \eqref{eq:gqg_1_1}, one gets explicit expression of the first term in Eq. \eqref{cross section in qg BF formula} as
\begin{align}
\label{q:gqg_1_2}
\frac{d \sigma^{gA\rightarrow g+X}}{d\rm P.S.}\bigg|_{gqg\, (1)} & =  
\int\! dr^{-} \ e^{ir^{-}(q^{+}-k^{+})} \frac{1}{2(N_{c}^{2}-1)}
\int\!d^{2}\z' \int\! d^{2}\z\ e^{-i(\q-\k)\cdot(\z-\z')} \int\! dz^{+}\int\! dz_{1}^{+}   \theta(z_{1}^{+}-z^{+})
\nn \\ 
&
\times
\Big\langle
U_{A}^{\dagger}\Big(\z',\frac{-r^{-}}{2}\Big)_{ab}
\ U_{A}\Big(\frac{L^{+}}{2},z_{1}^{+};\z,\frac{r^{-}}{2}\Big)_{aa'} 
\overline{\Psi}_{\beta}\Big(z_{1}^{+}, \z,\frac{r^{-}}{2}\Big)
 \gamma^{-}
 \nn \\
 & \hspace{0.5cm}
 \times
\Big[ \big(-igt^{a'}\big) \ U_{F}\Big(z_{1}^{+},z^{+};\z,\frac{r^{-}}{2}\Big) \ \big(-igt^{b'}\big) \Big]_{\beta\alpha} 
 \Psi_{\alpha}\Big(\underline{z},\frac{r^-}{2}\Big) 
 \ U_{A}\Big(z^{+},\frac{-L^{+}}{2};\z,\frac{r^{-}}{2}\Big)_{b'b} \
\Big\rangle
\end{align}
One can follow exactly the same steps and procedure to compute the second term in Eq. \eqref{cross section in qg BF formula}. The final result for this term can be written as 
\begin{align}
\label{q:gqg_2}
\frac{d \sigma^{gA\rightarrow g+X}}{d\rm P.S.}\bigg|_{gqg\, (2)} 
&=  
\int\! dr^{-} \ e^{ir^{-}(q^{+}-k^{+})} \frac{1}{2(N_{c}^{2}-1)}
\int\!d^{2}\z' \int\! d^{2}\z \ e^{-i(\q-\k)\cdot(\z-\z)} 
\int\! dz'^{+}\int\! dz_{1}^{+} \ \theta(z_{1}^{+}-z'^{+})
\nn \\
&
\times
\bigg\langle U_{A}^{\dagger}\Big(z'^{+},\frac{-L^{+}}{2};\z',\frac{-r^{-}}{2}\Big)_{b'b}
\ \overline{\Psi}_{\alpha}\Big(\underline{z'},\frac{-r^-}{2}\Big)
 \Big[ \big(igt^{b'}\big) \ U_{F}^{\dagger}\Big(z_{1}^{+},z'^{+};\z',\frac{-r^{-}}{2}\Big)\ \big(igt^{a'}\big) \Big]_{\beta\alpha}  
 \ \gamma^{-}
\nn \\ 
&
\times      
\Psi_{\beta}\Big(z_{1}^{+},\z',\frac{-r^{-}}{2}\Big) 
\ U_{A}^{\dagger}\Big(\frac{L^{+}}{2},z_{1}^{+};\z',\frac{-r^{-}}{2}\Big)_{aa'}
\ U_{A}\Big(\z,\frac{r^{-}}{2}\Big)_{ab}
\bigg\rangle
\end{align}
Finally summing these two terms, Eqs. \eqref{q:gqg_1_2} and \eqref{q:gqg_2}, the final result for the first sub contribution to the quark background contribution (first term in Eq. \eqref{eq:q_backg_total_gen}) to the forward gluon production cross section in gluon-nucleus scattering at NEik accuracy reads 
\begin{align}
\label{cross-section gqg}    
\frac{d \sigma^{gA\rightarrow g+X}}{d\rm P.S.}\bigg|_{gqg} &=  
\int\! dr^{-} \ e^{ir^{-}(q^{+}-k^{+})} \frac{1}{2(N_{c}^{2}-1)}
\int\!d^{2}\z' \int\! d^{2}\z \ e^{-i(\q-\k)\cdot(\z-\z')} 
\nn \\
& \hspace{-1.8cm}
\times
\bigg\{ \int\! dz^{+}\int\! dz_{1}^{+} \
\theta(z_{1}^{+}-z^{+})
\bigg\langle 
U_{A}^{\dagger}\Big(\z',\frac{-r^{-}}{2}\Big)_{ab}
\ U_{A}\Big(\frac{L^{+}}{2},z_{1}^{+};\z,\frac{r^{-}}{2}\Big)_{aa'} 
\overline{\Psi}_{\beta}\Big(z_{1}^{+},\z,\frac{r^{-}}{2}\Big) 
\nn \\
& \hspace{-1.8cm}
\times
\gamma^{-}
\Big[ \big(-igt^{a'}\big) \ U_{F}\Big(z_{1}^{+},z^{+};\z,\frac{r^{-}}{2}\Big) \ \big(-igt^{b'}\big) \Big]_{\beta\alpha} 
\ \Psi_{\alpha}\Big(\underline{z},\frac{r^-}{2}\Big) 
\ U_{A}\Big(z^{+},\frac{-L^{+}}{2};\z,\frac{r^{-}}{2}\Big)_{b'b}
\bigg \rangle 
\nn \\
&\hspace{-1.8cm}
+ \int\! dz'^{+}\int\! dz_{1}^{+} \ \theta(z_{1}^{+}-z'^{+}) 
\bigg\langle  U_{A}^{\dagger}\Big(z'^{+},\frac{-L^{+}}{2};\z',\frac{-r^{-}}{2}\Big)_{b'b}
\ \overline{\Psi}_{\alpha}\Big(\underline{z'},\frac{-r^-}{2}\Big)
\Big[ \big(igt^{b'}\big) \ U_{F}^{\dagger}\Big(z_{1}^{+},z'^{+};\z',\frac{-r^{-}}{2}\Big) \ \big(igt^{a'}\big)\Big]_{\beta\alpha} 
\nn \\
& \hspace{-1.8cm}
\times
 \gamma^{-} \ \Psi_{\beta}\Big(z_{1}^{+},\z',\frac{-r^{-}}{2}\Big)
\ U_{A}^{\dagger}\Big(\frac{L^{+}}{2},z_{1}^{+},\z',\frac{-r^{-}}{2}\Big)_{aa'}
\ U_{A}\Big(\z,\frac{r^{-}}{2}\Big)_{ab} \bigg\rangle
\bigg\}
\end{align}
Let us now consider the second case (see right panel in Fig. \ref{fig:gluon in q BF case 1 q}) where the associated propagator can be written as  
\begin{align}
M^{\mu \nu}_{F}(x,y)\big|_{g{\bar q}g} &= \int \! d^{4} z \int \! d^{4} z' \ G^{\mu \mu'}_{F}(x,z')_{aa'}|^{\rm IA}_{\rm Eik} 
\ \overline{\Psi}^{-}_{\beta}(z)
\ \Big[\big(-igt^{b'} \gamma_{\nu'}\big) \ S_{F}(z,z')^{{\rm II},{\bar q}}_{\rm Eik} \ \big(-igt^{a'}\gamma_{\mu'}\big) \Big]_{\beta\alpha}
\nn \\
&
\times 
\Psi^{-}_{\alpha}(z') \ G^{\nu' \nu}_{F}(z,y)_{b'b}|^{\rm BI}_{\rm Eik}
\end{align}
Substituting the explicit expressions for before-to-inside gluon propagator given in Eq. \eqref{eq:before-inside_gluon}, inside-to-inside antiquark propagator given in Eq. \eqref{eq:inside-inside-antiquark} and inside-to-after gluon propagator given in Eq. \eqref{eq:inside-after_gluon}, performing the integrations over $p^{+}, \z', z'^{-}$ and simplifying the gamma matrix structure similar to the previous case, one obtains 
\begin{align}
\label{eq:prop_g_q_bar_g}
M^{\mu\nu}_{F}(x,y)\big|_{g{\bar q}g} &=
\int\!\frac{d^{3}q}{(2\pi)^{3}} \frac{\theta(q^{+})}{2q^{+}} 
e^{-x\cdot \check{q}} 
\int\! \frac{d^{3}\uk}{(2\pi)^{3}} \frac{\theta(k^{+})}{2k^{+}} \ e^{iy\check{k}} 
\int\! d^{2}\z \ e^{-i\z\cdot(\q-\k)} 
\int\! dz^{-}\ e^{i(q^{+}-k^{+})z^{-}} 
\int\! dz^{+}\int\! dz'^{+} \ \theta(z'^{+}-z^{+})
\nn \\
&
\times
\Big[ g^{\mu i}-\frac{\eta^{\mu}\q^{i}}{q^{+}}\Big] 
\ U_{A}(x^{+},z'^{+};\z,z^{-})_{aa'} \ \overline{\Psi}_{\beta}(z) 
\ \Big[ \frac{-\gamma^{j}\gamma^{i}\gamma^{-}}{2}\Big] 
\Big[\big(-ig t^{b'}\big)\ U_{F}^{\dagger}(z'^{+},z^{+};\z,z^{-}) \big(-ig t^{a'}\big)\Big]_{\beta\alpha}
\nn \\
&
\times
\Psi_{\alpha}(z'^{+},\z,z^{-}) \ U_{A}(z^{+},y^{+};\z,z^{-})_{b'b} 
\ \Big[g^{j\nu}-\frac{\eta^{\nu}\k^{j}}{k^{+}}\Big]
\end{align}
Using this mixed propagator, the S-matrix element for gluon scattering can be computed via the LSZ-type reduction formula which reads      
\begin{align}
\label{eq:LSZ_g-qbar-g}
S_{g_{2}\leftarrow g_{1}}\big|_{g{\bar q}g} &= \lim_{x^{+}\rightarrow +\infty} (-1) (2p_{2}^{+}) \int\! d^{2}\x \int\! dx^{-}\ e^{ix\check{p}_{2}}
\epsilon_{\mu}^{\lambda_{2}} (p_{2})^{*} 
\nn \\
&
\times
 \lim_{y^{+}\rightarrow -\infty}(-1) (2p_{1}^{+})\int\! d^{2}\y \int\! dy^{-} \ e^{-iy\check{p}_{1}} \epsilon_{\nu}^{\lambda_{1}}(p_{1}) \ M^{\mu \nu}_{F}(x,y)\big|_{g{\bar q}g}
\end{align}
Inserting the explicit expression for the mixed propagator given in Eq. \eqref{eq:prop_g_q_bar_g} into the LSZ-type reduction formula in Eq. \eqref{eq:LSZ_g-qbar-g} one gets 
\begin{align}
S_{g_{2}\leftarrow g_{1}}\big|_{g{\bar q}g} &= 
\big({\epsilon^{i}_{\lambda_{2}}}^{*}\epsilon^{j}_{\lambda_{1}}\big)
\int\! d^{2}\z \ e^{-i\z\cdot(\q-\k)} 
\int\! dz^{-} \ e^{i(q^{+}-k^{+})z^{-}} 
\int\! dz^{+}\int\!dz'^{+} \ \theta(z'^{+}-z^{+})  
\ U_{A}\Big(\frac{L^{+}}{2},z'^{+};\z,z^{-}\Big)_{aa'}
\nn \\
&
\times
 \overline{\Psi}_{\beta}(z) \ \Big[\frac{\gamma^{j}\gamma^{-}\gamma^{i}}{2}\Big] \
\Big[\big(-igt^{b'})\ U_{F}^{\dagger}(z'^{+},z^{+};\z,z^{-})\ (-igt^{a'})\Big]_{\beta\alpha}
\ \Psi_{\alpha}(z'^{+},\z,z^{-}) 
\ U_{A}\Big(z^{+},\frac{-L^{+}}{2};\z,z^{-}\Big)_{b'b} 
\end{align}
One can now extract the $z^-$ dependent scattering amplitude from the S-matrix element and it reads 
\begin{align}
\label{Scattering amplitude g antiquark g}
i\mathcal{M}_{ab}^{\lambda_1 \lambda_2}(z^{-},\underline{k},\underline{q})\big|_{g{\bar q}g} &= 
\frac{1}{2k^{+}}({\epsilon^{i}_{\lambda_{2}}}^{*}\epsilon^{j}_{\lambda_{1}})
\int\! d^{2}\z \ e^{-i\z\cdot(\q-\k)} \int\! dz^{+}\int\!dz'^{+} \ \theta(z'^{+}-z^{+})  
\ U_{A}\Big(\frac{L^{+}}{2},z'^{+};\z,z^{-}\Big)_{aa'}
\nn \\
& \hspace{-2cm}
\times 
\overline{\Psi}_{\beta}(z) 
\ \Big[ \frac{\gamma^{j}\gamma^{-}\gamma^{i}}{2}\Big]  
\ \Big[\big(-ig t^{b'}\big) \ U_{F}^{\dagger}(z'^{+},z^{+};\z,z^{-})\ \big(- ig t^{a'}\big)\Big]_{\beta\alpha}
%\\&\times
\ \Psi_{\alpha}(z'^{+},\z,z^{-}) 
\ U_{A}\Big(z^{+},\frac{-L^{+}}{2};\z,z^{-}\Big)_{b'b} 
\end{align}
It is now straight forward to compute the cross section using the analog expression given in Eq. \eqref{cross section in qg BF formula}.  As in the previous case, for the NEik amplitude one uses Eq. \eqref{Scattering amplitude g antiquark g} that describes the case where the incoming gluon first converts into an antiquark which then converts back into a gluon before it exists medium. This amplitude is multiplied by the eikonal part of the gluon production amplitude computed from before-to-after gluon propagator and given  in Eq. \eqref{Scattering amplitude gluon on gluon}. Analogous  to the previous case, it is more convenient to separate the computation of this contribution into two parts corresponding. Let us now consider the first term:  
\begin{align}
\label{eq:antiquark_1_1}
\frac{d \sigma^{gA\rightarrow g+X}}{d\rm P.S.}\bigg|_{g{\bar q}g \, (1)} &= 
2k^{+} \ \int\! dr^{-} \ e^{ir^{-}(q^{+}-k^{+})} \frac{1}{2(N_{c}^{2}-1)}
%\sum
\sum_{\lambda_1 \lambda_2}\sum_{a,b} 
\frac{(2q^{+})}{2k^{+}(q^{+}+k^{+})}
\int\!d^{2}\z' \int\! d^{2}\z\ e^{-i(\q-\k)\cdot(\z-\z')}
\nn  \\
& \hspace{-1cm}
\times
\int\! dz^{+}\int\! dz_{1}^{+} \  \theta(z_{1}^{+}-z^{+}) 
\ ({\epsilon_{\lambda_{1}}^{k}}^{*}\epsilon_{\lambda_2}^{k})
\ U_{A}^{\dagger}\Big(\z',\frac{-r^{-}}{2}\Big)_{ab} 
\ ({\epsilon^{i}_{\lambda_{2}}}^{*}\epsilon^{j}_{\lambda_{1}})
\ U_{A}\Big(\frac{L^{+}}{2},z_{1}^{+};\z,\frac{r^{-}}{2}\Big)_{aa'} 
\overline{\Psi}_{\beta}\Big(\underline{z},\frac{r^-}{2}\Big)      
\nn \\
& \hspace{-1cm}
\times
\Big[\frac{\gamma^{j}\gamma^{-}\gamma^{i}}{2}\Big] 
\Big[\big(-ig t^{b'}\big) \ U_{F}^{\dagger}\Big(z_{1}^{+},z^{+};\z,\frac{r^{-}}{2}\Big)\ \big(-ig t^{a'}\big)\Big]_{\beta\alpha}
\Psi_{\alpha}\Big(z_{1}^{+},\z,\frac{r^{-}}{2}\Big)
\ U_{A}\Big(z^{+},\frac{-L^{+}}{2};\z,\frac{r^{-}}{2}\Big)_{b'b}
\end{align}
The summation over the gluon polarizations in the numerator can be performed trivially as  
\begin{align}
N_{2} &= \sum_{\lambda_2\lambda_1} 
{\epsilon_{\lambda_{1}}^{k}}^{*}\epsilon_{\lambda_2}^{k} 
{\epsilon^{i}_{\lambda_{2}}}^{*}\epsilon^{j}_{\lambda_{1}}
\overline{\Psi}_{\beta}\Big(\underline{z},\frac{r^-}{2}\Big) 
\Big[\frac{\gamma^{j}\gamma^{-}\gamma^{i}}{2}\Big] 
\Psi_{\alpha}\Big(z_{1}^{+},\z,\frac{r^{-}}{2}\Big)
= 
\delta^{ki}\delta^{kj} \ 
\overline{\Psi}_{\beta}\Big(\underline{z},\frac{r^-}{2}\Big) 
\Big[\frac{\gamma^{j}\gamma^{-}\gamma^{i}}{2}\Big] 
\Psi_{\alpha}\Big(z_{1}^{+},\z,\frac{r^{-}}{2}\Big)
\nn \\
&
=
\delta^{ij} \
\overline{\Psi}_{\beta}\Big(\underline{z},\frac{r^-}{2}\Big) 
\Big(\frac{-\gamma^{-}}{2}\Big) \ \frac{1}{2}
\big\{\gamma^{j},\gamma^{i}\big\} \
\Psi_{\alpha}\Big(z_{1}^{+},\z,\frac{r^{-}}{2}\Big)
 =
\delta^{ij} \
\overline{\Psi}_{\beta}\Big(\underline{z},\frac{r^-}{2}\Big) 
\Big(\frac{-\gamma^{-}}{2}\Big) \ g^{ij}
\ \Psi_{\alpha}\Big(z_{1}^{+},\z\frac{r^{-}}{2}\Big)
\nn \\
&
=
\overline{\Psi}_{\beta}\Big(\underline{z},\frac{r^-}{2}\Big) \ \gamma^{-}
\Psi_{\alpha}\Big(z_{1}^{+},\z,\frac{r^{-}}{2}\Big)          
\end{align}
Substituting this result into Eq. \eqref{eq:antiquark_1_1} and averaging over the target fields yields to the following result for the first contribution 
\begin{align}
\label{eq:g_antiq_g_1_fin}
 {\frac{d \sigma^{gA\rightarrow g+X}}{d\rm P.S.}}\bigg|_{g{\bar q}g\, (1)} &=  
 \int\! dr^{-} \ e^{ir^{-}(q^{+}-k^{+})}\frac{1}{2(N_{c}^{2}-1)}
\int\!d^{2}\z' \int\! d^{2}\z\ e^{-i(\q-\k)\cdot(\z-\z')} 
\int\! dz^{+}\int\! dz_{1}^{+}
\ \theta(z_{1}^{+}-z^{+}) 
\nn \\
& \hspace{-1cm}
\times
\bigg\langle
U_{A}^{\dagger}\Big(\z',\frac{-r^{-}}{2}\Big)_{ab} 
\ U_{A}\Big(\frac{L^{+}}{2},z_{1}^{+};\z,\frac{r^{-}}{2}\Big)_{aa'} 
\ \overline{\Psi}_{\beta}\Big(\underline{z},\frac{r^-}{2}\Big) 
\ \gamma^{-} 
\ \Big[ \big(-ig t^{b'}\big)\ U_{F}^{\dagger}\Big(z_{1}^{+},z^{+};\z,\frac{r^{-}}{2}\Big)\ \big(-ig t^{a'}\big)\Big]_{\beta\alpha}
\nn \\
& \hspace{-1cm}
\times
\Psi_{\alpha}\Big(z_{1}^{+},\z,\frac{r^{-}}{2}\Big)
\ U_{A}(z^{+},\frac{-L^{+}}{2};\z,\frac{r^{-}}{2}\Big)_{b'b} \
\bigg\rangle
\end{align}
A similar computation can be performed for the second contribution where the NEik corrections are considered at the complex conjugate amplitude and the amplitude is taken at eikonal order. The result reads  
\begin{align}
\label{eq:g_antiq_g_2_fin}
{\frac{d \sigma^{gA\rightarrow g+X}}{d\rm P.S.}}\bigg|_{g{\bar q}g\, (2)} &= 
\int\! dr^{-} \ e^{ir^{-}(q^{+}-k^{+})} \frac{1}{2(N_{c}^{2}-1)}
\int\!d^{2}\z'\int\! d^{2}\z\ e^{-i(\q-\k)\cdot(\z-\z')} 
\int\! dz'^{+}\int\! dz_{1}^{+}
\ \theta(z_{1}^{+}-z'^{+})
\nn \\
&
\times
\bigg\langle 
U_{A}^{\dagger}\Big(z'^{+},\frac{-L^{+}}{2};\z',\frac{-r^{-}}{2}\Big)_{b'b}
\ \overline{\Psi}_{\alpha}\Big(z_{1}^{+},\z',\frac{-r^{-}}{2}\Big)\ 
\Big[\big(ig t^{a'}\big)\ U_{F}\Big(z_{1}^{+},z'^{+};\z',\frac{-r^{-}}{2}\Big)\ \big(ig t^{b'}\big)\Big]_{\beta\alpha} 
\ \gamma^{-} 
\nn \\
&
\times
\Psi_{\beta}\Big(\underline{z'},\frac{-r^-}{2}\Big)
\ U_{A}^{\dagger}\Big(\frac{L^{+}}{2},z_{1}^{+};\z',\frac{-r^{-}}{2}\Big)_{aa'}
\ U_{A}\Big(\z,\frac{r^{-}}{2}\Big)_{ab}  \
\bigg \rangle
\end{align}
Finally, the total contribution that originates from the process where the incoming gluon interacts with background antiquark field and converts into an antiquark which subsequently interacts with a quark background field and converts into a gluon before exiting the target medium (see right panel of Fig. \ref{fig:gluon in q BF case 1 q}), is given by the sum of the two  sub contributions given in Eqs. \eqref{eq:g_antiq_g_1_fin} and \eqref{eq:g_antiq_g_2_fin}  and it reads 
\begin{align}
\label{eq:fin_g_antiqiuark_g_NEik}
{\frac{d \sigma^{gA\rightarrow g+X}}{d\rm P.S.}}\bigg|_{g{\bar q}g} &=  
\int\! dr^{-} \ e^{ir^{-}(q^{+}-k^{+})} \frac{1}{2(N_{c}^{2}-1)}
\int\!d^{2}\z'\int\! d^{2}\z\ e^{-i(\q-\k)\cdot(\z-\z')} 
\nn \\ 
& \hspace{-1.5cm}
\times 
\Bigg\{ 
\int\! dz^{+}\int\! dz_{1}^{+} \ 
\theta(z_{1}^{+}-z^{+}) \
\bigg\langle 
U_{A}^{\dagger}\Big(\z',\frac{-r^{-}}{2}\Big)_{ab} 
\ U_{A}\Big(\frac{L^{+}}{2},z_{1}^{+};\z,\frac{r^{-}}{2}\Big)_{aa'} 
 \overline{\Psi}_{\beta}\Big(\underline{z},\frac{r^-}{2}\Big) \ \gamma^{-}
\nn \\
& \hspace{-1.5cm}
\times 
\Big[\big(-ig t^{b'}\big) \ U_{F}^{\dagger}\Big(z_{1}^{+},z^{+};\z,\frac{r^{-}}{2}\Big)\ \big(-ig t^{a'}\big)\Big]_{\beta\alpha} \ 
 \Psi_{\alpha}\Big(z_{1}^{+},\z,\frac{r^{-}}{2}\Big)
 \ U_{A}\Big(z^{+},\frac{-L^{+}}{2};\z,\frac{r^{-}}{2}\Big)_{b'b} \ 
 \bigg\rangle
 \nn \\
 & \hspace{-1.5cm}
+ 
\int\! dz'^{+}\int\! dz_{1}^{+} \ \theta(z_{1}^{+}-z'^{+})
\ U_{A}^{\dagger}\Big(z'^{+},\frac{-L^{+}}{2};\z',\frac{-r^{-}}{2}\Big)_{b'b}
\ \overline{\Psi}_{\alpha}\Big(z_{1}^{+},\z',\frac{-r^{-}}{2}\Big)
\nn \\
& \hspace{-1.5cm}
\times
\Big[\big(ig t^{a'}\big) \ U_{F}\Big(z_{1}^{+},z'^{+};\z',\frac{-r^{-}}{2}\Big)\ \big( ig t^{b'}\big)\Big]_{\beta\alpha}
\gamma^{-}   
\Psi_{\beta}\Big(\underline{z'},\frac{-r^-}{2}\Big)
\ U_{A}^{\dagger}\Big(\frac{L^{+}}{2},z_{1}^{+};\z',\frac{-r^{-}}{2}\Big)_{aa'}
\  U_{A}\Big(\z,\frac{r^{-}}{2}\Big)_{ab} 
\bigg\rangle
 \bigg\}
\end{align}
%
%
 %%%%%%%%%%%%%%%%%%%%%%%%%%%%%%%%%%%%%%%%%%%%%%%%%%%%%%%%%%%%
 %%%%%%%%%%%%%%%%%%%%%%%%%%%%%%%%%%%%%%%%%%%%%%%%%%%%%%%%%%%%
\subsection{Total forward gluon production cross section in gluon-nucleus scattering at NEik accuracy}
%%%%%%%%%%%%%%%%%%%%%%%%%%%%%%%%%%%%%%%%%%%%%%%%%%%%%%%%%%%%
%%%%%%%%%%%%%%%%%%%%%%%%%%%%%%%%%%%%%%%%%%%%%%%%%%%%%%%%%%%%
Finally we can write the total forward gluon production cross section in gluon-nucleus scattering at NEik accuracy which was generically written in Eq. \eqref{eq:q_backg_total_gen}. The first term corresponding the gluon background contribution is given in Eq. \eqref{g_background_NEik_fin} and the second term corresponding to the quark background contribution is indeed sum of two terms given in Eqs. \eqref{cross-section gqg} and \eqref{eq:fin_g_antiqiuark_g_NEik}. Thus, the final result can be written as 
\begin{align}
\label{eq:X_sec_g_prod_in_gA_Fin_Gen_Eik}
{\frac{d \sigma^{gA\rightarrow g+X}}{d\rm P.S.}}\bigg|_{\text{NEik}} \! \! \! \! &=   
 \frac{1}{(N_{c}^{2}-1)} \ \int\! dr^{-}  e^{ir^{-}(q^{+}-k^{+})} \!
\int \! d^{2}\z' \  \int \! d^{2}\z\ e^{-i(\q -\k)\cdot(\z- \z')} 
\nn \\
& \hspace{-2cm}
\times
\bigg\{ 2q^+ \Big\langle \tr\Big[ U_{A}^{\dagger}\Big(\z', \frac{-r^-}{2}\Big) \ U_{A}\Big(\z, \frac{r^-}{2}\Big)\Big] \Big\rangle
%\nn \\
%& 
+ 
\Big\langle \tr\Big\{ 
U_{A}^{\dagger}\Big(\z', \frac{-r^-}{2}\Big) 
\Big[  - \frac{(\q^{j}\!+\!\k^{j})}{2} 
U_{A;j}^{(1)}\Big(\z,\frac{r^{-}}{2}\Big)
-i U_{A}^{(2)}\Big(\z,\frac{r^{-}}{2}\Big)\Big\} \Big\rangle
\nn \\
& \hspace{-2cm}
+ \Big\langle\tr \Big\{  
\Big[-\frac{(\q^{j}\!+\! \k^{j})}{2}  U_{A;j}^{(1) \dagger}\Big(\z',\frac{-r^{-}}{2}\Big) 
+i  U_{A}^{(2)\dagger}\Big(\z',\frac{-r^{-}}{2}\Big)\Big] 
U_{A}\Big(\z, \frac{r^-}{2}\Big) \Big\} \Big\rangle
\nn \\
& \hspace{-2cm}
+
\int\! dz^{+}\int\! dz_{1}^{+} \ \theta(z_{1}^{+}-z^{+})
\bigg\langle U_{A}^{\dagger}\Big(\z',\frac{-r^{-}}{2}\Big)_{ab}
 \ U_{A}\Big(\frac{L^{+}}{2},z_{1}^{+};\z,\frac{r^{-}}{2}\Big)_{aa'}
  \ \overline{\Psi}_{\beta}\Big(z_{1}^{+},\z,\frac{r^{-}}{2}\Big) 
\  \frac{\gamma^{-}}{2}
 \nn \\
 & \hspace{-2cm}
 \times
 \Big[ \big(-igt^{a'}\big)\ U_{F}\Big(z_{1}^{+},z^{+};\z,\frac{r^{-}}{2}\Big) \ \big(-igt^{b'}\big)\Big]_{\beta\alpha}
 \ \Psi_{\alpha}\Big(\uz,\frac{r^-}{2}\Big) 
 U_{A}\Big(z^{+},\frac{-L^{+}}{2};\z,\frac{r^{-}}{2}\Big)_{b'b} \bigg\rangle
 \nn \\
 & \hspace{-2cm}
 + \int\! dz^{+}\int\! dz_{1}^{+} \ \theta(z_{1}^{+}-z^{+})
 \bigg\langle 
 U_{A}^{\dagger}\Big(\z',\frac{-r^{-}}{2}\Big)_{ab} 
\  U_{A}\Big(\frac{L^{+}}{2},z_{1}^{+};\z,\frac{r^{-}}{2}\Big)_{aa'} 
\   \overline{\Psi}_{\beta}\Big(\uz,\frac{r^-}{2}\Big)
\ \frac{\gamma^{-}}{2}
\nn \\ 
& \hspace{-2cm}
\times 
 \Big[\big(-ig t^{b'}\big) \ U_{F}^{\dagger}\Big(z_{1}^{+},z^{+};\z,\frac{r^{-}}{2}\Big)\ \big(-ig t^{a'}\big)\Big]_{\beta\alpha}
  \ \Psi_{\alpha}\Big(z_{1}^{+},\z,\frac{r^{-}}{2}\Big)
\ U_{A}\Big(z^{+},\frac{-L^{+}}{2};\z,\frac{r^{-}}{2}\Big)_{b'b} 
\bigg\rangle
 \nn \\
 & \hspace{-2cm}
 +\int\! dz'^{+}\int\! dz_{1}^{+} \ \theta(z_{1}^{+}-z'^{+}) 
\bigg\langle 
U_{A}^{\dagger}\Big(z'^{+},\frac{-L^{+}}{2};\z',\frac{-r^{-}}{2}\Big)_{b'b}
 \  \overline{\Psi}_{\alpha}\Big(\underline{z'},\frac{-r^-}{2}\Big)
 \Big[(igt^{b'})\ U_{F}^{\dagger}\Big(z_{1}^{+},z'^{+};\z',\frac{-r^{-}}{2}\Big)\ (igt^{a'})\Big]_{\beta\alpha} 
 \nn \\
 & \hspace{-2cm}
 \times
\frac{\gamma^{-}}{2} \ \Psi_{\beta}\Big(z_{1}^{+},\z',\frac{-r^{-}}{2}\Big) 
\ U_{A}^{\dagger}\Big(\frac{L^{+}}{2},z_{1}^{+};\z',\frac{-r^{-}}{2}\Big)_{aa'}
\ U_{A}\Big(\z,\frac{r^{-}}{2}\Big)_{ab} \bigg\rangle
\nn \\
& \hspace{-2cm}
+ \int\! dz'^{+}\int\! dz_{1}^{+} \ \theta(z_{1}^{+}-z'^{+}) 
\bigg\langle U_{A}^{\dagger}\Big(z'^{+},\frac{-L^{+}}{2};\z',\frac{-r^{-}}{2}\Big)_{b'b} 
\ \overline{\Psi}_{\alpha}\Big(z_{1}^{+},\z',\frac{-r^{-}}{2}\Big)
\Big[\big(ig t^{a'}\big)\ U_{F}\Big(z_{1}^{+},z'^{+};\z',\frac{-r^{-}}{2}\Big)\ \big( ig t^{b'}\big)\Big]_{\beta\alpha}
\nn \\
& \hspace{-2cm}
\times
\frac{\gamma^{-} }{2} \  \Psi_{\beta}\Big(\underline{z'},\frac{-r^-}{2}\Big)
\ U_{A}^{\dagger}\Big(\frac{L^{+}}{2},z_{1}^{+};\z',\frac{-r^{-}}{2}\Big)_{aa'}
\ U_{A}\Big(\z,\frac{r^{-}}{2}\Big)_{ab} 
 \bigg\rangle \bigg\}
\end{align}

The forward gluon production cross section in gluon-nucleus scattering at NEik accuracy given in Eq. \eqref{eq:X_sec_g_prod_in_gA_Fin_Gen_Eik} includes the $z^-$ dependence of both the gluon and quark background fields. Thus, it goes beyond the static target field approximation and includes the dynamics of the target. 
In order fully isolate the strictly Eikonal and the strictly NEik contributions, one should now perform a gradient expansion of the color operators around $r^-=0$. One should keep the zeroth and the first order terms in this expansion for the generalized eikonal contribution in Eq.~\eqref{eq:X_sec_g_prod_in_gA_Fin_Gen_Eik}, and only the zeroth order terms in this gradient expansion for all the other contribution in Eq.~\eqref{eq:X_sec_g_prod_in_gA_Fin_Gen_Eik}, since they are already of order NEik. Thanks to this expansion, it is then possible to perfom the integration over $r^-$, and one finds
%
%Since the $z^-$ dependence is an extra suppression factor, one can perform a Taylor expansion around $z^-=0$ in the gluon production cross section expression given in Eq. \eqref{eq:X_sec_g_prod_in_gA_Fin_Gen_Eik} to arrive at a strictly eikonal and explicitly NEik contributions. This can be performed in the following way. In the Taylor expansion of the Generalized Eikonal term (the first term in Eq. \eqref{eq:X_sec_g_prod_in_gA_Fin_Gen_Eik}) one keeps both the zeroth order and the first order terms in the expansion. The zeroth order term in the expansion of the Generalized Eikonal term would correspond the strict eikonal contribution while the first order terms in the expansion of the Generalized Eikonal terms would correspond to the explicit NEik corrections associated with the dynamics of the target. On the other hand, one can also perform a Taylor expansion around $z^-=0$ for the NEik terms that appear with the $z^-$ dependence (all the terms in Eq. \eqref{eq:X_sec_g_prod_in_gA_Fin_Gen_Eik} except the first term) and keep only the zeroth order term in the expansion. Those terms would correspond to the explicit NEik contributions to the gluon production cross section in forward gluon-nucleus scattering. Any term beyond zeroth order in the expansion of this contribution would be a higher order eikonal contribution which is beyond the accuracy of our computation. After all said and done, the result of the expansion around $r^-=0$ can be written as 
%
%
\begin{align}
{\frac{d \sigma^{gA\rightarrow g+X}}{d\rm P.S.}}\bigg|_{\text{NEik}} 
=&\,    (2 \pi)\delta(q^{+}-k^{+})\, \frac{1}{(N_{c}^{2}-1)}
\int \! d^{2}\z' \  \int \! d^{2}\z\ e^{-i(\q -\k)\cdot(\z- \z')} 
%\nn \\
%& 
%\times
\bigg\{ 2q^+ \Big\langle\tr\Big\{ U_{A}^{\dagger}(\z') U_{A}(\z)
%+ \big[\partial_{z^{-}}U_{A}^{\dagger}(\z')\big]U_{A}(\z)
%+ U_{A}^{\dagger}(\z') \big[\partial_{z^{-}}U_{A}(\z)\big]
\Big\}\Big\rangle
\nn \\
&
+ 
\Big\langle\tr\ \Big\{ U_{A}^{\dagger}(\z') \Big[ - \frac{(\q^{j}\!+\! \k^{j})}{2} U_{A;j}^{(1)}(\z)
    -i U_{A}^{(2)}(\z)\Big] \Big\}\Big\rangle
+ \Big\langle\tr \ \Big\{ \Big[ - \frac{(\q^{j}\!+\! \k^{j})}{2} U_{A;j}^{(1) \dagger}(\z') 
   +i  U_{A}^{(2)\dagger}(\z')\Big] U_{A}(\z) \Big\}  \Big\rangle
\nn  \\
& 
+
\int\! dz^{+}\int\! dz_{1}^{+}
\ \theta(z_{1}^{+}-z^{+}) 
\Big\langle U_{A}^{\dagger}(\z')_{ab} \  U_{A}\Big(\frac{L^{+}}{2},z_{1}^{+},\z\Big)_{aa'}  
\overline{\Psi}_{\beta}(z_{1}^{+},\z) \ \frac{\gamma^{-}}{2}
\nn \\
& \hspace{4cm}
\times
 \Big[\big(-igt^{a'}\big)\ U_{F}(z_{1}^{+},z^{+},\z)\ \big(-igt^{b'}\big)\Big]_{\beta\alpha} \Psi_{\alpha}(\uz) 
\  U_{A}\Big(z^{+},\frac{-L^{+}}{2},\z\Big)_{b'b}
\Big\rangle
 \nn \\
 & 
 +  \int\! dz^{+}\int\! dz_{1}^{+}\ \theta(z_{1}^{+}-z^{+})\
\Big\langle 
U_{A}^{\dagger}(\z')_{ab} 
\  U_{A}\Big(\frac{L^{+}}{2},z_{1}^{+},\z\Big)_{aa'} 
\ \overline{\Psi}_{\beta}(\uz) \  \frac{\gamma^{-}}{2} 
\nn \\
& \hspace{4cm}
\times 
\Big[\big(-ig t^{b'}\big) \ U_{F}^{\dagger}(z_{1}^{+},z^{+},\z)\ \big(-ig t^{a'}\big)\Big]_{\beta\alpha}
\Psi_{\alpha}(z_{1}^{+},\z)
\ U_{A}\Big(z^{+},\frac{-L^{+}}{2},\z\Big)_{b'b} \ 
\Big\rangle
\nn \\
&
+
\int\! dz'^{+}\int\! dz_{1}^{+}
\ \theta(z_{1}^{+}-z'^{+}) 
\Big\langle U_{A}^{\dagger}\Big(z'^{+},\frac{-L^{+}}{2},\z'\Big)_{b'b}
\ \overline{\Psi}_{\alpha}(\underline{z'})
\Big[ \big(igt^{b'}\big)\ U_{F}^{\dagger}(z_{1}^{+},z'^{+},\z') \ \big(igt^{a'}\big)\Big]_{\beta\alpha} 
\nn \\
&  \hspace{4cm}
\times
\frac{\gamma^{-}}{2} \ \Psi_{\beta}(z_{1}^{+},\z') 
\  U_{A}^{\dagger}\Big(\frac{L^{+}}{2},z_{1}^{+},\z'\Big)_{aa'}
\  U_{A}(\z)_{ab} \Big\rangle
\nn \\
&
+ \int\! dz'^{+}\int\! dz_{1}^{+} \ \theta(z_{1}^{+}-z'^{+})
 \Big\langle U_{A}^{\dagger}\Big(z'^{+},\frac{-L^{+}}{2},\z'\Big)_{b'b}
 \ \overline{\Psi}_{\alpha}(z_{1}^{+},\z')
\ \Big[\big(ig t^{a'}\big) \ U_{F}(z_{1}^{+},z'^{+},\z') \ \big(ig t^{b'}\big)\Big]_{\beta\alpha}
\nn \\
&  \hspace{4cm}
\times
\frac{\gamma^{-} }{2} \ \Psi_{\beta}(\underline{z'})
\ U_{A}^{\dagger}\Big(\frac{L^{+}}{2},z_{1}^{+},\z'\Big)_{aa'}
\ U_{A}(\z)_{ab} \Big\rangle 
\bigg\}
\nn\\
&\, +  2q^+ (2 \pi)\delta'(q^{+}-k^{+})\, \frac{i}{2}\, \frac{1}{(N_{c}^{2}-1)}
\int \! d^{2}\z' \  \int \! d^{2}\z\ e^{-i(\q -\k)\cdot(\z- \z')} 
\nn \\
& 
\times
\bigg\{  
  \Big\langle\tr\Big\{U_{A}^{\dagger}(\z') \big[\partial_{z^{-}}U_{A}(\z,z^-)\big]\Big\}\Big\rangle
- \Big\langle\tr\Big\{ \big[\partial_{z^{-}}U_{A}^{\dagger}(\z',z^-)\big]U_{A}(\z)\Big\rangle\bigg\}\bigg|_{z^-=0}
\, .
\end{align}
The last contribution, proportional to a derivative of a Dirac delta, is the NEik correction beyond the static limit. It seems rather formal at this stage. But remember that a realistic application of this result, at the hadron level, would require a convolution with the gluon distribution in the projectile hadron, and with either a fragmentation function or a jet function.

%%%%%%%%%%%%%%%%%%%%%%%%%%%%%%%%%%%%%%%%%%%%%%%%%%%%%%%%%%%%%%%%%%%%%%%%%%%%%%%%%%%%%%%%%%%%%%%%%%%%%%%%%%%%%%%%%%%%%%%%%%%%%%%%%%%%%%%%%%%%%%%%%%%%%%%%%%%%%%%%%%%%%
 \section{Forward quark production in quark-nucleus scattering at NEik accuracy}  
 \label{sec:quark_in_quark-nucleus} 
%%%%%%%%%%%%%%%%%%%%%%%%%%%%%%%%%%%%%%%%%%%%%%%%%%%%%%%%%%%%%%%%%%%%%%%%%%%%%%%%%%%%%%%%%%%%%%%%%%%
Another application of the results that are computed in Sec. \ref{Sec:before-to-after-gluon_prop} and \ref{sec:Various_parton_prop} is to study single inclusive quark production at  forward rapidity in quark-nucleus scattering at NEik accuracy. As in the previous section, here we also study partonic level cross section. 

Similar to the gluon production studied in Sec. \ref{sec:gluon_production_in_gA}, the quark production cross section at NEik accuracy gets contributions from scattering on a purely gluonic background field, and an extra contribution with quark background effects on top of the gluon background field. The total cross section is written as a sum of these two contributions:
%
%. 
\begin{align}
\label{eq:q_prod_X_sec_Sche}
\frac{d\sigma^{qA\rightarrow q+ X}}{d\rm P.S.}\bigg|_{\rm NEik} &= 
\frac{  d\sigma^{qA\rightarrow q+ X}}{d\rm P.S.}\bigg|_{g \,\rm backg.}
+
\frac{  d\sigma^{qA\rightarrow q+ X}}{d\rm P.S.}\bigg|_{q \,\rm backg.}
\end{align}
In the rest of this section, we compute and discuss these contributions separately.  
%
%
%%%%%%%%%%%%%%%%%%%%%%%%%%%%%%%%%%%%%%%%%%%%%%%%%%%%%%%%%%
\subsection{Forward quark production cross section in quark-nucleus scattering at NEik accuracy: gluon background contribution}
%%%%%%%%%%%%%%%%%%%%%%%%%%%%%%%%%%%%%%%%%%%%%%%%%%%%%%%%%%
%
%  
Let us first discuss the forward gluon background contribution to the quark production in quark-nucleus scattering at NEik. This contribution originates from relaxing the assumptions adopted for eikonal approximation for the gluon background. The computation of this contribution can be performed by following the same steps as in the case of gluon production presented in Sec. \ref{sec:gluon_production_in_gA}. On the other hand, this cross section has been computed previously in \cite{Altinoluk:2020oyd}. However, in that manuscript a static target was assumed and the effects of the dynamical target were neglected. Since, the computations of a cross section for dynamic and static target are effectively very similar, here we only present the results of \cite{Altinoluk:2020oyd} for the gluon contribution to the quark production cross section at NEik but restoring the $z^-$ dependence of the background field to go beyond the static target field limit which reads 
%   
 %  
%\begin{align}
%\label{NEik cross section q in g BF}
%\frac{  d\sigma^{qA\rightarrow q+ X}}{d\rm P.S.} \bigg|_{g\, \rm backg.}=& 
% \frac{1}{N_{c}} \ \int\! dr^{-} \ e^{ir^{-}(q^{+}-k^{+})} 
% \int \! d^{2}\z' \  \int \! d^{2}\z\ e^{-i(\q -\k)\cdot(\z- \z')} \bigg\{ 
% 2k^{+} 
% \ 
% \Big\langle \tr\Big[ U_{F}^{\dagger}\Big(\z', \frac{-r^-}{2}\Big) U_{F}\Big(\z, \frac{r^-}{2}\Big)\Big] \Big\rangle
% \nn \\
% &    
%+ 
%  \int_{\frac{-L^{+}}{2}}^{\frac{L^{+}}{2}} \! dz^{+}  
% \Big\langle \tr  \ \Big\{ 
%  U_{F}^{\dagger}\Big(\z', \frac{-r^-}{2}\Big) 
%  \ U_{F}\Big({\frac{L^{+}}{2}},z^{+}; \z, \frac{r^-}{2}\Big) 
%  \nn \\
%  & \hspace{2.5cm}
%  \times 
%  \Big[ - \frac{(\q^{j}+ \k^{j})}{2}   \Big(\overrightarrow{D}_{\z^{j}}^{F}-  \overleftarrow{D}_{\z^{j}}^{F} \Big) 
%  -i\Big( \overleftarrow{D}_{\z^{j}}^{F} \overrightarrow{D}_{\z^{j}}^{F} \Big) \Big]  
%\ U_{F}\Big(z^{+},-{\frac{L^{+}}{2}}; \z, \frac{r^-}{2}\Big) 
% \Big\} \Big\rangle
% \nn \\
% & 
%+
%\int_{\frac{-L^{+}}{2}}^{\frac{L^{+}}{2}} \! dz'^{+}  
%\Big\langle \tr \ \Big\{   
%U_{F}^{\dagger}\Big(z'^{+},-{\frac{L^{+}}{2}}; \z', \frac{-r^-}{2}\Big)  
%\  \Big[ \frac{(\q^{l}+ \k^{l})}{2}   \Big(\overrightarrow{D}_{\z'^{l}}^{F}-  \overleftarrow{D}_{\z'^{l}}^{F} \Big)
%+i\Big( \overleftarrow{D}_{\z'^{l}}^{F} \overrightarrow{D}_{\z'^{l}}^{F} \Big) \Big]
%\nn \\
%&  \hspace{2.5cm}
%\times
% U_{F}^{\dagger}\Big({\frac{L^{+}}{2}},z'^{+}; \z', \frac{-r^-}{2}\Big) \ U_{F}\Big(\z, \frac{r^-}{2}\Big) \Big\} \Big\rangle \bigg\}    
% \end{align}
%
%
%
\begin{align}
\label{NEik cross section q in g BF}
\frac{  d\sigma^{qA\rightarrow q+ X}}{d\rm P.S.}\bigg|_{g\, \rm backg.} =& 
 \frac{1}{N_{c}} \ \int\! dr^{-} \ e^{ir^{-}(q^{+}-k^{+})} 
\int \! d^{2}\z' \  \int \! d^{2}\z \ e^{-i(\q -\k)\cdot(\z- \z')} \bigg\{
\ 2q^+ \Big\langle \tr \Big[  U_{F}^{\dagger}\Big(\z', \frac{-r^-}{2}\Big) U_{F}\Big(\z, \frac{r^-}{2}\Big) \Big]\Big\rangle
\nn  \\
& 
+ 
\Big\langle\tr \Big\{ 
\ U_{F}^{\dagger}\Big(\z', \frac{-r^-}{2}\Big) 
\Big[ - \frac{(\q^{j}\!+\! \k^{j})}{2} U_{F,j}^{(1)} \Big(\z,\frac{r^-}{2}\Big)
-iU_{F}^{(2)}\Big(\z,\frac{r^-}{2}\Big) \Big]\Big\}\Big\rangle
 \nn \\ 
 &   
 + 
\Big\langle  \tr \Big\{ \Big[
- \frac{(\q^{l}\!+\! \k^{l})}{2} U_{F,l}^{(1)\dagger}\Big(\z',\frac{-r^{-}}{2}\Big)  
+i U_{F}^{(2)\dagger}\Big(\z',\frac{-r^{-}}{2}\Big) \Big] 
\ U_{F}\Big(\z, \frac{r^-}{2}\Big)  \Big\} \Big\rangle    \bigg\}
\end{align}
where the $r^-$ is defined as the difference between the minus coordinates in the amplitude and complex conjugate amplitude. We have used the relations \eqref{eq:def_Dec_Wil_1} and \eqref{eq:def_Dec_Wil_2} to write this expression in a more compact form.

%Note that, the compact expressions of the Wilson lines $U^{(1)}_R$ and $U^{(2)}_R$ are written with field strength insertion. 

%%%%%%%%%%%%%%%%%%%%%%%%%%%%%%%%%%%%%%%%%%%%%%%%%%%%%%%%%%%%%%%%%%%%%%%%%%%%%%%%%%%%%%%%%%%%%%
\subsection{Forward quark production cross section in quark-nucleus scattering at NEik accuracy: quark background contribution}
\begin{figure}
\centering
\includegraphics[scale=0.5]{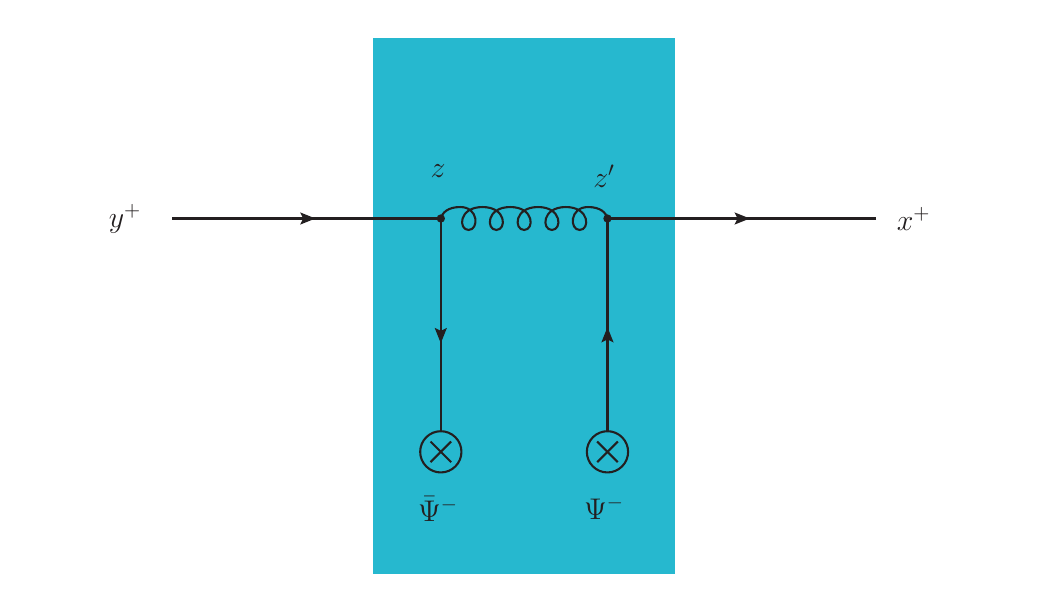}
\caption{Diagram for quark production from an incoming quark temporarily converted into a gluon, via two interactions with the quark background field}
\label{fig: quark with q bf}
\end{figure}

%%%%%%%%%%%%%%%%%%%%%%%%%%%%%%%%%%%%%%%%%%%%
The second contribution to the forward quark production cross section in quark-nucleus scattering at NEik accuracy is the quark background contribution. This contribution originates from the following case. The incoming quark interacts with antiquark background field and turns into a gluon inside the target. Then, this gluon interacts again with a quark background field and turns back into a quark which then exits the target medium (see Fig. \ref{fig: quark with q bf}). 
The computation of this contribution is analogous to the quark background contribution in gluon production and we follow the same strategy to compute it.  We start with writing the mixed propagator at NEik accuracy for this process which reads 
\begin{align}
M_{F}(x,y)\big|^{qgq}_{\rm NEik} &= 
\int\! d^{4}z \int\!d^{4}z' \ 
\Big[ S_{F}(x,z')\big|^{{\rm IA},\, q}_{\rm Eik} \ \big(-igt^{a}\gamma_{\mu}\big) \Big]_{\beta \beta'}
\ \Psi^{-}_{\beta'}(z')\ 
\Big[ G^{\mu \nu}_{F}(z', z)\big|_{\rm Eik}^{\rm II} \Big]_{ab}
\nn \\
&
\times 
\overline{\Psi^{-}}(z)_{\alpha'}\ 
\Big[ \big(-igt^{b}\gamma_{\nu}\big)\ S_{F}(z,y)\big|_{\rm Eik}^{{\rm BI},\, q} \Big]_{\alpha'\alpha}
\end{align}
Substituting the explicit expressions for the before-to-inside quark propagator Eq. \eqref{BI quark}, inside-to-inside gluon propagator Eq. \eqref{eq:inside-inside_gluon} and the inside-to-after quark propagator Eq. \eqref{q IA}, one obtains
\begin{align}
M_{F}(x,y)\big|^{qgq}_{\rm NEik} &= \int\! d^{4}z \int\!d^{4}z' 
\ \int\! \frac{d^{3}\uq}{(2\pi)^3} \frac{\theta(q^+)}{2q^{+}} 
\int\! \frac{dp^{+}}{2\pi} \frac{\theta(p^+)}{2p^+} 
\int\! \frac{d^{3}\uk}{(2\pi)^3} \frac{\theta(k^+)}{2k^+}
\ e^{-i x\cdot \check{q}}
\ e^{iz'^{-}q^{+}} \ e^{-i\z'\cdot\q} \ e^{-i(z'^{-} - z^{-})p^{+}} 
\nn \\
&
\times 
\ e^{-iz^{-}k^{+}}\ e^{i\z\cdot\k}
\ e^{i y\cdot \check{k}}
\ (\check{\slashed q} +m)
\ \Big[ U_{F}(x^{+},z'^{+};\z', z^{-}) (igt^{a})\Big]_{\beta\beta'} 
\ \Big[ \frac{\gamma^{i} \gamma^{+} \gamma^{-}}{2} \Big] 
\ \Psi_{\beta'}(z')
\nn \\
&
\times
\theta(z'^{+}\!-\!z^{+})
\int \! d^{2}\z_{1} \ \delta^{2}(\z'-\z_1) \ \delta^{2}(\z_{1}-\z) 
\ U_{A}(z'^{+},z^{+};\z_{1}, z^{-})_{ab} 
\nn \\
&
\times
\ \overline{\Psi}_{\alpha'}(z) 
\Big[ \frac{\gamma^{-}\gamma^{+}\gamma^{i}}{2}\Big] (\check{\slashed k}+ m)    
\ \Big[ (igt^{b})\ U_{F}(z^{+},y^{+};\z,z^{-})\Big]_{\alpha'\alpha}  
    \end{align}
One can now perform the integrations the trivial integrations over $\z_1$ and $\z$. Moreover, integration over $z^{\prime -}$ can be performed and it gives a delta function of $q^+-p^+$. Performing the trivial integration over $p^+$, one obtains the following expression for the mixed propagator:   
\begin{align}
\label{eq:mixed_prop_gqg}
M_{F}(x,y)\big|^{qgq}_{\rm NEik} &=  
\int\! \frac{d^{3}\uq}{(2\pi)^3} 
\frac{\theta(q^+)}{(2q^{+})^2} 
\ e^{-ix\check{q}} 
\int\! \frac{d^{3}\uk}{(2\pi)^3} 
\frac{\theta(k^+)}{2k^+} 
\ e^{iy\check{k}}
\int\! d^{2}\z e^{-i\z\cdot(\q- \k)} \int\! dz^{-} \ e^{i(q^{+}-k^{+})z^{-}}
\nn \\
&
\times
\int\! dz^{+} \int\!dz'^{+} \ \theta(z'^{+}-z^{+}) 
\ (\check{\slashed q} +m)
\Big[U_{F}(x^{+},z'^{+};\z, z^{-})\ (igt^{a})\Big]_{\beta\beta'} 
\Big[ \frac{\gamma^{i} \gamma^{+} \gamma^{-}}{2} \Big] 
\ \Psi_{\beta'}(z'^{+},\z,z^{-})
\nn \\
&
\times 
U_{A}(z'^{+},z^{+};\z, z^{-})_{ab} 
\ \overline{\Psi}_{\alpha'}(z)
\Big[ \frac{\gamma^{-}\gamma^{+}\gamma^{i}}{2} \Big] 
(\check{\slashed k}+ m) 
\  \Big[ (igt^{b}) U_{F}(z^{+},y^{+};\z,z^{-}) \Big]_{\alpha'\alpha} 
\end{align}
As usual, the S-matrix element for quark scattering can be written in terms of the mixed propagator via LSZ-type reduction formula as 
\begin{align}
\label{eq:S_matrix_gqg}
S_{q(\check{p_{2}},h',\beta)\leftarrow q(\check{p_{1}},h,\alpha)}(x,y) &= 
\lim_{x^{+} \rightarrow \infty} \lim_{y^{+}\rightarrow-\infty}\ 
\int\! d^{2}\x \int\! dx^{-} \int\! d^{2}\y \int\! dy^{-} 
\nn \\
&
\times 
e^{ix\cdot \check{p_{2}}} \ e^{-iy\cdot \check{p}_{1}} \ \overline{u}(\check{p}_{2},h') \ \gamma^{+} 
\ M_{F}(x,y)_{\beta\alpha}\big|^{qgq}_{\rm NEik} \ \gamma^{+} \ u(\check{p}_{1},h)
 \end{align}
Substituting the explicit expression of the mixed propagator given in Eq. \eqref{eq:mixed_prop_gqg} into Eq. \eqref{eq:S_matrix_gqg}, one obtains 
\begin{align}
S_{q(\check{p_{2}},h',\beta)\leftarrow q(\check{p_{1}},h,\alpha)}(x,y) &= 
\frac{1}{2p_{2}^{+}2p_{2}^{+}2p_{1}^{+}} 
\int\! dz^{-}\ e^{i(p_{2}^+ - p_{1}^{+})z^{-}} 
\int\! d^{2}\z e^{-i\z\cdot(\p_{2}- \p_1)} 
\ \int\!dz^{+} \int\! dz'^{+} \ \theta(z'^{+}-z^{+}) 
\nn \\
& \hspace{-2cm}
\times
(-g^{2})\ \Big[ U_{F}\Big( \frac{L^{+}}{2},z'^{+};\z, z^{-}\Big)\  t^{a} \Big]_{\beta\beta'}
\ \overline{u}(\check{p}_{2},h') \ \gamma^{+} \ (\check{\slashed p}_{2} +m) 
\Big[\frac{\gamma^i \gamma^{+}\gamma^{-}}{2} \Big] 
\ \Psi_{\beta'}(z'^+,\z, z^{-})
\nn \\
& \hspace{-2cm}
\times 
\Big[ U_{A}(z'^{+},z^{+};\z, z^{-})\Big]_{ab} 
\overline{\Psi}_{\alpha'}(z) 
\Big[\frac{ \gamma^{-}\gamma^{+}\gamma^{i}}{2} \Big] 
\ (\check{\slashed p}_{1}+ m) \ \gamma^{+} \ u(\check{p}_{1}, h)
\Big[ t^{b} U_{F}\Big(z^{+}, \frac{-L^{+}}{2};\z,z^{-}\Big)\Big]_{\alpha'\alpha}
\end{align}
Using the relation $\gamma^{+}(\check{\slashed p}_{2}+m)\gamma^{+} = 2p_{2}^{+} \gamma^{+}$ to simplify the gamma matrix structure and renaming $p_{2} = q$ and $p_{1} = k$, one gets the final expression for the S-matrix element for quark scattering as 
\begin{align}
S_{q(\check{q},h',\beta)\leftarrow q(\check{k},h,\alpha)}(x,y) &= 
\frac{1}{2q^{+}}  \int\! dz^{-}\ e^{iz^-(q^+ - k^{+})}    
\int\! d^{2}\z \ e^{-i\z\cdot(\q-\k)} 
\ \int\!dz^{+} \int\! dz'^{+} \ \theta(z'^{+}-z^{+}) 
\nn \\
&
\times
(-g^{2})\ \Big[ U_{F}\Big( \frac{L^{+}}{2},z'^{+};\z, z^{-}\Big) \ t^{a}\Big]_{\beta\beta'}
 \ \overline{u}(\check{q},h') 
\  \Big[ \frac{\gamma^{+}\gamma^{i}\gamma^{-}}{2} \Big]
 \ \Psi_{\beta'}(z'^+,\z, z^{-})
 \nn \\
 &
 \times 
\Big[ U_{A}(z'^{+},z^{+};\z, z^{-})\Big]_{ab}
\ \overline{\Psi}_{\alpha'}(z) 
\ \Big[\frac{ \gamma^{-}\gamma^{i}\gamma^{+}}{2}\Big]
\ u(\check{k}, h)
\ \Big[ t^{b}U_{F}\Big(z^{+}, \frac{-L^{+}}{2};\z,z^{-}\Big)\Big]_{\alpha'\alpha}
 \end{align}
One can extract the scattering amplitude from the $z^-$-dependent S-matrix element as (see Appendix B of \cite{Altinoluk:2022jkk} for a detailed discussion of the $z^-$ dependent S-matrix and its relation to scattering amplitude)
%(z^{-}, \underline{k}, \underline{q})
%
\begin{align}
\label{Scattering amp qgq neik}
i\mathcal{M}^{h'h}_{\beta\alpha}(z^{-}, \underline{k}, \underline{q})\big|^{\rm NEik}_{q\,  \rm backg.} &= 
\frac{1}{2q^{+}2k^{+}}   \int\! d^{2}\z \ e^{-i\z\cdot(\q-\k)} \ \int\!dz^{+} \int\! dz'^{+} \  \theta(z'^{+}-z^{+}) 
(-g^{2})\ \Big[ t^{a} U_{F}\Big( \frac{L^{+}}{2},z'^{+};\z, z^{-}\Big)\Big]_{\beta\beta'} 
\ \overline{u}(\check{q},h') 
\nn \\
& \hspace{-1cm}
\times
\Big[\frac{\gamma^{+}\gamma^{i}\gamma^{-}}{2}\Big] \ \Psi_{\beta'}(z'^+,\z, z^{-})
U_{A}(z'^{+},z^{+};\z, z^{-})_{ab}
\ \overline{\Psi}_{\alpha'}(z) 
\Big[\frac{ \gamma^{-}\gamma^{i}\gamma^{+}}{2}\Big]
\ u(\check{k}, h) 
\  \Big[ t^{b} U_{F}\Big(z^{+}, \frac{-L^{+}}{2};\z,z^{-}\Big)\Big]_{\alpha'\alpha}
\end{align}
Note that the amplitude is given in Eq. \eqref{Scattering amp qgq neik} is at NEik order since there are two quark background interactions inside the medium. Therefore, when computing the quark background contribution to the forward quark production cross section in quark-nucleus scattering, one should only account for the interference of quark background contribution to the quark production amplitude and the Eikonal piece of the quark production amplitude in a gluon background field to stay at NEik accuracy at the level of the cross section.  The quark scattering amplitude in a gluon background field at Eikonal order is very well known and it was also computed in \cite{Altinoluk:2020oyd} as  
\begin{align}
\label{Scattering amp q eik}
i\mathcal{M}^{h'h}_{\beta\alpha}(z^{-}, \underline{k}, \underline{q})\big|^{\rm Eik}_{g \, \rm backg.} &= 
\frac{1}{2k^{+}} \int\! d^{2}\z \ e^{-i\z\cdot(\q-\k)} 
\ \overline{u}(\check{q},h') 
\ U_{F} ( \z,z^{-})_{\beta\alpha} 
\ \gamma^{+} \ u(\check{k},h)
\end{align}
Schematically, the quark contribution to the  $z^-$ dependent quark production cross section can be written as \cite{Altinoluk:2022jkk}
\begin{align}
\label{eq:q_background_1}
\frac{d\sigma^{qA\rightarrow q+ X}}{d\rm P.S.}\bigg|_{q\, \rm backg.} &= 
2k^{+} \int\! dr^{-} \ e^{ir^{-}(q^{+}-k^{+})} \sum_{h,h'}\sum_{\alpha, \beta} 
\Big[ \Big\langle 
{\mathcal{M}_{F}\Big(\frac{-r}{2},\underline{k}, \underline{q}\Big)^{\dagger}\Big|_{\rm Eik} 
\mathcal{M}_{F}\Big(\frac{r^{-}}{2}, \underline{k}, \underline{q}\Big)\Big|_{\rm NEik}}
\Big\rangle
\nn \\
& \hspace{6cm}
+ 
\Big\langle
{\mathcal{M}_{F}\Big(\frac{-r}{2},\underline{k}, \underline{q}\Big)^{\dagger}\Big|_{\rm NEik} 
\mathcal{M}_{F}\Big(\frac{r^{-}}{2}, \underline{k}, \underline{q}\Big)\Big|_{\rm Eik}}
\Big\rangle\Big]
 \end{align}
where the NEik amplitude is the quark production amplitude in quark background field given in Eq. \eqref{Scattering amp qgq neik} and the Eikonal amplitude is the quark production amplitude in gluon background field given in Eq. \eqref{Scattering amp q eik}.  The cross section can be computed in a more convenient way when written as two separate sub contributions corresponding to the first and second terms in Eq. \eqref{eq:q_background_1}. After substituting the explicit expressions for the Eikonal and NEik amplitudes into the quark production cross section formula, the first term reads 
\begin{align}
\label{eq:q_back_X_sec_q_prod_1}
\frac{d\sigma^{qA\rightarrow q+ X}}{d\rm P.S.}\bigg|_{q\, \rm backg. \, (1)} &= 
\frac{1}{2q^{+}2k^{+}} \int\! dr^{-} e^{ir^{-}(q^{+}-k^{+})}\frac{1}{2N_{c}} 
\int\! d^{2}\z \int\! d^{2}\z' \ e^{-i(\z-\z')\cdot(\q-\k)} 
\int\! dz^{+} \int\! dz^{+}_{1} \ \theta(z_1^+ - z^+)
\nn \\
& \hspace{-1.7cm}
\times 
\sum_{h, h'}\sum_{\alpha, \beta} \ (-g^{2})
\overline{u}(\check{k},h)
\ \Big[U_{F}^{\dagger}\Big(\z', \frac{-r^{-}}{2}\Big) \ t^{a}\Big]_{\beta\alpha} 
\gamma^{+} \ u(\check{q},h') 
\ U_{F}\Big(\frac{L^{+}}{2},z^{+}_{1};\z,\frac{r^{-}}{2}\Big)_{\beta\beta'}
\ \overline{u}(\check{q},h') 
\Big[\frac{\gamma^{+}\gamma^{i}\gamma^{-}}{2}\Big] 
\nn \\
& \hspace{-1.7cm}
\times
\Psi_{\beta'}\Big(z_{1}^{+},\z,\frac{r^{-}}{2}\Big)
\ U_{A}\Big(z_{1}^{+},z^{+};\z,\frac{r^{-}}{2}\Big)_{ab} 
\overline{\Psi}_{\alpha'}\Big(\underline{z},\frac{r^-}{2}\Big)
\Big[ \frac{\gamma^{-}\gamma^{i}\gamma^{+}}{2}\Big]
\ u(\check{k},h) 
\Big[ t^{b}\ U_{F}\Big(z^{+},\frac{-L^{+}}{2};\z,\frac{r^{-}}{2}\Big)\Big]_{\alpha'\alpha}
\end{align}
The summation over the quark helicities can be performed easily and the Dirac structure can be organized as follows: 
 \begin{align}
 N_{1} &= \sum_{hh'} \overline{u}(\check{k},h) \gamma^{+} 
 u(\check{q},h') \overline{u}(\check{q},h') 
 \Big[\frac{\gamma^{+}\gamma^{i}\gamma^{-}}{2}\Big] 
 \Psi_{\beta'}\Big(z_{1}^{+},\z,\frac{r^{-}}{2}\Big)
 \overline{\Psi}_{\alpha'}\Big(\underline{z},\frac{r^-}{2}\Big)
 \Big[\frac{\gamma^{-}\gamma^{i}\gamma^{+}}{2}\Big] 
 u(\check{k},h)
 \nn \\
 &
 = - 
 \sum_{hh'} \overline{\Psi}_{\alpha'}\Big(\underline{z},\frac{r^-}{2}\Big)
 \Big[\frac{\gamma^{-}\gamma^{i}\gamma^{+}}{2}\Big] 
 u(\check{k},h) \overline{u}(\check{k},h) 
 \gamma^{+} 
 u(\check{q},h') \overline{u}(\check{q},h') 
 \Big[\frac{\gamma^{+}\gamma^{i}\gamma^{-}}{2}\Big] 
 \Psi_{\beta'}\Big(z_{1}^{+},\z,\frac{r^{-}}{2}\Big)
 \nn \\
 &
 = -
 \overline{\Psi}_{\alpha'}\Big(\underline{z},\frac{r^-}{2}\Big)
 \Big[ \frac{\gamma^{-}\gamma^{i}\gamma^{+}}{2}\Big]
  (\slashed k + m) \gamma^{+} (\slashed q + m)
  \Big[\frac{\gamma^{+}\gamma^{i}\gamma^{-}}{2}\Big] 
  \Psi_{\beta'}\Big(z_{1}^{+},\z,\frac{r^{-}}{2}\Big)
  \nn \\
&
= - 
2q^{+}2k^{+}\  \overline{\Psi}_{\alpha'}\Big(\underline{z},\frac{r^-}{2}\Big)
\ \gamma^{-} 
\Psi_{\beta'}\Big(z_{1}^{+},\z,\frac{r^{-}}{2}\Big)
 \end{align}   
Here, the overall minus sign originates from anticommutation of the quark background field insertions. Using this result in the first term of the quark background contribution to the quark production cross section given in Eq. \eqref{eq:q_back_X_sec_q_prod_1} in cross-section and simplifying it  further, one obtains 
\begin{align}
\label{cross section q in q-g BF1}
\frac{d\sigma^{qA\rightarrow q+ X}}{d\rm P.S.}\bigg|_{q\,  \rm backg. \, (1)} &=  
\int\! dr^{-} e^{ir^{-}(q^{+}-k^{+})}\frac{1}{2N_{c}} \ g^{2} \int\! d^{2}\z \int\! d^{2}\z' \ e^{-i(\z-\z')\cdot(\q-\k)} 
\int\! dz^{+} \int\! dz^{+}_{1} \ \theta(z_{1}^{+}-z^{+})  
\nn \\
&
\times 
\Big\langle 
U_{F}^{\dagger}\Big(\z', \frac{-r^{-}}{2}\Big)_{\beta\alpha} 
\Big[ U_{F}\Big(\frac{L^{+}}{2},z^{+}_{1};\z,\frac{r^{-}}{2}\Big) t^{a}\Big]_{\beta\beta'}
\ \overline{\Psi}_{\alpha'}\Big(\underline{z},\frac{r^-}{2}\Big)
\ \gamma^{-} 
\ \Psi_{\beta'}\Big(z_{1}^{+},\z,\frac{r^{-}}{2}\Big)
\nn  \\
&
\times  
 U_{A}\Big(z_{1}^{+},z^{+};\z,\frac{r^{-}}{2}\Big)_{ab}  
\ \Big[ t^{b}\ U_{F}\Big(z^{+},\frac{-L^{+}}{2};\z,\frac{r^{-}}{2}\big)\Big]_{\alpha'\alpha}\ \Big\rangle
 \end{align}
A similar calculation follows for the second term of the quark background contribution to the quark production cross section (second term in Eq. \eqref{eq:q_background_1}) and the final result reads 
\begin{align}
\label{cross section q in q-g BF2}
\frac{d\sigma^{qA\rightarrow q+ X}}{d\rm P.S.}\bigg|_{q\, \rm backg. \, (2)} &=  
\int\! dr^{-} e^{ir^{-}(q^{+}-k^{+})}\frac{1}{2N_{c}} \ g^{2}
\int\! d^{2}\z \int\! d^{2}\z' \ e^{-i(\z-\z')\cdot(\q-\k)} 
 \int\! dz'^{+} \int\! dz^{+}_{1}  \ 
 \nn \\
 &
\times\ \theta(z_{1}^{+}-z'^{+}) \   
\Big\langle \Big[ U_{F}^{\dagger}\Big(z'^{+},\frac{-L^{+}}{2};\z',\frac{-r^{-}}{2}\Big) \ t^{b}\big]_{\alpha'\alpha} 
U_{A}^{\dagger}\Big(z_{1}^{+},z'^{+};\z',\frac{-r^{-}}{2}\Big)_{ab}
\ 
\nn \\
&
\times 
\overline{\Psi}_{\beta'}\Big(z_{1}^{+},\z',\frac{-r^{-}}{2}\Big)\gamma^{-} \ \Psi_{\alpha'}\Big(\underline{z'},\frac{-r^-}{2}\Big)
\  \Big[ t^{a} \ U_{F}^{\dagger}\Big(\frac{L^{+}}{2},z^{+}_{1};\z',\frac{-r^{-}}{2}\Big)\Big]_{\beta\beta'}
\ U_{F}\Big(\z, \frac{r^{-}}{2}\Big)_{\beta\alpha}  \  \Big\rangle
 \end{align}
%
%

%%%%%%%%%%%%%%%%%%%%%%%%%%%%%%%%%%%%%%%%%%%%%%%%%%%%%%%%%%
\subsection{Total forward quark production cross section in quark-nucleus scattering at NEik accuracy}
%%%%%%%%%%%%%%%%%%%%%%%%%%%%%%%%%%%%%%%%%%%%%%%%%%%%%%%%%% 
The total forward quark production cross section in quark-nucleus scattering given in Eq. \eqref{eq:q_prod_X_sec_Sche} together with the gluon background contribution given in Eqs. \eqref{NEik cross section q in g BF} and the quark background contributions given in Eqs. \eqref{cross section q in q-g BF1} and \eqref{cross section q in q-g BF2}  can be written as 
\begin{align}
\frac{d\sigma^{qA\rightarrow q+ X}}{d\rm P.S.}\bigg|_\text{NEik} &= 
 \frac{1}{N_{c}} \ \int\! dr^{-} \ e^{ir^{-}(q^{+}-k^{+})} 
\int \! d^{2}\z' \  \int \! d^{2}\z\ e^{-i(\q -\k)\cdot(\z- \z')}
 \bigg\{
 2q^+ \Big\langle\tr \Big[ U_{F}^{\dagger}\Big(\z', \frac{-r^-}{2}\Big) U_{F}\Big(\z, \frac{r^-}{2}\Big)\Big] \Big\rangle
 \nn \\
 & 
+ %\int \! dz^{+} \
\Big\langle \tr\ \Big\{ U_{F}^{\dagger}\Big(\z', \frac{-r^-}{2}\Big) \Big[
-\frac{(\q^j\!+\!\k^j)}{2}\ U_{F; j}^{(1)}\Big(\z,\frac{r^-}{2}\Big) - iU_{F}^{(2)}\Big(\z, \frac{r^-}{2}\Big)\Big]\Big\}\Big\rangle
\nn \\
& 
+
%\int \! dz'^{+}    
\Big\langle  \tr \ \Big\{ 
\Big[ -\frac{(\q^j\!+\!\k^j)}{2}\ \ U_{F; j}^{(1)\dagger}\Big(\z,\frac{-r^-}{2}\Big)
+ iU_{F}^{(2)\dagger}\Big(\z, \frac{-r^-}{2}\Big)\Big]\, U_F\Big(\z, \frac{r^-}{2}\Big)\Big\}\Big\rangle
 \nn \\
 &
+  \int\! dz^{+} \int\! dz^{+}_{1} \ \theta(z_{1}^{+}-z^{+})  \ \frac{g^{2}}{2} 
 \Big\langle 
 U_{F}^{\dagger}\Big(\z', \frac{-r^{-}}{2}\Big)_{\beta\alpha} 
 \ \Big[ U_{F}\Big(\frac{L^{+}}{2},z^{+}_{1};\z,\frac{r^{-}}{2}\Big) t^{a}\Big]_{\beta\beta'}  
\  \overline{\Psi}_{\alpha'}\Big(\underline{z},\frac{r^-}{2}\Big) 
 \nn \\
 & \hspace{3.5cm}
 \times
 \ \gamma^{-}
\Psi_{\beta'}\Big(z_{1}^{+},\z,\frac{r^{-}}{2}\Big)
  U_{A}\Big(z_{1}^{+},z^{+};\z,\frac{r^{-}}{2}\Big) \ \Big[ t^{b} \  U_{F}\Big(z^{+},\frac{-L^{+}}{2};\z,\frac{r^{-}}{2}\Big)\Big]_{\alpha'\alpha} \ \Big\rangle
 \nn \\
 &
+ 
\int\! dz'^{+} \int\! dz^{+}_{1} \ \theta(z_{1}^{+}-z'^{+}) \ \frac{g^{2}}{2} 
 \Big\langle 
 \Big[ U_{F}^{\dagger}\Big(z'^{+},\frac{-L^{+}}{2};\z',\frac{-r^{-}}{2}\Big) \ t^{b}\Big]_{\alpha'\alpha} 
 \ U_{A}^{\dagger}\Big(z_{1}^{+},z'^{+};\z',\frac{-r^{-}}{2}\Big)_{ab} 
 \nn \\
 & \hspace{2.0cm}
 \times
 \  \overline{\Psi}_{\beta'}\Big(z_{1}^{+},\z',\frac{-r^{-}}{2}\Big)
 \gamma^{-} \ 
 \Psi_{\alpha'}\Big(\underline{z'},\frac{-r^-}{2}\Big)
 \  \Big[ t^{a} \ U_{F}^{\dagger}\Big(\frac{L^{+}}{2},z^{+}_{1};\z',\frac{-r^{-}}{2}\Big)\Big]_{\beta\beta'}
 \ U_{F}\Big(\z, \frac{r^{-}}{2}\Big)_{\beta\alpha} \Big\rangle 
 \bigg\}
  \end{align}
Like in the gluon to gluon channel in the previous section, we can now isolate the stictly eikonal and the strictly NEik terms by performing a gradient expansion around $r^-=0$, and keeping the zeroth and first order terms from the generalized eikonal contribution, and only the zeroth order term from the other contributions, which are already of order NEik. In such a way, we get
%
%
%As usual, one can also get the static target field limit of the quark production cross section by performing a Taylor expansion around $z^-=0$ and keeping only the zeroth order term in the NEik contributions while keeping both zeroth and first terms in the expansion in the eikonal contribution. The result simply reads 
 %
%
\begin{align}
\frac{d\sigma^{qA\rightarrow q+ X}}{d\rm P.S.}\bigg|_\text{NEik} &= 
  (2 \pi)\delta(q^{+}-k^{+})\, \frac{1}{N_{c}}
\int \! d^{2}\z' \  \int \! d^{2}\z\ e^{-i(\q -\k)\cdot(\z- \z')}
%\nn \\
% & \hspace{-2cm}
% \times
 \bigg\{ 
 2k^{+} \Big\langle \tr \big[ U_{F}^{\dagger}(\z') U_{F}(\z)\big] \Big\rangle
% + 
% 2k^{+}  \Big\langle \tr \Big\{ U_{F}^{\dagger}(\z') \big[\partial_{z^{-}}U_{F}(\z)\big]\Big\} \Big\rangle
% +  
% 2k^{+} \Big\langle \tr \Big\{ \big[\partial_{z^{-}}U_{F}^{\dagger}(\z')\big] U_{F}(\z)\Big\} \Big\rangle
 \nn \\
 & \hspace{-2cm}
 + 
%\int  \! dz^{+}  
\Big\langle \tr \ \Big\{  U_{F}^{\dagger}(\z')   
\Big[ - \frac{(\q^{j}\!+\! \k^{j})}{2} U_{F;j}^{(1)}(\z)  -iU_{F}^{(2)}(\z) \Big]\Big\}\Big\rangle  
+
%\int dz'^{+} 
\Big\langle \tr \Big\{ \Big[ - \frac{ (\q^{j}\!+\! \k^{j})}{2} U_{F;j}^{(1)\dagger}(\z')  +iU_{F}^{(2)\dagger}(\z')\Big] U_F(\z)\Big\}\Big\rangle   
\nn \\
 & \hspace{-2cm}
 +  \int\! dz^{+} \int\! dz^{+}_{1} \theta(z_{1}^{+}-z^{+}) \frac{g^{2}}{2} 
 \Big\langle U_{F}^{\dagger}(\z') \
  U_{F}\Big(\frac{L^{+}}{2},z^{+}_{1},\z\Big) t^{a} 
 \  \overline{\Psi}(\underline{z})\ \gamma^{-}
\Psi(z_{1}^{+},\z)   
U_{A}(z_{1}^{+},z^{+},\z)_{ab} \ t^{b}   
U_{F}\Big(z^{+},\frac{-L^{+}}{2},\z\Big) \Big\rangle
 \nn \\
 & \hspace{-2cm}
+ 
\int\! dz'^{+} \int\! dz^{+}_{1} \theta(z_{1}^{+}-z'^{+}) \frac{g^{2}}{2}   
\Big\langle U_{F}^{\dagger}\Big(z_{1}^{+},\frac{-L^{+}}{2},\z')\ t^{b} 
\ U_{A}^{\dagger}(z_{1}^{+},z'^{+},\z')_{ab}
\overline{\Psi}(z_{1}^{+},\z') \gamma^{-} 
\Psi(\underline{z'})\  t^{a} 
U_{F}^{\dagger}\Big(\frac{L^{+}}{2},z^{+}_{1},\z'\Big) 
U_{F}(\z) \Big\rangle  
 \bigg\}
 \nn\\
&\hspace{-2cm}\, +  2q^+ (2 \pi)\delta'(q^{+}-k^{+})\, \frac{i}{2}\, \frac{1}{N_{c}}
\int \! d^{2}\z' \  \int \! d^{2}\z\ e^{-i(\q -\k)\cdot(\z- \z')} 
\nn \\
& 
\times
\bigg\{  
  \Big\langle\tr\Big\{U_{F}^{\dagger}(\z') \big[\partial_{z^{-}}U_{F}(\z,z^-)\big]\Big\}\Big\rangle
- \Big\langle\tr\Big\{ \big[\partial_{z^{-}}U_{F}^{\dagger}(\z',z^-)\big]U_{F}(\z)\Big\rangle\bigg\}\bigg|_{z^-=0}
\, .
  \end{align}
 %
 %   

 %%%%%%%%%%%%%%%%%%%%%%%%%%%%%%%%%%%%%%%%%%%%%%%%%%%%%
\section{Forward gluon production in quark-nucleus scattering at NEik accuracy}
\label{sec:g_prod_in_qA}
 %%%%%%%%%%%%%%%%%%%%%%%%%%%%%%%%%%%%%%%%%%%%%%%%%%%%%
 %%%%%%%%%%%%%%%%%%%%%%%%%%%%%%%%%%%%%%%%%%%%%%%%%%%%%
Apart from the forward gluon production in gluon-nucleus scattering (Sec.~\ref{sec:gluon_production_in_gA}) and the forward quark production in quark-nucleus scattering (Sec.~\ref{sec:quark_in_quark-nucleus}), at NEik accuracy one actually has two more mechanism for gluon or quark production. The first one is the forward gluon production in quark nucleus scattering and the second one is the forward quark production in gluon-nucleus scattering. Both of these two mechanisms originate from quark background scattering in the amplitude and in the complex conjugate amplitude, so that the obtained cross section is at NEik order.  

Let us start with the computation of the forward gluon production cross section in quark-nucleus scattering at NEik accuracy. In this mechanism, the incoming quark interact with antiquark background field and it converts into a gluon both in the amplitude and in the complex conjugate amplitude. The mixed propagator for this case simply reads 
\begin{figure}
  \centering
  \includegraphics[width=0.5\linewidth]{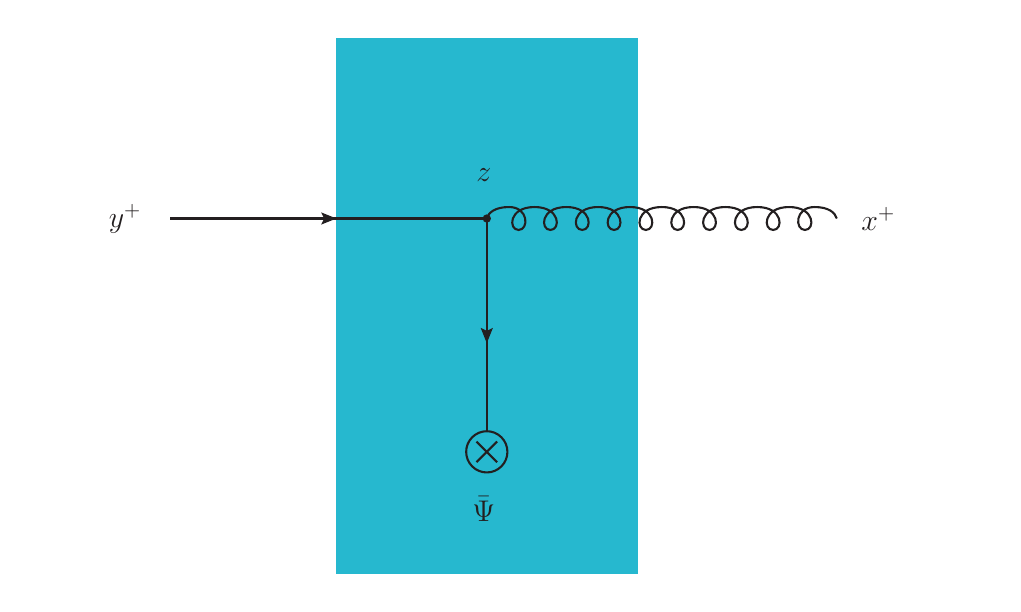}
  \caption{Diagram for quark to gluon conversion due to the quark background field of the target}
  \label{fig: case 1}
\end{figure}
%%%%%%%%%%%%%%%%%%%%%%%%%%%%%%%%%%%%%%%%%%%%%%%%%%%%%%%%%%%%%
%
%
\begin{align}
\label{General case 1}
M^{\mu}_{a,\alpha}(x,y) \big|_{q\, {\rm to}\, g } = \int \! d^{4}z \ G^{\mu \mu'}_{F}(x,z)_{ab}\big|^{\rm IA}_{\rm Eik} \ \overline{\Psi}^{-}_{\beta}(z) 
\ \Big[ (-igt^{b} \gamma_{\mu'}) \ S_{F}(z,y)\big|^{\rm BI}_{\rm Eik}\Big]_{\beta \alpha}
\end{align}
Substituting the explicit expressions of before-to-inside quark propagator Eq. \eqref{BI quark} and inside-to-after gluon propagator Eq. \eqref{eq:inside-after_gluon} into the mixed propagator and using the following relations for the Dirac matrices 
\begin{align}
\gamma^{0}\gamma^{-}\gamma^{+} \gamma_{\mu'} \gamma^{+} \gamma^{i} &=  \gamma^{0}\gamma^{-} \big( \{\gamma^{+}, \gamma_{\mu'}\} - \gamma_{\mu'} \gamma^{+}\big)\gamma^{+} \gamma^{i}
\nn \\
&= \gamma^{0}\gamma^{-}\gamma^{+} \gamma^{i}\big( 2g^{+}_{\mu'}\big)
\end{align}
to simplify the resulting expression, one obtains 
%
%
%\begin{align}
%M^{\mu}_{a,\alpha}(x,y)\big|_{g\, {\rm to}\,q} &= 
%\int \! d^{4}z 
%\  \int \! \frac{d^{3}\uq}{(2\pi)^{3}} \ \frac{\theta(q^{+})}{2q^+} 
%\ e^{-ix \cdot \check{q}} \ e^{iz^-q^+}  
%\  U_{A}(x^{+},z^{+},\z, z^{-})_{ab}  
%\nn \\
%& 
%\times  
%\Big[ 
%- g^{\mu}_{j} g^{j \mu'} + \frac{i g^{\mu}_{j}  \eta^{\mu'}}{q^{+}} \overleftarrow{D}_{\z^j}^{A} 
%+ \frac{g^{\mu}_{j}  \q^{j} \eta^{\mu'}}{q^{+}} + \frac{\eta^{\mu} g^{\mu'}_{j} \q^{j}}{q^{+}} 
%+ \frac{i \eta^{\mu}\eta^{\mu'}}{q^{+}q^{+}}\q^{j} \overleftarrow{D}_{\z^{j}}^{A}
%+ \frac{\eta^{\mu}\eta^{\mu'}}{q^{+}q^{+}}\q^{j} \q^{j}  
%\Big] 
%\ e^{-i\q\cdot \z} 
%\nn \\ 
%& 
%\times  
%\overline{\Psi}_{\beta}(z) \ \frac{ \gamma^{-} \gamma^{+}}{2} 
%\ \big[-igt^{b}_{\beta\beta'} \gamma_{\mu'}\big]
%\int \! \frac{d^{3}\uk}{(2 \pi)^3} \frac{\theta(k^+)}{2k^+} \ e^{-iz^-k^+} \ e^{iy \cdot \check{k}} \
%\ e^{i\z\cdot\k} (\check{\sk} + m) 
% \ U_{F}(z^+,y^+,\z,z^{-})_{\beta' \alpha}
%\end{align}
%
%
%Here, after simplification factor multiple of $\overrightarrow{D}_{z^i}^{F}$ goes to zero. Now, contracting $\mu'$ indices and using relation $\gamma^+ \gamma^+ = 0$, we get,
%
%
\begin{align}
\label{eq:prop_q_to_g}
M^{\mu}_{a,\alpha}(x,y)\big|_{q\, {\rm to}\,g} &=
\int \! \frac{d^{3}\uq}{(2\pi)^{3}} 
\frac{\theta(q^{+})}{2q^+} 
\  \int \! \frac{d^{3}\uk}{(2 \pi)^3}
\ \frac{ \theta(k^+)}{2k^+}
\ e^{-ix \cdot \check{q}} \ e^{iy \cdot \check{k}}
 \int \! d^4z \ e^{iz^-(q^+-k^{+})}  \ e^{-i\z\cdot(\q-\k)} 
 \ U_{A}(x^{+},z^{+},\z,z^{-})_{ab}
 \nn \\
 &
 \times 
\Big[g^{\mu}_{j} + \frac{\eta^{\mu} \q^{j}}{q^{+}}\Big] 
\ \overline{\Psi}_{\beta}(z) \Big[\ \frac{\gamma^{-}\gamma^{+} \gamma^{j}}{2}\Big]   
\ (\check{\sk} + m) 
 \ \Big[\big(igt^{b}\big) \ U_{F}(z^+,y^+,\z,z^{-})\Big]_{\beta \alpha}  
\end{align}
%
%
%
%
%This is the expression for mixed propagator for quark propagator before and inside the target and gluon propagator is from inside to after the target with the target being quark and gluon background field.\\
%%%%%%%%%%%%%%%%%%%%%%%%%%%%%%%%%%%%%%%%%%%%%%%%%%%%%%%%%%%%%%%%%%%%%%%%%%%%%%%%%%%%%%%%%%%%%%%%%%%%%%%%%%%%%%%%%%%%%
%
We can now use the obtained mixed propagator to compute the S-matrix element. The S-matrix element for gluon production in quark-nucleus scattering, where the incoming quark has helicity $h$, four momenta $\check{k}$ and color $\alpha$ and the outgoing gluon has polarization $\lambda$, four momenta $\check{q}$ and color $a$, is given by the LSZ-type reduction formula which reads 
\begin{align}
\label{eq:S-matrix_gen_qg}
 S_{g(\check{q}, \lambda,a) \leftarrow q(\check{k}, h, \alpha)} &= 
 \lim_{x^{+}\to +\infty} ( -1)\ (2q^{+})\int d^{2}\x \int dx^{-}\, e^{ix\cdot \check{q}}\, \epsilon^{\lambda}_{\mu}(\underline{q})^{*} 
 \nn \\
 &\times 
 \lim_{y^{+} \to -\infty} \int \! d^{2}\y \int \! dy^{-} \ e^{-iy\check{k}}\ M^{\mu}_{a,\alpha}(x,y)\big|_{g\, {\rm to}\, q} \ \gamma^{+} \ u(\check{k},h)
    \end{align}
After substituting the expression for mixed propagator, Eq. \eqref{eq:prop_q_to_g}, into Eq. \eqref{eq:S-matrix_gen_qg} and performing the trivial integrations, one obtains the S-matrix element as 
\begin{align}
\label{s matrix gq}
S_{g(\check{q}, \lambda,a) \leftarrow q(\check{k}, h, \alpha)} &= 
 - \frac{1}{2k^{+}}
 \epsilon^{\lambda}_{\mu}(\underline{q})^{*}  \ 
  \int \! dz^+   \int \! d^{2}\z  \ e^{-iz\cdot(\q-\k)} 
  \int\! dz^{-} \ e^{iz^-(q^{+}-k^{+})}
 \  U_{A}\Big(\frac{L^{+}}{2},z^{+},\z,z^{-}\Big)_{ab}
  \nn  \\
  &
  \times 
 \Big[ g^{\mu}_{j} + \frac{\eta^{\mu} \q^{j}}{q^{+}}\Big] 
 \ \overline{\Psi}_{\beta}(z) \ \Big[\frac{\gamma^{-}\gamma^{+} \gamma^{j}}{2} \Big]
 \ (\check{\sk} + m) 
 \ \Big[ \big(igt^{b}\big) \ U_{F}\Big(z^+,\frac{-L^{+}}{2},\z,z^{-}\Big)\Big]_{\beta \alpha}  \gamma^{+} u(\check{k},h)
  \end{align}
As discussed in previous sections, we can extract the $z^-$ dependent scattering amplitude from the S-matrix element Eq. \eqref{s matrix gq} which simply reads 
\begin{align}
i \mathcal{M}^{\lambda,h}_{a,\alpha}(z^{-}, \underline{k}, \underline{q})\big|_{q\, {\rm to}\, g} &=    
\frac{1}{(2k^{+})^2}
\epsilon^{\lambda}_{\mu}(\underline{q})^{*}  \ 
\int \! dz^+   \int \! d^{2}\z  \ e^{-i\z\cdot(\q-\k)}
\ U_{A}\Big(\frac{L^{+}}{2},z^{+},\z,z^{-}\Big)_{ab}
\ \Big[-g^{\mu}_{j} - \frac{\eta^{\mu} \q^{j}}{q^{+}}\Big]
\nn \\ 
&
\times 
 \ \overline{\Psi}_{\beta}(z) \ \Big[ \frac{\gamma^{-}\gamma^{+} \gamma^{j}}{2}\Big]  \  
(igt^{b})\ (\check{\sk} + m) 
\ U_{F}\Big(z^+,\frac{-L^{+}}{2},\z,z^{-}\Big) 
\gamma^{+} u(\check{k},h)    
\end{align}
Using the following simplifications for the gamma matrix structure 
\begin{align}
\gamma^{+}\gamma^{j}(\check{\sk}+ m)\gamma^{+} &= \gamma^{+} \gamma^{j} \big[\big\{\sk, \gamma^{+}\big\} -\gamma^{+}\check{\sk} \big] + (\gamma^{+}\gamma^{j}\gamma^{+})m
               = 2\gamma^{+} \gamma^{j}k^{+} - \gamma^{+} \gamma^{j}\gamma^{+}(\check{\sk}  -m) \\       
 \gamma^{+} \gamma^{j}\gamma^{+} & = \gamma^{+}\big[\big\{\gamma^{j},\gamma^{+}\big\}- \gamma^{+}\gamma^{j}\big] = \gamma^{+}(2g^{j+}) -0 =0
\end{align}
one obtains the scattering amplitude as 
\begin{align}
\label{amplitude gq}
i \mathcal{M}^{\lambda,h}_{a,\alpha}(z^{-}, \underline{k}, \underline{q})\big|_{q\, {\rm to}\, g} &=    
\frac{1}{2k^{+}}
\ \int \! dz^+   \int \! d^{2}\z  \ e^{-i\z\cdot(\q-\k)} 
\ U_{A}\Big(\frac{L^{+}}{2},z^{+},\z,z^{-}\Big)_{ab}  
\  {\epsilon_{\lambda}^{j}}^{*}  
\ \overline{\Psi}_{\beta}(z) 
\ \Big[\frac{\gamma^{-}\gamma^{+} \gamma^{j}}{2} \Big] 
\nn \\
&
\times 
\Big[\big(igt^{b}\big)\ U_{F}\Big(z^+,\frac{-L^{+}}{2},\z,z^{-}\Big)\Big]_{\beta \alpha}\ u(\check{k},h)    
    \end{align}
The scattering amplitude can now be used to compute the $z^-$ dependent cross section in the same way as discussed in previous sections which simply reads 
\begin{align}
{\frac{d \sigma^{qA \rightarrow g+X}}{d\rm P.S.}} &= 2k^{+} \int\! dr^{-} \ e^{ir^{-}(q^{+}-k^{+})}
   \frac{1}{2N_{c}}\sum_{h,\lambda} \sum_{a,\alpha} 
 \Big\langle   
 {\mathcal{M}\Big(\frac{-r^{-}}{2},  \underline{k}, \underline{q}\Big)^{\dagger} \ \mathcal{M}\Big(\frac{r^{-}}{2},  \underline{k}, \underline{q}\Big)}
 \Big\rangle
\end{align}
Plugging the explicit expression for the scattering amplitude given in Eq. \eqref{amplitude gq} into the cross section one gets
\begin{align}
\label{eq:qtog_Xsection_1}
\frac{d\sigma^{qA\rightarrow g+X}}{d\rm P.S.} \bigg|_\text{NEik} &=  
 \frac{g^{2}}{2N_{c}}  
\int\! dr^{-} \ e^{ir^{-}(q^{+}-k^{+})} 
\int\! d^{2}\z \int \! d^{2}\z'  \ e^{i(\q-\k)\cdot(\z'-\z)}
\ \sum_{h,\lambda}\sum_{\alpha,a}   
\ {\epsilon_{\lambda}^{j}}^{*}   \epsilon_{\lambda}^{j'} 
\frac{1}{ 2k^{+}}  \int \! dz'^+   \int \! dz^+ \ \overline{u}(\check{k},h)
\nn \\
&
\times
\Big\langle
\ \Big[ U_{F}^{\dagger}\Big(\frac{-L^{+}}{2},z'^+,\z', \frac{-r^{-}}{2}\Big)  \ t^{b'}\Big]_{\beta' \alpha} 
\ \Big[\frac{\gamma^{j'}\gamma^{+}\gamma^{-}}{2}\Big] 
\ \Psi_{\beta'}\Big(\underline{z}',\frac{-r^{-}}{2}\Big) 
\ U_{A}^{\dagger}\Big(\frac{L^{+}}{2},z'^{+},\z',\frac{-r^{-}}{2}\Big)_{ab'}
\nn \\
&
\times  
 U_{A}\Big(\frac{L^{+}}{2},z^{+},\z,\frac{r^{-}}{2}\Big)_{ab}
\ \overline{ \Psi}_{\beta}\Big(\underline{z},\frac{r^{-}}{2}\Big) 
\ \Big[ \frac{\gamma^{-}\gamma^{+} \gamma^{j}}{2} \Big] \ u(\check{k},h)
\ \Big[ t^{b}\ U_{F}\Big(z^+,\frac{-L^{+}}{2},\z,\frac{r^{-}}{2}\Big)\Big]_{\beta \alpha}
\ \Big\rangle
\end{align}
In order to simplify the expression for the cross section, one should perform the summation over the quark helicity and gluon polarizations, which yields to  
\begin{align}
\label{eq:N_1_q-to-g}
 N_1&= \sum_{h,\lambda} {\epsilon_{\lambda}^{j}}^{*}   \epsilon_{\lambda}^{j'} 
 \Big[ 
 \overline{u}(\check{k},h)   
 \frac{\gamma^{j'}\gamma^{+} \gamma^{-}}{2} \  
 \Psi_{\beta'}\Big(\underline{z}',\frac{-r^{-}}{2}\Big)   
 \overline{ \Psi}_{\beta}\Big(\underline{z},\frac{r^{-}}{2}\Big)
\ \frac{\gamma^{-}\gamma^{+} \gamma^{j}}{2} 
u(\check{k},h)
 \Big] 
 \nn \\
 &=
 \delta^{j'j} \   
 \overline{ \Psi}_{\beta}\Big(\underline{z},\frac{r^{-}}{2}\Big)
 \ \frac{\gamma^{-}\gamma^{j} \gamma^{+}}{2} 
 (\slashed k + m)  
 \frac{\gamma^{+}\gamma^{j'} \gamma^{-}}{2} 
 \ \Psi_{\beta'}\Big(\underline{z}',\frac{-r^{-}}{2}\Big)          
\nn \\
&= 
k^{+} \delta^{j'j}\  \overline{ \Psi}_{\beta}\Big(\underline{z},\frac{r^{-}}{2}\Big)
\  \gamma^{j} \gamma^{-}\gamma^{j'} 
\ \Psi_{\beta'}\Big(\underline{z}',\frac{-r^{-}}{2}\Big)
\nn \\
&= 
2k^{+}  \ \overline{ \Psi}_{\beta}\Big(\underline{z},\frac{r^{-}}{2}\Big)
\ \gamma^{-} \ \Psi_{\beta'}\Big(\underline{z}',\frac{-r^{-}}{2}\Big)
\end{align}
where we have used $\gamma^{+}(\slashed k + m )\gamma^{+} = 2k^{+}\gamma^{+}$ and $\gamma^{-}\gamma^{+}\gamma^{-} = 2\gamma^{-}$. 
Substituting the resulting expression back in the cross section given in Eq. \eqref{eq:qtog_Xsection_1}, one obtaines the final expression for the gluon production cross section in quark-nucleus scattering at NEik accuracy as 
\begin{align}
\label{eq:qtog_Xsection_withzminus_fin}
\frac{d\sigma^{qA\rightarrow g+X}}{d\rm P.S.} \bigg|_\text{NEik}  &=   
 \frac{g^{2}}{2N_{c}}   
\int\! dr^{-} \ e^{ir^{-}(q^{+}-k^{+})} 
\int\! d^{2}\z \ \int \! d^{2}\z'  \ e^{i(\q-\k)\cdot(\z'-\z)}  
\int \! dz'^+   \int \! dz^+ 
\nn \\
& 
\times
\Big\langle 
\ \Big[ U_{F}^{\dagger}\Big(\frac{-L^{+}}{2},z'^+,\z' ,\frac{-r^{-}}{2}\Big)\ t^{b'}\Big]_{\beta' \alpha} 
\overline{\Psi}_{\beta}\Big(\underline{z},\frac{r^{-}}{2}\Big) 
\ \gamma^{-} \ \Psi_{\beta'}\Big(\underline{z}',\frac{-r^{-}}{2}\Big) 
\nn \\
&
\times
U_{A}^{\dagger}\Big(\frac{L^{+}}{2},z'^{+},\z',\frac{-r^{-}}{2}\Big)_{ab'} 
\ U_{A}\Big(\frac{L^{+}}{2},z^{+},\z,\frac{r^{-}}{2}\Big)_{ab}   
\ \Big[t^{b}\ U_{F}\Big(z^+,\frac{-L^{+}}{2},\z,\frac{r^{-}}{2}\Big)\Big]_{\beta \alpha}\
\Big\rangle
\end{align}
Since the two insertions of the quark background field make this contribution NEik already, it is safe to neglect non-static effects, which would here be relevant only at NNEik order. Hence, one should gradient expand the quark fields and Wilson lines around $r^-=0$ and take into account only the zeroth order terms in the expansion. 
Then, the only $r^-$ dependence remains in the phase and the integration over $r^-$ gives a Dirac delta function in $q^+-k^+$, so that one obtains the gluon production cross section in forward quark-nucleus scattering at strict NEik order as 
\begin{align}
\label{eq:qtog_Xsection_static_fin}
\frac{d\sigma^{qA\rightarrow g+X}}{d\rm P.S.} \bigg|_\text{NEik} &=   
 \frac{g^{2}}{2N_{c}}   (2\pi)\delta(q^+\!-\!k^+)
%\int\! dr^{-} \ e^{ir^{-}(q^{+}-k^{+})} 
\int\! d^{2}\z \ \int \! d^{2}\z'  \ e^{i(\q-\k)\cdot(\z'-\z)}  
\int \! dz'^+   \int \! dz^+ 
%\nn \\
%& 
%\times
\Big\langle 
\ \Big[ U_{F}^{\dagger}\Big(\frac{-L^{+}}{2},z'^+\Big)\ t^{b'}\Big]_{\beta' \alpha} 
\overline{\Psi}_{\beta}(\underline{z}) 
\nn \\
&
\times
\gamma^{-} \ \Psi_{\beta'}(\underline{z}') 
\ U_{A}^{\dagger}\Big(\frac{L^{+}}{2},z'^{+},\z'\Big)_{ab'} 
\ U_{A}\Big(\frac{L^{+}}{2},z^{+},\z\Big)_{ab}   
\ \Big[t^{b}\ U_{F}\Big(z^+,\frac{-L^{+}}{2},\z\Big)\Big]_{\beta \alpha}\
\Big\rangle
\end{align}
%
%
%%%%%%%%%%%%%%%%%%%%%%%%%%%%%%%%%%%%%%%%%%%%%%%%%%%%%%%%%%%%%%%%%%%%%%%%%%%%%%%%%%%%%%%%%%%%%%%%%%%%%%%%%%%%%%%%%%%%%%%%%%%%%%%%%%%%%%%%%%%%%%%%%%%%%%%%%%%%%%%%%%%%%%%%%%%
\section{Forward quark production in gluon-nucleus scattering at NEik accuracy}
\label{sec:q_prod_in_gA}
\begin{figure}
  \centering
  \includegraphics[width=0.5\linewidth]{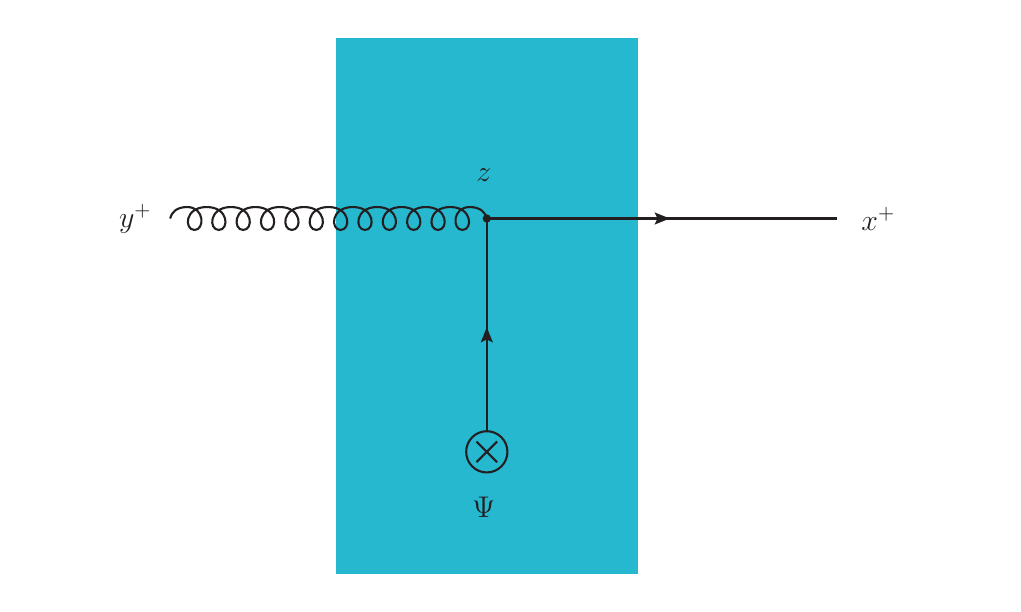}
  \caption{Diagram for quark to gluon conversion due to the quark background field of the target}
  \label{fig: case 2}
\end{figure}

%%%%%%%%%%%%%%%%%%%%%%%%%%%%%%%%%%%%%%%%%%%%%%%%%%%%
%%%%%%%%%%%%%%%%%%%%%%%%%%%%%%%%%%%%%%%%%%%%%%%%%%%% 
The last process that we are interested in studying is the forward quark production in gluon-nucleus scattering at NEik accuracy. In this process, the incoming gluon interacts with the quark background field of the target, converts into a quark and this quark exits the target medium (see Fig. \ref{fig: case 2}). The computation of the forward quark production in gluon-nucleus scattering at NEik accuracy closely follow the forward gluon production in quark-nucleus scattering studied in Sec. \ref{sec:g_prod_in_qA}. As in the previous case, we start with the mixed propagator for this case which reads  
\begin{align}
\label{eq:mixed_prop_q_to_g_gen}
M^{\;\;\nu}_{\beta,b}(x,y)\big|_{g\, {\rm to}\, q} = \int \! d^{4}z  
\ \Big[S_{F}(x,z)\big|^{{\rm IA},q}_{\rm Eik} \ \big(-igt^{a} \gamma_{\nu'}\big) \Big]_{\beta\alpha}  
\  \Psi^{-}_{\alpha}(z) 
\ \Big[ G^{\nu' \nu}_{F}(z,y)\big|^{\rm BI}_{\rm Eik}\Big]_{ab}
\end{align}
Substituting the before-to-inside gluon propagator given in Eq. \eqref{eq:before-inside_gluon} and inside-to-after quark propagator given in Eq. \eqref{q IA} in the mixed propagator Eq. \eqref{eq:mixed_prop_q_to_g_gen} and simplifiying the Dirac matrix structure one obtains 
 \begin{align}
 \label{eq:q_to_g_prop}
M^{\;\;\nu}_{\beta,b}(x,y)\big|_{g\, {\rm to}\, q} &=   
 \int\! \frac{d^{3}\uq}{(2\pi)^{3}} \frac{\theta(q^{+})}{2q^{+}}\ e^{-ix\check{q}} 
 \ \int \! \frac{d^{3}\uk}{(2\pi)^{3}} \ e^{iy\check{k}} \frac{\theta(k^{+})}{2k^{+}}
 \int d^4z \ e^{iz^{-}(q^{+}-k^{+})} \ e^{-i\z\cdot(\q-\k)} 
 \nn \\
 &
 \times
 (\check{\slashed q}+m) \Big[U_{F}(x^{+},z^{+},\z,z^{-}) \ (-igt^{a})\Big]_{\beta\alpha}\ 
  \Big[-g^{i\nu} + \frac{\eta^{\nu}\k^{i}}{k^{+}}\Big]
  \Big[ \frac{\gamma_{i}\gamma^{+}\gamma^{-}}{2} \Big]
  \Psi_{\alpha}(z) \
  U_{A}(z^{+},y^{+},\z,z^{-})_{ab}
\end{align}
We can now write the S-matrix element by using the obtained mixed propagator. The S-matrix element for quark production in froward gluon-nucleus scattering with incoming gluon with polarization $\lambda$, four momenta  $\check{k}$, color $b$ and the outgoing quark with helicity $h$, four momenta $\check{q}$ and color $\beta$ can be obtained from the mixed propagator given in Eq. \eqref{eq:q_to_g_prop} via LSZ-type reduction formula as 
\begin{align}
\label{eq:LSZ_q-to-g}
S_{q(\check{q},h,\beta)\leftarrow g(\check{k},b,\lambda)}&= 
\lim_{x^{+} \to \infty} \int\! d^{2}\x \int\! dx^{-} \ e^{ix \check{q}} \ \overline{u}(q,h) \gamma^{+} 
\nn \\
&
\times 
\lim_{y^{+} \to -\infty}(-1)(2k^{+})\
\int\! d^{2}\y \int\! dy^{-} \ e^{iy\check{k}}  \ M^{\;\;\nu}_{\beta,b}(x,y)\big|_{q\, {\rm to}\, g} \ \epsilon^{\lambda}_{\nu}(k)    
\end{align}
Substituting the explicit expression of the mixed propagator given in Eq. \eqref{eq:q_to_g_prop} into Eq. \eqref{eq:LSZ_q-to-g}, performing trivial integrations, renaming $p_{2} \equiv q$ and $p_{1} \equiv k$ for convenience one obtains the expression for the S-matrix which reads 
\begin{align}
 S_{q(\check{q},h,\beta)\leftarrow g(\check{k},b,\lambda)}&=      
 \int\!dz^{-} \ e^{iz^{-}(q^{+}-k^{+})}
\int \! d^{2}\z \ e^{-i\z\cdot(\q-\k)} 
\int \! dz^{+} \
\Big[ U_{F}\Big(\frac{L^{+}}{2},z^{+},\z,z^{-}\Big)\, (igt^{a}) \Big]_{\beta\alpha}
\nn \\
& 
\times 
\overline{u}(q,h) \ \Big[ \frac{\gamma^{+}\gamma_{i}\gamma^{-}}{2}\Big] 
\Psi_{\alpha}(z) \ \epsilon_{\lambda}^{i}
\ U_{A}\Big(z^{+},\frac{-L^{+}}{2},\z,z^{-}\Big)_{ab}
\end{align}
The $z^-$ dependent scattering amplitude for this process thus reads 
\begin{align}
\label{eq:scattering_amp_g_to_q}
 i \mathcal{M}^{\;\;\lambda}_{\beta,b}(z^{-}, \underline{k}, \underline{q}) &=     
 \frac{1}{2k^{+}}\ \int \! d^{2}\z 
 \ e^{-i\z\cdot(\q-\k)} \int \! dz^{+} \
\Big[ U_F\Big(\frac{L^{+}}{2},z^{+},\z,z^{-}\Big)\, (igt^{a}) \Big]_{\beta\alpha} 
\nn \\
& 
\times
\overline{u}(q,h) \  \Big[ \frac{\gamma^{+}\gamma_{i}\gamma^{-}}{2} \Big] 
\Psi_{\alpha}(z) \  \epsilon_{\lambda}^{i}
\ U_{A}\Big(z^{+},\frac{-L^{+}}{2},\z,z^{-}\Big)_{ab}     
\end{align}
The $z^-$ dependent quark production cross section in forward gluon-nucleus scattering can be obtained as before via 
\begin{align}
\frac{d\sigma^{gA\rightarrow q+X}}{d\rm P.S.} &=  
2k^{+}\int\! dr^{-} \ e^{ir^{-}(q^{+}-k^{+})} \frac{1}{2(N_{c}^{2}-1)} 
\sum_{h,\lambda}\sum_{a,\beta} 
\Big\langle
\mathcal{M}^{\lambda} \Big( \frac{-r^-}{2},\underline{k}, \underline{q}\Big)^{\dagger} 
\  \mathcal{M}^{\lambda}\Big( \frac{r^-}{2},\underline{k}, \underline{q}\Big)\
\Big\rangle
\end{align}
where one should substitute the scattering amplitude given in Eq.\eqref{eq:scattering_amp_g_to_q}. Upon this substitution, the quark production cross section in forward gluon-nucleus scattering can be written as 
\begin{align}
\label{eq:Xsec_g-to-q_1}
\frac{d\sigma^{gA\rightarrow q+X}}{d\rm P.S.} \bigg|_\text{NEik}&=   
2k^{+}\int\! dr^{-} \ e^{ir^{-}(q^{+}-k^{+})} 
 \frac{1}{2(N_{c}^{2}-1)} 
\sum_{h, \lambda}\sum_{a,\beta} 
\frac{1}{2k^{+}}
\int \! d^{2}\z' \ e^{i\z'\cdot(\q-\k)} \int \! dz'^{+}  
\nn \\
&
\times  
\Big\langle
U_{A}^{\dagger}\Big(z'^{+},\frac{-L^{+}}{2},\z',\frac{-r^{-}}{2}\Big)_{a'b}
\big( {\epsilon_{\lambda}^{i'}}^{*}\big) 
\ \overline{\Psi}_{\alpha'}\Big(z',\frac{-r^-}{2}\Big)
\ \Big[ \frac{\gamma^{-}\gamma^{+}\gamma_{i'}}{2}\Big] \   u(q,h) 
\nn \\
&
\times
\Big[ \big(gt^{a'} \big) \ U_{F}^{\dagger}\Big(\frac{L^{+}}{2},z'^{+},\z', \frac{-r^{-}}{2}\Big) \Big]_{\beta\alpha'} 
\frac{1}{2k^{+}} \ \int \! d^{2}\z \ e^{-i\z\cdot(\q-\k)} 
\int \! dz^{+}
 \ \Big[ U_{F}\Big(\frac{L^{+}}{2},z^{+},\z,\frac{r^{-}}{2}\Big) \ \big(gt^{a} \big)\Big]_{\beta\alpha}
 \nn \\
 &
 \times
\overline{u}(q,h) \Big[ \frac{\gamma_{i}\gamma^{+}\gamma^{-}}{2} \Big] 
\Psi_{\alpha}\Big(z,\frac{r^-}{2}\Big) \   \epsilon_{\lambda}^{i}
\ U_{A}\Big(z^{+},\frac{-L^{+}}{2},\z,\frac{r^{-}}{2}\Big)_{ab} \ 
\Big\rangle
\end{align}
Summation over the quark helicity and gluon polarization can be performed in a similar way as in Eq. \eqref{eq:N_1_q-to-g}. For this process, we have 
\begin{align}
\label{eq:num_g-to-q}
N_{1}&=  
\sum_{ \lambda, h}{\epsilon_{\lambda}^{i'}}^{*}\epsilon_{\lambda}^{i}  \overline{\Psi}(z') \ \frac{\gamma^{-}\gamma^{+}\gamma_{i'}}{2} \   u(q,h) \ \overline{u}(q,h) \frac{\gamma_{i}\gamma^{+}\gamma^{-}}{2} \Psi(z) 
\nn \\
&=
\delta^{i'i}  \ \overline{\Psi}(z')\frac{\gamma^{-}\gamma^{+}\gamma_{i'}}{2} \ (\slashed{\check{q}}+m) \frac{\gamma_{i}\gamma^{+}\gamma^{-}}{2} \Psi(z)  
=
\delta^{i'i} \overline{\Psi}(z')   \ \frac{\gamma^{-}\gamma_{i'}}{2} \ 2q^{+}\frac{\gamma^{+}\gamma_{i}\gamma^{-}}{2} \Psi(z)   
\nn \\
&
= 
\delta^{i'i} q^{+}  \overline{\Psi}(z') \ \bigg( \frac{-\gamma^{-}}{2} \ \big(\frac{\{ \gamma_{i'},\gamma_{i}\}}{2}\big)\gamma^{+}\gamma^{-} \bigg) \Psi(z)
= 2 q^{+} \  \overline{\Psi}(z') \gamma^{-}\ \Psi(z)
\end{align}
Finally, substituting Eq. \eqref{eq:num_g-to-q} into the cross section given in Eq. \eqref{eq:Xsec_g-to-q_1}, one gets the expression for the quark production cross section in forward gluon-nucleus scattering as 
%
%But we know that,
%$
%\sum_{\lambda}  \varepsilon^i_{\lambda} \varepsilon_{\lambda}^{j\, *}
%=\, \delta^{ij}$
%\, 
%and $\sum_{h}u(q)\overline{u}(q)=(\slashed{\check{q}}+m)$
%
%\begin{eqnarray*}
%    \begin{aligned}
%        N_{1} &=\delta^{i'i}  \ \overline{\Psi}(z')\frac{\gamma^{-}\gamma^{+}\gamma_{i'}}{2} \ (\slashed{\check{q}}+m) \frac{\gamma_{i}\gamma^{+}\gamma^{-}}{2} \Psi(z) \\
%         &= \delta^{i'i} \overline{\Psi}(z')   \ \frac{\gamma^{-}\gamma_{i'}}{2} \ 2q^{+}\frac{\gamma^{+}\gamma_{i}\gamma^{-}}{2} \Psi(z) \\
%         &= \delta^{i'i} q^{+} \ \overline{\Psi}(z') \ \big( \frac{-\gamma^{-}\gamma_{i'}}{2} \ \gamma_{i}\gamma^{+}\gamma^{-} \big)\Psi(z) \\
%         &= \delta^{i'i} q^{+} \  \overline{\Psi}(z') \bigg( \frac{-\gamma^{-}}{2} \ \big(\frac{\gamma_{i'}\gamma_{i}+\gamma_{i}\gamma_{i'}}{2}\big)\gamma^{+}\gamma^{-} \bigg)\Psi(z) \\
%        &= \delta^{i'i} q^{+}  \overline{\Psi}(z') \ \bigg( \frac{-\gamma^{-}}{2} \ \big(\frac{\{ \gamma_{i'},\gamma_{i}\}}{2}\big)\gamma^{+}\gamma^{-} \bigg) \Psi(z)\\
%         &= \delta^{i'i} q^{+}  \overline{\Psi}(z')  \bigg( \frac{-\gamma^{-}}{2} \ \big(\frac{2g_{i'i}}{2}\big)\gamma^{+}\gamma^{-} \bigg)\Psi(z)\\
%         &= \delta^{i'i}\delta_{i'i} q^{+}  \   \overline{\Psi}(z') \gamma^{-}\Psi(z)\\  
%         &= 2 q^{+} \  \overline{\Psi}(z') \gamma^{-}\ \Psi(z)
%        \end{aligned}
%\end{eqnarray*}
%
%
\begin{align}
\label{cross section g-q z minus}        
\frac{d\sigma^{gA\rightarrow q+X}}{d\rm P.S.} \bigg|_\text{NEik}&=   
\frac{2q^{+}}{2k^{+}}\int\! dr^{-} \ e^{ir^{-}(q^{+}-k^{+})}  \frac{1}{2(N_{c}^{2}-1)} 
\int \! d^{2}\z'  \int \! d^{2}\z \ e^{i(\z'-\z)\cdot(\q-\k)} 
\int \! dz'^{+} \int \! dz^{+} 
\nn \\
&
\times  
\Big\langle
U_{A}^{\dagger}\Big(z'^{+},\frac{-L^{+}}{2},\z',\frac{-r^{-}}{2}\Big)_{a'b}
\ \Big[ (gt^{a'} )\ U_{F}^{\dagger}\Big(\frac{L^{+}}{2},z'^{+},\z', \frac{-r^{-}}{2}\Big)\Big]_{\beta\alpha'} 
\ \overline{\Psi}_{\alpha'}\Big(z',\frac{-r^-}{2}\Big) \  \gamma^{-}
\nn \\
&
\times 
\Psi_{\alpha}\Big(z,\frac{r^-}{2}\Big) 
\ \Big[U_{F}\Big(\frac{L^{+}}{2},z^{+},\z,\frac{r^{-}}{2}\Big) \ (gt^{a} )\Big]_{\beta\alpha}
\ U_{A}\Big(z^{+},\frac{-L^{+}}{2},\z,\frac{r^{-}}{2}\Big)_{ab} \
\Big\rangle
 \end{align}
Again, this process starts only at NEik order, so that we can neglects corrections beyond the static limit, taking $r^-=0$ in the Wilson lines and quark background field insertions.   All in all, one gets the  forward quark production cross section in gluon-nucleus scattering at strict NEik accuracy as  
\begin{align}
\label{cross section g-q z minus_static_target}        
\frac{d\sigma^{gA\rightarrow q+X}}{d\rm P.S.} \bigg|_\text{NEik}&=   
 \frac{1}{2(N_{c}^{2}-1)} (2\pi)\delta(q^+\!-\!k^+)
\int \! d^{2}\z'  \int \! d^{2}\z \ e^{i(\z'-\z)\cdot(\q-\k)} 
\ \int \! dz'^{+} \int \! dz^{+} 
\Big\langle
U_{A}^{\dagger}\Big(z'^{+},\frac{-L^{+}}{2},\z'\Big)_{a'b}
\nn \\
&
\times  
\ \Big[ (gt^{a'} )\ U_{F}^{\dagger}\Big(\frac{L^{+}}{2},z'^{+},\z'\Big)\Big]_{\beta\alpha'} 
\ \overline{\Psi}_{\alpha'}(z') \  \gamma^{-}
%\nn \\
%&
%\times 
\Psi_{\alpha}(z) 
\ \Big[U_{F}\Big(\frac{L^{+}}{2},z^{+},\z\Big) \ (gt^{a} )\Big]_{\beta\alpha}
\ U_{A}\Big(z^{+},\frac{-L^{+}}{2},\z\Big)_{ab} \
\Big\rangle
\, .
 \end{align}
\section{Summary and outlook}
\label{sec:outlook} 
%%%%%%%%%%%%%%%%%%%%%%%%%%%%%%%%%%%%%%%%%%%%%%%%%%%%%
%%%%%%%%%%%%%%%%%%%%%%%%%%%%%%%%%%%%%%%%%%%%%%%%%%%%%

In this paper, we presented a comprehensive study of both gluon and quark propagators at full NEik accuracy in a dynamical gluon background field, and included the effects of a quark background field at NEik accuracy as well. We start with the derivation of the before-to-after gluon propagator in a dynamical gluon propagator at NEik accuracy. The NEik corrections to before-to-after gluon propagator stemming from relaxing the shockwave approximation were first computed in \cite{Altinoluk:2014oxa}. In this paper, we perform the derivation of those corrections in a more systematic way. Moreover, we also derive the NEik corrections beyond the static target field limit as well as the corrections originating from the interaction with the transverse component of the background gluon field. On the other hand, before-to-after quark propagator at NEik accuracy was computed in \cite{Altinoluk:2020oyd,Altinoluk:2021lvu}. Therefore, in this paper we just present those results for completeness. 

Apart from before-to-after quark and gluon propagators, in this work we perform the derivations of before-to-inside, inside-to-inside and inside-to-after quark and gluon propagators at eikonal accuracy. Even though these propagators are computed at eikonal order, 
they typically contribute only at NEik order to observables since either one or both legs of these propagators are inside the medium. 

Finally, by using all these propagators, we compute single inclusive quark and single inclusive gluon  forward production cross sections in quark-nucleus and gluon-nucleus scatterings at NEik accuracy. In the computation of gluon production in  gluon-nucleus scattering  and quark production in quark-nucleus scattering both gluon background and quark background contributions are taken into account. On the other hand, in the computation of gluon production in  quark-nucleus scattering and quark production in  gluon-nucleus scattering pure gluon background does not contribute since one needs at least one t-channel quark exchange in order to convert an incoming quark into a gluon or an incoming gluon into a quark.  

Before-to-after, before-to-inside, inside-to-inside and inside-to-after quark and gluon propagators computed in this paper can serve as  building blocks for computing many observables at NEik accuracy. As a natural continuation of this work and as an immediate application of the results derived here, we are planing to study forward dijet production in  pA collisions at NEik accuracy. It would be interesting to understand the interplay between the kinematic twist corrections and the NEik corrections in the back-to-back limit in that context. 

Another interesting observable that can be computed by using the propagators derived in this paper is single inclusive jet production in deep inelastic scattering (SIDIS) at NEik accuracy. As recently shown in \cite{Altinoluk:2024vgg}, in the strict eikonal limit, CGC computations only provide gluon-splitting contribution to the quark TMD. However, at NEik accuracy, by including a t-channel quark exchange, one may achieve to a complete connection with quark TMD from the CGC calculations. 

Last but not least, one can also study the back-to-back limit of photon+jet production at forward rapidity in pA collisions at NEik accuracy. At eikonal order, this observable is known to probe dipole gluon TMD without taking the back-to-back limit since photon does not interact with the target \cite{Dominguez:2011wm}. However, it is not clear whether this remains  true at NEik accuracy. Therefore, one should compute the photon+jet production at NEik order in a pure gluon background to see if this statement holds and what type of corrections one gets to the gluon TMD at NEik order. Moreover, by including a t-channel quark exchange in the photon+jet production at NEik order, one can also get access to quark TMDs which we are planing include in our future study.

\acknowledgements{%We thank ... for useful discussions. 
TA is supported in part by the National Science Centre (Poland) under the research Grant No. 2023/50/E/ST2/00133 (SONATA BIS 13). GB and SM are supported in part by the National Science Centre (Poland) under the research Grant No. 2020/38/E/ST2/00122 (SONATA BIS 10). 
%This work has been performed in the framework of MSCA RISE 823947 ``Heavy ion collisions: collectivity and precision in saturation physics'' (HIEIC) and has received funding from the European Union's Horizon 2020 research and innovation programme under grant agreement No. 824093.}

\appendix

%%%%%%%%%%%%%%%%%%%%%%%%%%%%%%%%%%%%%%%%%%%%%%%%%%%%%
%%%%%%%%%%%%%%%%%%%%%%%%%%%%%%%%%%%%%%%%%%%%%%%%%%%%%
%%%%%%%%%%%%%%%%%%%%%%%%%%%%%%%%%%%%%%%%%%%%%%%%%%%%%
%%%%%%%%%%%%%%%%%%%%%%%%%%%%%%%%%%%%%%%%%%%%%%%%%%%%%
%%%%%%%%%%%%%%%%%%%%%%%%%%%%%%%%%%%%%%%%%%%%%%%%%%%%%

%%%%%%%%%%%%%%%%%%%%%%%%%%%%%%%%%%%%%%%%%%%%%%%%%%
%%%%%%%%%%%%%%%%%%%%%%%%%%%%%%%%%%%%%%%%%%%%%%%%%%
%%%%%%%%%%%%%%%%%%%%%%%%%%%%%%%%%%%%%%%%%%%%%%%%%%

\section{Derivation of the NEik corrections to the gluon propagator in a pure $\A^-$ background}
\label{app:derivation_NEik_pure_A-}

%%%%%%%%%%%%%%%%%%%%%%%%%%%%%%%%%%%%%%%%%%%%%%%%%%
%%%%%%%%%%%%%%%%%%%%%%%%%%%%%%%%%%%%%%%%%%%%%%%%%%
%%%%%%%%%%%%%%%%%%%%%%%%%%%%%%%%%%%%%%%%%%%%%%%%%%
This section is devoted to the details of the derivation of the medium corrections to the gluon propagator at NEik accuracy in gluon $\A^-$ background field, in particular NEik contributions beyond the shockwave limit. The steps in this derivation is closely related with the derivation of NEik corrections to a quark propagator performed in \cite{Altinoluk:2020oyd}. The new feature in the derivation of the medium corrections to the gluon propagator at NEik accuracy presented in the rest of this section is the inclusion of NEik corrections beyond the static approximation of the target, by keeping an overall $z^-$ dependence of the background field. 

The medium corrections to a gluon propagator in a pure $\A^-$ background before performing the shockwave approximation is given in Eq. \eqref{Without shockwave approx G}, up to NNEik corrections beyond the static limit. 
Focusing on the case $x^{+} > y^{+}$, it reads
%The NEik corrections to the gluon propagator in a pure $\A^-$ background stems from relaxing the shockwave approximation for the target fields and considering a finite longitudinal width $L^+$ of the target along $z^+$ direction. By keeping this width in Eq. \eqref{Without shockwave approx G} and focusing on the case where $x^{+} > y^{+}$, one gets the medium corrections to a gluon propagator in a pure $\A^-$ background including all Eik and NEik terms as 
%
%
\begin{align}
\delta G_{F}^{\mu\nu}(x,y)\Big|_{\text{pure} \ \A^-, \, z^-}  &= 
         \int \! \frac{d^{3}\underline{q}}{(2\pi)^3} \int \! \frac{d^{3}\uk}{(2\pi)^{3}}  \ \frac{\theta(q^+) \theta(k^+)}{q^+ + k^+}\ \bigg[ -g^{\mu \nu} + \frac{\check{k}^{\mu} \eta^{\nu}}{k^{+}} + \frac{\eta^{\mu} \check{q}^{\nu}}{q^{+}} - \frac{\eta^{\mu} \eta^{\nu}}{q^{+} k^{+}} (\check{q} \cdot \check{k})   \bigg] \nn \\
&\hspace{-3cm} \times 
\int\! dz^{-} \ e^{-iq^{+}(x^{-}-z^{-})} \ e^{-ik^{+}(z^{-}-y^{-})} \sum_{N=1}^{+\infty} \  
\Bigg[ \prod_{n=1}^{N} \int \! d^{3}\underline{z_{n}} \Bigg] 
\Bigg[\mathcal{P}_{n} \prod_{n=1}^{N} \Big(-igA^{-}(\underline{z_{n}},z^-) \cdot T\Big)\Bigg] \nn \\
& \hspace{-3cm} \times 
\Bigg[ \prod_{n=0}^{N}\theta(z_{n+1}^{+}-z_{n}^{+}) \Bigg] 
\ e^{i\q \cdot (\x-\z_N )} \ e^{i\k \cdot(\z_{1} - \y)} 
\ e^{\frac{-i(x^{+} - z_{N}^{+})\q^{2}}{2q^{+}}} \ e^{\frac{-i(z_{1}^{+} - y^{+})\k^{2}}{2k^{+}}} \nn \\
& \hspace{-3cm} \times
\Bigg[ \prod_{n=1}^{N-1}  \int \! \frac{d^2\p_{n}}{(2 \pi)^2} 
\ e^{i\p_n\cdot(\z_{n+1} - \z_{n})}  \ e^{\frac{-i(z^{+}_{n+1} -z_{n}^{+})\p_{n}^{2}}{(q^+ + k^+)}}\Bigg] + \text{NNEik}
\, ,
\end{align}
or, after regrouping terms and integrating over $\p_n$, 
\begin{align}
\label{eq:delta_G_gen_1}
\delta G_{F}^{\mu\nu}(x,y)\Big|_{\text{pure} \ \A^-, \, z^-}
&= 
\int \! \frac{d^{3}\underline{q}}{(2\pi)^3} \int \! \frac{d^{3}\uk}{(2\pi)^{3}}  
 \frac{\theta(q^+) \theta(k^+)}{q^+ \!+\! k^+}
 \bigg[ -g^{\mu \nu} \!+\! \frac{\check{k}^{\mu} \eta^{\nu}}{k^{+}} \!+\! \frac{\eta^{\mu} \check{q}^{\nu}}{q^{+}} \!-\! \frac{\eta^{\mu} \eta^{\nu}}{q^{+} k^{+}} (\check{q} \!\cdot \! \check{k}) \bigg]
e^{-i\check{q}\cdot x}  e^{i\check{k} \cdot y} \!\!
\int\! dz^{-}  e^{iz^{-}(q^{+}-k^{+})} \nn \\
&\hspace{-3cm} \times 
\  \sum_{N=1}^{+\infty} 
\  \Bigg[ \prod_{n=1}^{N} \int \! d^{3}\underline{z_{n}} \Bigg] 
\Bigg[\mathcal{P}_{n} \prod_{n=1}^{N} \Big(-igA^{-}(\underline{z_{n}},z^-) \cdot T\Big)\Bigg] 
\ e^{-i\z_{N }\cdot \q} \ e^{i\z_{1}\cdot \k} 
\ e^{i z_{N}^{+}\check{q}^{-}} \ e^{-iz_{1}^{+}\check{k}^{-}}
\nn \\
& \hspace{-3cm} \times 
\Bigg[ \prod_{n=0}^{N}\theta(z_{n+1}^{+}-z_{n}^{+}) \Bigg] 
% \nn \\
%& \hspace{-3cm} \times
 \Bigg[ \prod_{n=1}^{N-1} \frac{(-i)(q^+ \!+\! k^+)}{4\pi (z_{n+1}^{+}\!-\!z_{n}^{+})} 
 \ e^{\frac{i(q^+ + k^+)(\z_{n+1} - \z_n)^{2}}{4(z^{+}_{n+1} - z_{n}^{+})}}\Bigg] + \text{NNEik}
 \, .
\end{align}
The idea of the shockwave approximation, which is part of the Eikonal approximation, is that in the limit of large boost of the target, the target becomes infinitely Lorentz contracted along the $x^+$ direction, so that partons scattering on that target have no time (along $x^+$) to have a noticeable transverse motion while they cross the whole target. Hence, in that limit, all of the interactions with the background field along the same propagator happens at the same transverse position $\z$, see Eq.~\eqref{delta_G_bef_Wilson}.
The NEik corrections beyond the shockwave limit stem from the difference between the true transverse positions of the interaction vertices with the background field and the approximate position $\z$.

For that calculation, we follow the same steps as in Ref.~\cite{Altinoluk:2020oyd}, where the case of the quark propagator was considered. By comparison, one can realize that for $k^+=q^+$ the last two lines of Eq. \eqref{eq:delta_G_gen_1} are the same as the last three lines of Eq. (A1) of Ref.~\cite{Altinoluk:2020oyd} for the case of $x^+>y^+$, up the change of color representation. Hence the same procedure can be followed to collect the field insertions to write the result in terms of Wilson lines. The result can be read off from Eq. (A14) of \cite{Altinoluk:2020oyd}, and for the gluon propagator in the case of $x^+>y^+$ it reads 
\begin{align}
\label{eq:delta_G_gen_2}
\delta G_{F}^{\mu\nu}(x,y)\Big|_{\text{pure} \ \A^-, \, z^-}^{\rm Eik} +
\delta G_{F}^{\mu\nu}(x,y)\Big|_{\text{pure} \ \A^-, \, z^-}^{\rm NEik} 
&= 
\int \! \frac{d^{3}\underline{q}}{(2\pi)^{3}} \int \! \frac{d^{3}\underline{k}}{(2\pi)^{3}}  
\frac{\theta(q^+) \theta(k^{+})}{q^{+} + k^{+}} 
\bigg[ -g^{\mu\nu} + \frac{\check{k}^{\mu} \eta^{\nu} }{k^{+}} + \frac{\eta^{\mu} \check{q}^{\nu}}{q^{+}}
-\frac{\eta^{\mu}\eta^{\nu}}{q^{+}k^{+}} (\check{q} \cdot \check{k}) \bigg] \nn \\
& \hspace{-3cm} \times  
\ e^{-ix \cdot \check{q}} \ e^{iy \cdot \check{k}}
\int\! dz^{-} \ e^{-iz^{-}(q^{+}-k^{+})} \int \!  d^{2} \z \ e^{-i\z\cdot(\q - \k)} 
\Bigg\{ \bigg[ U_{A}(x^{+},y^{+}; \z,z^{-}) -1 \bigg]\nn \\
& \hspace{-3cm}+ 
\frac{\q^{j}+ \k^{j}}{q^{+} + k^{+}} \int_{y^{+}}^{x^{+}} 
\! dv^{+}  v^{+}  \ U_{A}(x^{+},v^{+}; \z, z^{-}) 
\bigg[ -ig \ T\cdot \frac{d}{d\z^{j}} \A^{-}(v^{+},\z, z^{-})\bigg]
U_{A}(v^{+},y^{+}; \z, z^{-}) \nn \\
& \hspace{-3cm}-
\frac{i}{q^{+} + k^{+}} \int_{y^{+}}^{x^{+}} \! dv^{+} \int_{v^{+}}^{x^{+}} \! dw^{+} \ (w^{+}-v^{+}) 
\ U_{A}(x^{+},w^{+}; \z,z^{-}) \ 
\bigg[ -ig\ T \cdot \frac{d}{d\z^{j}} \A^{-}(w^{+},\z, z^{-})\bigg] \nn \\
& \hspace{-3cm} \times 
\ U_{A}(w^{+},v^{+}; \z,z^{-})  \bigg[ -ig \ T \cdot \frac{d}{d\z^{j}} \A^{-}(v^{+},\z,z^{-})\bigg]
\ U_{A}(v^{+},y^{+}; \z,z^{-})
 \Bigg\} + \text{NNEik}
\end{align}
Eq. \eqref{eq:delta_G_gen_2} can be further simplified and written in a more compact form. First of all, using the definition of a Wilson line Eq. \eqref{def:Wilson_R}, one has (for $x^+>y^+$)
\begin{align}
\frac{d}{d\z^{j}} \ U_{A}(x^{+},y^{+}; \z,z^{-}) = \int_{y^{+}}^{x^{+}} \! dz^{+} 
\ U_{A}(x^{+},z^{+}; \z,z^{-}) \bigg[ -ig \ T \cdot \frac{d}{d\z^{j}} \A^{-}(z^{+},\z,z^{-})\bigg] 
\ U_{A}(z^{+},y^{+}; \z,z^{-})
\end{align}
which can be used to simply the bilocal term in Eq. \eqref{eq:delta_G_gen_2} as 
\begin{align}
\label{eq:GP_biloc_fin}
&
 \int_{y^{+}}^{x^{+}} \! dv^{+} \ \int_{v^{+}}^{x^{+}} \! dw^{+} \ (w^{+}-v^{+}) 
\ U_{A}(x^{+},w^{+}; \z,z^{-})  
\bigg[ -ig \ T \cdot \frac{d}{d\z^{j}} \A^{-}(w^{+},\z,z^{-})\bigg]
\ U_{A}(w^{+},v^{+}; \z,z^{-}) \nn \\
&\times 
\bigg[ -ig \ T \cdot \frac{d}{d\z^{j}} \A^{-}(v^{+},\z,z^{-})\bigg] 
\ U_{A}(v^{+},y^{+}; \z,z^{-})
= \int_{y^{+}}^{x^{+}} \! dz^{+}
\ U_{A}(x^{+},z^{+}; \z,z^{-}) \overleftarrow{\frac{d}{d\z^{j}}}\ \overrightarrow{\frac{d}{d\z^{j}}} 
\ U_{A}(z^{+},y^{+}; \z,z^{-})
\, ,
\end{align}
for $x^+>y^+$. 

Moreover, in the case of propagation from before the background field to after the background field, meaning $x^+>L^+/2$ and $y^+<-L^+/2$ (with $L^+$ the width of the target), 
one can simplify the local term in Eq. \eqref{eq:delta_G_gen_2} by noting that the integration range is effectively restricted to the support of the background field, $-L^+/2<v^+<L^+/2$. One has 
\begin{align}
v^{+} = \frac{1}{2}\bigg[ \int_{z^{+}_{\text{min}}}^{v^{+}} \! dz^{+} - \int_{v^{+}}^{z^{+}_{\text{max}}} \! dz^{+}  \bigg] + \frac{(z^{+}_{\text{max}} + z^{+}_{\text{min}})}{2} 
\end{align}
where $z^{+}_{\text{max}} = L^+/2$ and $z^{+}_{\text{min}} = -L^+/2$, so that 
\begin{align}
\label{def:v_plus}
v^{+} = \frac{1}{2}\bigg[ \int_{\frac{-L^{+}}{2}}^{v^{+}} \! dz^{+} - \int_{v^{+}}^{\frac{L^{+}}{2}} \! dz^{+}  \bigg]
\end{align}
Thus, by using Eq. \eqref{def:v_plus}, we can write the local term in Eq. \eqref{eq:delta_G_gen_2} as 
\begin{align}
\label{eq:GP_loc_fin}
&
\int \!  d^{2} \z \ e^{-i\z\cdot(\q - \k)} \ \int_{\frac{-L^{+}}{2}}^{\frac{L^{+}}{2}} \! dv^{+} 
\ v^{+} 
\ U_{A}\bigg({\frac{L^{+}}{2}},v^{+}; \z, z^{-}\bigg) 
\bigg[ -ig \ T \cdot \frac{d}{d\z^{j}} \A^{-}(v^{+},\z,z^{-})\bigg]
 U_{A}\bigg(v^{+},-{\frac{L^{+}}{2}}; \z, z^{-}\bigg) \nn \\
 & =
\frac{1}{2}  \int \!  d^{2} \z \ e^{-i\z(\q - \k)} 
\ \int_{\frac{-L^{+}}{2}}^{\frac{L^{+}}{2}} \! dz^{+} 
\bigg[\frac{d}{d\z^{j}} 
U_{A}\bigg({\frac{L^{+}}{2}},z^{+}; \z, z^{-}\bigg)\bigg] 
\ U_{A}\bigg(z^{+},-{\frac{L^{+}}{2}}; \z, z^{-}\bigg) \nn  \\
& - 
\frac{1}{2} \int \!  d^{2} \z \ e^{-i\z\cdot(\q- \k)} 
\ \int_{\frac{-L^{+}}{2}}^{\frac{L^{+}}{2}} \! dz^{+} 
\ U_{A}\bigg({\frac{L^{+}}{2}},z^{+}; \z, z^{-}\bigg)
\bigg[\frac{d}{d\z^{j}} U_{A}\bigg(z^{+},-{\frac{L^{+}}{2}}; \z, z^{-}\bigg)\bigg] \nn \\
& = 
-\frac{1}{2} \int \!  d^{2} \z \ e^{-i\z\cdot(\q - \k)} 
\ \int_{\frac{-L^{+}}{2}}^{\frac{L^{+}}{2}} \! dz^{+} 
\bigg[ U_{A}\bigg({\frac{L^{+}}{2}},z^{+}; \z, z^{-}\bigg) 
\left(\overrightarrow{\frac{d}{d\z^{j}}} - \overleftarrow{\frac{d}{d\z^{j}}} \right)
U_{A}\bigg(z^{+},-{\frac{L^{+}}{2}}; \z, z^{-}\bigg) \bigg]
\, ,
\end{align}
where the derivatives act only on the Wilson lines, not on the phases.
Finally, substituting Eqs. \eqref{eq:GP_biloc_fin} and \eqref{eq:GP_loc_fin} into Eq. \eqref{eq:delta_G_gen_2}, one  gets the medium corrections to a before-to-after gluon propagator with positive light-cone momentum at NEik accuracy in a pure $\A^-$ background field which is given in Eq. \eqref{A-NEik}.

%%%%%%%%%%%%%%%%%%%%%%%%%%%%%%%%%%%%%%%%%%%%%
%%%%%%%%%%%%%%%%%%%%%%%%%%%%%%%%%%%%%%%%%%%%%
%%%%%%%%%%%%%%%%%%%%%%%%%%%%%%%%%%%%%%%%%%%%%%%%%%%
\section{Derivation of inside-to-inside quark propagator}
\label{App:inside-to-inside-quark}
%%%%%%%%%%%%%%%%%%%%%%%%%%%%%%%%%%%%%%%%%%%%%%%%%%%%%
 In this Appendix we present the derivation the inside-to-inside quark propagator at eikonal order, i.e. when the beginning and the end points of the quark propagator are between $-L^+/2$ and $L^+/2$.  As it we will explain in detail in this appendix, the inside-to-inside quark propagator at eikonal order gets contribution from the interaction of with $\A^-$ and $\A^j$ components of the background gluon field.     
        
 %%%%%%%%%%%%%%%%%%%%%%%%%%%%%%%%%%%%%%%%%%%%%%%%%%%%%%%
\subsection {Inside-to-inside quark propagator in a pure $\A^-$ background at eikonal order}
%%%%%%%%%%%%%%%%%%%%%%%%%%%%%%%%%%%%%%%%%%%%%%%%%%%%%
Our starting point for the computation of this contribution is the quark propagator at generalized eikonal order, in a background field containing only a $\A^-$ component,  which is computed in \cite{Altinoluk:2022jkk} as 
\begin{align}
S_{F}(x,y) |_{\text{Eik},\, \A^{-}} &=  \delta^{3}(\underline{x} - \underline{y}) \gamma^+ \ \int \! \frac{dk^+}{2 \pi} \frac{i}{2k^+} \ e^{-ik^{+}(x^{-} - y^{-})} 
\nn \\
&+
\int \! \frac{d^{3}\up}{(2 \pi)^3} \ \int \! \frac{d^{3}\uk}{(2 \pi)^3}\ e^{-ix \cdot \check{p} + iy \cdot \check{k}} \int \! dz^- \ e^{iz^-(p^+ - k^+)} \ \int \! d^{2}\z \ e^{-i\z\cdot (\p - \k)} \ \frac{(\check{\sp}+ m)}{2p^+} \gamma^{+}
\nn  \\
&
\times 
\bigg\{ \theta(x^{+} -y^{+}) \theta(p^+) \theta(k^+) \ U_{F}(x^+,y^+,\z,z^-)
\nn  \\
& 
- \theta(y^{+} -x^{+}) \theta(-p^+) \theta(-k^+) \ U_{F}^{\dagger}(x^+,y^+,\z,z^-) \bigg\} \frac{(\check{\sk} + m)}{2k^+}
    \end{align}
Let us concentrate on the case where  $x^+ > y^+$ case,  so that this expression of the quark propagator becomes 
\begin{align}
\label{eq:II_quark_A_minus_1}
S_{F}(x,y) |_{\text{Eik},\, \A^{-}} &= 
\int \! \frac{d^{3}\up}{(2 \pi)^3} \ \int \! \frac{d^{3}\uk}{(2 \pi)^3}\ e^{-ix \cdot \check{p} + iy \cdot \check{k}} \int \! dz^- \ e^{iz^-(p^+ - k^+)} \ \int \! d^{2}\z \ e^{-i\z\cdot(\p - \k)} \ \frac{\check{ \sp}+ m}{2p^+} \gamma^{+}
\nn  \\
&
\times  
\theta(p^+) \theta(k^+) \ U_{F}(x^+,y^+,\z,z^-) \frac{(\check{\sk} + m)}{2k^+} 
 \end{align}
Our aim is to compute the quark propagator where both ends of the propagator are inside the medium such that  $\frac{-L^+}{2}< x^+ < \frac{L^+}{2}$ and $\frac{-L^+}{2}< y^+ < \frac{L^+}{2}$, at the eikonal order. In that regime, $x^+$ and $y^+$ are of order $L^+$ which is power suppressed in the shockwave limit. Hence, at eikonal order, it is safe to neglect all terms proportional to  $x^+$ and $y^+$ in the phases.
% One should approximate the phase factor in $y^{+}$ and $x^{+}$ by one since the the rest of the terms that can come from the phase factor brings extra factor of suppression. 
 Moreover, one can also expand the gamma matrix structure term by term to simplify it. Using the relation,
\begin{align}
\gamma^+ \gamma^- \gamma^+ \gamma^- = \Big[ \{\gamma^+, \gamma^-\} - \gamma^- \gamma^+ \Big] \gamma^+ \gamma^- = 2g^{+-} \gamma^+ \gamma^- = 2 \gamma^+ \gamma^- 
\end{align}
one obtains
\begin{align}
\label{eq:gamma_Struc_II_A_minus}
        (\check{\slashed{p}} + m) \gamma^+ (\check{ \slashed{k} } + m) &= (\gamma^+ \check{p}^- + \gamma^- p^+ - \gamma^i \p^i + m) \gamma^+ (\gamma^+ \check{k}^- + \gamma^- k^+ - \gamma^j \k^j + m)
\nn \\        
&
= 2\gamma^- p^+ k^+ - \gamma^- \gamma^+ \gamma^j p^+ \k^j + \gamma^- \gamma^+ m p^+ - \gamma^i \gamma^+ \gamma^- \p^i k^+ 
\nn \\
& 
+ \gamma^i \gamma^+ \gamma^j \p^i \k^j - \gamma^i \gamma^+ m \p^i + \gamma^+ \gamma^- m k^+ - \gamma^+ \gamma^j m \k^j + m^2 \gamma^+
\nn \\
&
= 2\gamma^- p^+ k^+ - \gamma^- \gamma^+ \gamma^j p^+ \k^j + \gamma^- \gamma^+ m p^+ 
        + \gamma^+ \gamma^- m k^+ - \gamma^i \gamma^+ \gamma^- \p^i k^+
\nn \\
&
+ \gamma^i \gamma^+ \gamma^j \p^i \k^j - \gamma^i \gamma^+ m \p^i - \gamma^+ \gamma^j m \k^j + m^2 \gamma^+
 \end{align}
Substituting the simplified form of the gamma matrix structure given in Eq. \eqref{eq:gamma_Struc_II_A_minus} into Eq. \eqref{eq:II_quark_A_minus_1} and approximating the phase factor in $y^+$ and $x^+$ by one, we get 
\begin{align}
S_{F}(x,y) \big|_{\text{Eik},\, \A^{-}}^{\text{II,q}} &= 
\int \! \frac{d^{2}\p}{(2 \pi)^2} \ \int \! \frac{dp^+}{2\pi} 
\int \! \frac{d^{2}\k}{(2 \pi)^2}\ \int \! \frac{dk^+}{2\pi} 
\frac{ \theta(p^+)}{2p^+}  \frac{ \theta(k^+)}{2k^+} 
\ e^{-ix^-p^+} e^{i\x\cdot\p} \ e^{iy^{-}k^{+}} \ e^{-i\y\cdot\k} 
\nn \\
& 
\times
 \int \! dz^- \ e^{iz^-(p^+ - k^+)} 
 \int \! d^{2}\z \ e^{-i\z\cdot(\p - \k)} 
 \ \Big[  2\gamma^- p^+ k^+ - \gamma^- \gamma^+ \gamma^j p^+ \k^j   - \gamma^i \gamma^+ \gamma^- \p^i k^+ 
 \nn \\
 &  + \gamma^i \gamma^+ \gamma^j \p^i \k^j  +\gamma^- \gamma^+ m p^+ 
        + \gamma^+ \gamma^- m k^+
         - \gamma^i \gamma^+ m \p^i - \gamma^+ \gamma^j m \k^j + m^2 \gamma^+  \Big]
\nn  \\
&
 \times U_{F}(x^+,y^+,\z,z^-)
    \end{align}
Since we are interested in calculating the propagator at Eikonal order, we can perform a Taylor expansion of the Wilson line around an initial value of the minus coordinate $z^-=x^-$ and take the zeroth order term in the expansion. In that case, the only $z^-$ dependence appear in the phase and performing the integration over it gives 
\begin{align}
S_{F}(x,y) |_{\text{Eik},\, \A^{-}}^{\text{II,q}} &= 
\int \! \frac{d^{2}\p}{(2 \pi)^2} \ \int \! \frac{dp^+}{2\pi} 
\int \! \frac{d^{2}\k}{(2 \pi)^2}\ \int \! \frac{dk^+}{2\pi} 
\frac{ \theta(p^+)}{2p^+}  \frac{ \theta(k^+)}{2k^+} 
\ e^{-ix^-p^+} e^{i\x\cdot\p} \ e^{iy^{-}k^{+}} \ e^{-i\y\cdot\k}
\nn \\  
& 
\times
 2\pi \delta(p^+ - k^+)
 \int \! d^{2}\z \ e^{-i\z\cdot(\p - \k)} 
 \ \Big[  2\gamma^- p^+ k^+ - \gamma^- \gamma^+ \gamma^j p^+ \k^j   - \gamma^i \gamma^+ \gamma^- \p^i k^+ 
 \nn \\
 &  
 + \gamma^i \gamma^+ \gamma^j \p^i \k^j  +\gamma^- \gamma^+ m p^+ 
        + \gamma^+ \gamma^- m k^+
         - \gamma^i \gamma^+ m p^i - \gamma^+ \gamma^j m \k^j + m^2 \gamma^+  \Big]
\nn \\
&
   \times U_{F}(x^+,y^+,\z, x^{-})
\end{align}
One can now use the delta function to trivially integrate over $p^{+}$ to arrive at 
\begin{align}
S_{F}(x,y) |_{\text{Eik},\, \A^{-}}^{\text{II,q}} 
   &=  \int \! \frac{d^{2}\p}{(2 \pi)^2} \  \int \! \frac{d^{2}\k}{(2 \pi)^2}\ \int \! \frac{dk^+}{2\pi}  
 \theta(k^+) \ e^{-ix^-k^+} e^{i\x\cdot\p} \ e^{iy^{-}k^{+}} \ e^{-i\y\cdot\k} 
\nn \\ 
& 
\times
 \int \! d^{2}\z \ e^{-i\z\cdot(\p - \k)} 
 \ \Big[  \frac{\gamma^-}{2}  - \frac{1}{4k^+}(\gamma^- \gamma^+ \gamma^j  \k^j   + \gamma^i \gamma^+ \gamma^- \p^i )   + \frac{\gamma^i \gamma^+ \gamma^j \p^i \k^j}{4(k^+)^2} 
 \nn \\
 &
 + 
 \frac{m}{2k^+}
- \frac{\gamma^i \gamma^+ m \p^i}{4(k^+)^2} - \frac{\gamma^+ \gamma^j m \k^j}{4(k^+)^2} + \frac{m^2 \gamma^+}{4(k^+)^2}  \Big]
 U_{F}(x^+,y^+,\z, x^{-})
   \\
    \end{align}
Next step is to integrate over $\k$ and $\p$. For that purpose, one can write the $\k$ and $\p$ terms in the numarator as derivates acting on the $\k$ and $\p$ dependent phases as 
\begin{align}
S_{F}(x,y) |_{\text{Eik},\, \A^{-}}^{\text{II,q}} &= 
\int \! \frac{d^{2}\p}{(2 \pi)^2} \  \int \! \frac{d^{2}\k}{(2 \pi)^2}
\ \int \! \frac{dk^+}{2\pi}   \theta(k^+) \ e^{-ix^-k^+} \ e^{i\x\cdot\p} \ e^{iy^{-}k^{+}} \ e^{-i\y\cdot\k} 
\nn \\
& 
\times
 \int \! d^{2}\z \ e^{-i\z\cdot(\p - \k)} 
 \ \bigg[  \frac{\gamma^-}{2} + \frac{m}{2k^+}+ \frac{m^2 \gamma^+}{4(k^+)^2} 
 - \frac{1}{4k^+}\Big[
 \gamma^- \gamma^+ \gamma^j \Big(i \overleftarrow{\frac{d}{d\y^j}}\Big)  + \gamma^i \gamma^+ \gamma^-  \Big(-i \overleftarrow{\frac{d}{d\x^i}}\Big) \Big] 
 \nn \\
 &
 + \frac{\gamma^i \gamma^+ \gamma^j }{4(k^+)^2}  \Big(-i \overleftarrow{\frac{d}{d\x^i}}\Big) \Big(i \overleftarrow{\frac{d}{d\y^j}}\Big)
 - \frac{\gamma^i \gamma^+ m }{4(k^+)^2}  \Big(-i \overleftarrow{\frac{d}{d\x^i}}\Big) - \frac{\gamma^+ \gamma^j m }{4(k^+)^2} \Big(i \overleftarrow{\frac{d}{d\y^j}}\Big)   \bigg]
    U_{F}(x^+,y^+,\z, x^{-})
\end{align}
Integrations over $\p$ and $\k$ can be performed trivially and the result reads 
\begin{align}
\label{pure A -}
  S_{F}(x,y) |_{\text{Eik},A^{-}}^{\text{II,q}}    &= 
  \int \! \frac{dk^+}{2\pi}   \frac{\theta(k^+)}{2k^+} \ e^{-i(x^- - y^-)k^+}  
\nn \\
&
\times  \Bigg\{ 
  \int \! d^{2}\z \
\delta^2(\x - \z) \ \delta^2(\z- \y) \ U_{F}(x^+,y^+,\z, x^{-})
\Big[  \gamma^- k^+ + m + \frac{m^2 \gamma^+}{2k^+} \Big]
\nn \\
&
+ 
\bigg[ \frac{i}{2}\gamma^i \gamma^+ \gamma^- \overrightarrow{\frac{d}{d\x^i}} + \frac{i \gamma^i \gamma^+}{2k^+} m  \overrightarrow{\frac{d}{d\x^i}} \bigg]
\int \! d^{2}\z \
\delta^2(\x - \z) \ \delta^2(\z- \y) \ U_{F}(x^+,y^+,\z, x^{-})
\nn \\
& 
+
\int \! d^{2}\z \
\delta^2(\x - \z) \ \delta^2(\z- \y) \ U_{F}(x^+,y^+,\z, x^{-})
\bigg[ \frac{-i\gamma^- \gamma^+ \gamma^j }{2}  \overleftarrow{\frac{d}{d\y^j}} 
         - \frac{i\gamma^+ \gamma^j m }{2k^+}  \ \overleftarrow{\frac{d}{d\y^j}} \bigg]
\nn \\
&
+ 
\frac{\gamma^i \gamma^+ \gamma^j }{2k^+}  \overrightarrow{\frac{d}{d\x^i}} 
         \bigg[  \int \! d^{2}\z \
\delta^2(\x - \z) \ \delta^2(\z- \y) 
   \ U_{F}(x^+,y^+,\z, x^{-})\bigg] \overleftarrow{\frac{d}{d\y^j}}
   \Bigg\}
\end{align}
Eq. \eqref{pure A -} is the final expression for the pure $\A^-$ contribution to the inside-to-inside quark propagator at Eikonal accuracy. 
%%%%%%%%%%%%%%%%%%%%%%%%%%%%%%%%%%%%%%%%%%%%%%%%%%%%%%%%%%%%%
\subsection{Eikonal contribution to inside-to-inside quark propagator from $\A_{\perp}$ insertions}
\label{subsec:A_perp-to_II_quark}
Even though, the interaction with the transverse component of the background field is of order $(\gamma_t)^0$, and thus suppressed with respect to the $\A^-$ component, this suppression can be compensated. At eikonal order, the integration over the longitudinal coordinate gives a factor of $L^+$ which is of order $1/\gamma_t$ and it comes with $\A^-$ that is order $\gamma_t$, thus giving an order $(\gamma_t)^0$ contribution over all. Similarly, one can consider the instantaneous part of the quark propagator which gives an order $(\gamma_t)^0$ contribution when integrated over the longitudinal coordinate. If the one considers the insertion of a  transverse component of the background field that is of order $(\gamma_t)^0$ coupled to the instantaneous part of the quark propagator, one gets an overall order $(\gamma_t)^0$ term which is an eikonal contribution. There are three such contributions that would give terms at Eikonal order. 
%
%Generally, we use the obtained $A^-$  contribution of the corresponding case to derive contributions coming from transverse component insertion. But here, for the ease of computation and to easily get expressions such as; later on we can combine all of them to write in terms of covariant derivatives; we use a generalized eikonal order quark propagator.\\
\\
%%%%%%%%%%%%%%%%%%%%%%%%%%%%%%%%%%%%%%%%%%%%%%%%%%%%%%%%%%%

First contribution comes from inserting a transverse background field on the side of $x^+$. In this case, the propagator can be written as 
\begin{align}
\label{eq:contr_x+side}
S_{F}(x,y) \big|_{\text{Eik},\, \A_{\perp x}}^{\text{II,q}} = 
\int \! d^{4}w \   S_{F}(x,w)\big|_{\text{Eik},\, \A^{-}}^{\text{Inst.\, q}} \ (-ig \gamma^j t^a) A^{a}_{j}(\underline{w}) \   S_{F}(w,y) \big|_{\text{Eik},\, \A^{-}}^{\text{q}}
\end{align}
where we have an eikonal quark propagator in a pure $\A^-$ background from point $y$ to $w$. At point $w$, we insert the transverse background field and from point $w$ to $x$ we have the instantaneous quark propagator. When the explicit expressions for the quark propagators are inserted in Eq. \eqref{eq:contr_x+side}, one gets
\begin{align}
\label{eq:contr_x+side_2}
S_{F}(x,y) \big|_{\text{Eik},\, \A_{\perp}, x}^{\text{II,q}} &=   
\int \! d^4 w 
\delta^{3}(\underline{x} - \underline{w}) \gamma^+ \ \int \! \frac{dq^+}{2 \pi} \frac{i}{2q^+} \ e^{-iq^{+}(x^{-} - w^{-})}
\Big[ (-ig \gamma^j t^a) \A^{a}_{j}(\underline{w}) \Big]
\nn \\
& 
\times 
\int \! \frac{d^{3}\up}{(2 \pi)^3} \ \int \! \frac{d^{3}\uk}{(2 \pi)^3}
\ e^{-iw \cdot \check{p} + iy \cdot \check{k}} 
\int \! dz^- \ e^{iz^-(p^+ - k^+)} 
\ \int \! d^{2}\z \ e^{-i\z\cdot(\p - \k)} 
\ \frac{(\check{\slashed{p}}+ m)}{2p^+} \gamma^{+} 
\nn \\
&
\times  
\theta(p^+) \theta(k^+) \ U_{F}(w^+,y^+,\z,z^-)
\frac{(\check{\slashed{k}} + m)}{2k^+}         
\end{align}
One can simplify the gamma matrix structure by using the relation $ \gamma^+ \gamma^- \gamma^+ \gamma^- = 2 \gamma^+ \gamma^-$, and arrive at 
 \begin{align}
\gamma^+ (\check{\slashed{p}}+ m ) \gamma^+ (\check{\slashed{k}}) &=2\gamma^+ p^+ [\gamma^- k^+ - \gamma^l \k^l + m]
\end{align}
Substituting this simplified gamma matrix structure back into Eq. \eqref{eq:contr_x+side_2}, and dropping the phase factors $ e^{-iw^+ \check{p}^-}$ and $e^{ + iy^+ \check{k}^-}$ since we are interested in computing the propagator at eikonal order, we get 
\begin{align}
S_{F}(x,y) |_{\text{Eik},\, \A_{\perp x}}^{\text{II,q}} &=   
\int \! d^4 w \  \delta^{2}(\x - \w) \delta(x^+ - w^+)  
\ \int \! \frac{dq^+}{2 \pi} \frac{i}{2q^+} \ e^{-iq^{+}(x^{-} - w^{-})} 
\big[ig t^a \ A^{a}_{j}(\underline{w})\big] 
\nn \\
& 
\times \int \! \frac{dp^+}{2\pi} \int \! \frac{dk^+}{2\pi} \int \! 
\frac{d^{2}\p}{(2 \pi)^2} \ \int \! \frac{d^{2}\k}{(2 \pi)^2} 
\ e^{-iw^- p^+} \ e^{i \w\cdot\p} \ e^{iy^-k^+} \ e^{-i\y\cdot \k} \ \int \! dz^- \ e^{iz^-(p^+ - k^+)}
\nn  \\
&
\times
 \int \! d^{2}\z \ e^{-i\z(\p - \k)} \  \theta(p^+) \theta(k^+) 
 \bigg[  \gamma^j \gamma^+ \Big( \frac{\gamma^-}{2} - \frac{\gamma^l \k^l}{2k^+} + \frac{m}{2k^+}\Big)
\bigg] 
 U_{F}(w^+,y^+,\z,z^-)
\end{align}
Integrations over $\w$ and $w^+$ can be performed trivially and the result reads
\begin{align}
S_{F}(x,y) |_{\text{Eik},\, \A_{\perp x}}^{\text{II,q}} &=   
\int \! dw^- \ \int \! \frac{dq^+}{2 \pi} \frac{i}{2q^+} \int \! \frac{dp^+}{2\pi} 
\int \! \frac{dk^+}{2\pi} \int \! \frac{d^{2}\p}{(2 \pi)^2} 
\ \int \! \frac{d^{2}\k}{(2 \pi)^2} 
\nn \\
&
\times 
e^{-iq^{+}x^{-}} \ e^{-i q^+ w^{-}} \  e^{-iw^- p^+} \ e^{i \x\cdot \p} \ e^{iy^-k^+} \ e^{-i\y\cdot \k} 
\ \theta(p^+) \ \theta(k^+) \ 
\big[ ig t^a \ A^{a}_{j}(\underline{x})\big] 
\nn \\
& 
\times  \ \ \int \! dz^- \ e^{iz^-(p^+ - k^+)} 
\int \! d^{2}\z \ e^{-i\z\cdot(\p - \k)} 
\ \Big[  \gamma^j \gamma^+ \Big( \frac{\gamma^-}{2} - \frac{\gamma^l \k^l}{2k^+} + \frac{m}{2k^+}\Big)\Big]
U_{F}(x^+,y^+,\z,z^-)
\end{align}
We can again perform a gradient expansion of the Wilson line around the initial value $z^{-}=x^{-}$ and keep only the zeroth order term in the expansion to stay at Eikonal order. In that case, the integrations over $z^{-}$, $w^{-}$ can be performed trivially, one obtains
\begin{align}
S_{F}(x,y) |_{\text{Eik},\, \A_{\perp x}}^{\text{II,q}} &=    
\int \! \frac{dq^+}{2 \pi} \frac{i}{2q^+} \int \! \frac{dp^+}{2\pi} 
\int \! \frac{dk^+}{2\pi} \int \! \frac{d^{2}\p}{(2 \pi)^2} \ \int \! \frac{d^{2}\k}{(2 \pi)^2} 
\ e^{-iq^{+}x^{-}}  e^{i \x\cdot\p} \ e^{iy^-k^+} \ e^{-i\y\cdot\k} 
\nn \\
&
\times 
2 \pi \delta(p^+ - k^+) \theta(p^+) \ \theta(k^+) \ 
\big[ig t^a \ A^{a}_{j}(\underline{x})\big] \ 2\pi \delta(q^+ - p^+) 
\nn \\
&
\times
\int \! d^{2}\z \ e^{-i\z\cdot(\p - \k)} \   
\Big[  \gamma^j \gamma^+ \Big( \frac{\gamma^-}{2} - \frac{\gamma^l \k^l}{2k^+} + \frac{m}{2k^+}\Big)\Big] 
\ U_{F}(x^+,y^+,\z, x^{-})
\end{align}
It is now also straight forward to integrate over $q^+$, $p^+$ and $\p$, which yields to 
\begin{align}
S_{F}(x,y) \big|_{\text{Eik},\, \A_{\perp x}}^{\text{II,q}} &=    
\int \! \frac{dk^+}{2\pi} \  \frac{\theta(k^+)}{2k^+} 
\ e^{-i(x^- -y^-)k^+} \int \! \frac{d^{2}\k}{(2 \pi)^2} \int \! d^{2}\z
\ e^{-i\y\cdot\k}\   e^{i\z\cdot\k} \  \delta^2(\x - \z)
 \big[ig t^a \ \A^{a}_{j}(\underline{x})\big]
 \nn \\
 &
 \times 
\Big[  \frac{i \gamma^j \gamma^+ \gamma^-}{2} - \frac{i \gamma^j \gamma^+ \gamma^l \k^l}{2k^+} + \frac{i \gamma^j \gamma^+ m}{2k^+} \Big]
\ U_{F}(x^+,y^+,\z, x^{-})
\end{align}
Finally, rewriting $\k^l$ as a derivative acting on the phase and then integrating over it, we get the final expression for the first contribution as
\begin{align}
\label{transverse x}
S_{F}(x,y)\big|_{\text{Eik},\, \A_{\perp x}}^{\text{II,q}} &=    
\int \! \frac{dk^+}{2\pi} \  \frac{\theta(k^+)}{2k^+} \ e^{-i(x^- -y^-)k^+} \int \! d^{2}\z
\ \delta^2(\x - \z) \ \delta^2(\z- \y)
\big[ ig t^a \ \A^{a}_{j}(\underline{x})\big]  
\nn \\
&
\times 
U_{F}(x^+,y^+,\z, x^{-})
\Big[  \frac{i \gamma^j \gamma^+ \gamma^-}{2} + \frac{ \gamma^j \gamma^+ \gamma^l}{2k^+} \overleftarrow{\frac{d}{d\y^j}} + \frac{i \gamma^j \gamma^+ m}{2k^+}
\Big]
\end{align}

The next contribution comes from inserting the transverse component of the background field on the $y^+$ analogue to the first contribution. In this case the propagator can be written as 
\begin{align}
\label{eq:contr_y+_side_1}
S_{F}(x,y) \big|_{\text{Eik},\, \A_{\perp y}}^{\text{II,q}} = 
\int \! d^{4}w \   S_{F}(x,w)\big|_{\text{Eik},\, \A^{-}}^{\text{q}} 
\ \big(-ig \gamma^j t^a\big) \A^{a}_{j}(\underline{w}) \   S_{F}(w,y) |_{\text{Eik},\, \A^{-}}^{\text{Inst,q}}
\end{align}
where we have an instantaneous quark propagator from $y$ to $w$. At point $w$ the transverse component of the background field is inserted and then we have a quark propagator in a pure $\A^-$ background from $w$ to $x$. Inserting the explicit expressions of the quark propagators in eq. \eqref{eq:contr_y+_side_1}, one obtains 
\begin{align}
S_{F}(x,y) \big|_{\text{Eik},\, \A_{\perp y}}^{\text{II,q}} &= 
\int \! d^{4}w \ \int \! \frac{d^{3}\up}{(2 \pi)^3} \ \int \! \frac{d^{3}\uq}{(2 \pi)^3}
\ e^{-ix \cdot \check{q} + iw \cdot \check{p}} \int \! dz^- \ e^{iz^-(q^+ - p^+)} 
\ \int \! d^{2}\z \ e^{-i\z\cdot(\q - \p)} 
\nn \\
&
\times 
\frac{(\check{\slashed{q} }+ m)}{2q^+} \gamma^{+}
\ \theta(q^+) \ \theta(p^+) 
\ U_{F}(x^+,w^+,\z,z^-)
\frac{(\check{\slashed{p}} + m)}{2p^+} 
\big[ (-ig \gamma^j t^a) A^{a}_{j}(\underline{w})\big]
\nn \\
&
\times 
\delta^{3}(\underline{w} - \underline{y}) \gamma^+ 
\ \int \! \frac{dk^+}{2 \pi} \frac{i}{2k^+} \ e^{-ik^{+}(w^{-} - y^{-})} 
\end{align}
Simplifying the gamma matrix structure and following the same steps and performing the integrations as in the previous case, one gets the second contribution as 
\begin{align}
\label{transverse y}
S_{F}(x,y) \big|_{\text{Eik},\, \A_{\perp y}}^{\text{II,q}} &= 
\int \! \frac{dk^+}{2 \pi} \frac{\theta(k^+)}{2k^+} \ e^{-i(x^-y^-) k^+} 
\ \Big[ \frac{-i\gamma^- \gamma^+ \gamma^j}{2}  + \frac{\gamma^i \gamma^+ \gamma^j }{2k^+}\overrightarrow{\frac{d}{d\x^i}} - \frac{i\gamma^+ \gamma^j m}{2k^+})  \Big]  
\nn \\
&
\times 
\int \! d^{2}\z   \ \delta^2(\x-\z)\ \delta^2(\z -\y)
\ U_{F}(x^+,y^+,\z, x^{-})  \  \big[-ig  t^a \A^{a}_{j}(\underline{y}) \big]
\end{align}

The last contribution comes from inserting the transverse component of the background field both on the $x^+$ and $y^+$ sides. In that case, the propagator reads To compute this contribution, we will use expression, 
\begin{align}
S_{F}(x,y)\big|_{\text{Eik},\, \A_{\perp xy}}^{\text{II,q}} &= 
\int \! d^{4}w \ \int\! d^{4}w'   
S_{F}(x,w')\big|_{\text{Eik},\, \A^{-} }^{\text{Inst,q}} \ (-ig \gamma^{j'} t^a) \A^{a}_{j'}(\underline{w'}) 
\  S_{F}(w',w) \big|_{\text{Eik},\, \A^{-}}^{\text{q}} 
\nn \\
&
\times 
(-ig \gamma^j t^b) A^{b}_{j}(\underline{w}) 
\  S_{F}(w,y) \big|_{\text{Eik}\, \A^{-}}^{\text{Inst,q}}  
\end{align}
where we have an instantaneous quark propagator from point $y$ to $w$. At point $w$ transverse background field is inserted. Then, we have a quark propagator in a pure $\A^-$ background from point $w$ to $w'$ where another transverse background field is inserted. Finally, we have another instantaneous quark propagator from point $w'$ to $x$. This contribution can be computed by following similar steps and the final result can be written as 
\begin{align}
\label{transverse xy}
S_{F}(x,y) \big|_{\text{Eik},\, \A_{\perp xy}}^{\text{II,q}} &= 
\int \! \frac{dk^+}{2 \pi} \frac{\theta(k^+)}{2k^{+}} 
\  e^{-i(x^{-}- y^-)k^+} 
\big[ig t^a A^{a}_{i}(\underline{x})\big]  
\nn  \\
&
\times 
\int \! d^{2}\z \ \delta^2(\x - \z) \ \delta^2(\z - \y) 
\ U_{F}(x^+,y^+,\z,x^{-})
\big[-ig t^b \A^{b}_{j}(\underline{y})\Big]
\ \Big[\frac{\gamma^{i} \gamma^+ \gamma^j}{2k^+}\Big]       
\end{align}
%
%
%%%%%%%%%%%%%%%%%%%%%%%%%%%%%%%%%%%%%%%%%%%%%%%%%%%%%%%%%%%%%%%%%%%%%%%%%%%%%%%%%%%%%%%%%%%%%%%%%%%%%%%%%%%
\subsection{Total inside-to-inside quark propagator in a gluon background field at eikonal order}
The total inside-to-inside quark propagator at eikonal order for $x^+ > y^+$ can be written as sum of the contributions computed above. Schematically, it reads 
\begin{align}
 S_{F}(x,y) \big|_{\text{Eik}}^{\text{II,q}} =  
 S_{F}(x,y) \big|_{\text{Eik},\, \A^{-}}^{\text{II,q}} 
 +  
 S_{F}(x,y) \big|_{\text{Eik},\, \A_{\perp x}}^{\text{II,q}} 
 +  
 S_{F}(x,y) \big|_{\text{Eik},A_{\perp y}}^{\text{II,q}} 
 +  
 S_{F}(x,y) \big|_{\text{Eik},A_{\perp xy}}^{\text{II,q}}
\end{align}
where each contribution on the right hand side are given in Eqs. \eqref{pure A -}, \eqref{transverse x}, \eqref{transverse y} and \eqref{transverse xy} respectively. Using these explicit expressions for each contribution and also using the following relation 
\begin{align}
\overrightarrow{D}_{\x^i} \ f(x,y) \ \overleftarrow{D}_{\y^j} &= 
\Big(\overrightarrow{\partial}_{\x^i} + ig t^a \A_i^a(\underline{x})\Big) \ f(x,y) 
\ \Big(\overleftarrow{\partial}_{\y^j} - ig t^b A_j^b(\underline{y})\Big)
\nn \\
&
= \overrightarrow{\partial}_{\x^i} f(x,y) \overleftarrow{\partial}_{\y^j} + \overrightarrow{\partial}_{\x^i} f(x,y) \big[ - ig t^b \A_j^b(\underline{y})\big] + \big[ ig t^a \A_i^a(\underline{x})\big] f(x,y) \overleftarrow{\partial}_{\y^j}   
\nn \\
&+ 
\big[ ig t^a \A_i^a(\underline{x})\big] f(x,y) \big[ - ig t^b \A_j^b(\underline{y})\big] 
\end{align}    
one obtains Eq. \eqref{eq:inside-inside_quark} for the inside-to-inside quark propagator at eikonal order. Moreover, one can study the case $y^+>x^+$ using the same method, and find the result given in Eq. \eqref{eq:inside-inside-antiquark} for the inside-to-inside antiquark propagator at eikonal order.   
%
%

%%%%%%%%%%%%%%%%%%%%%%%%%%%%%%%%%%%%%%%%%%%%%%%%%%%%%%%%%%%%%%%%%%%%%%%%%%%%%%%%%%%%%%%%%%%%%%%%%%%%%%%%%%%%%%%%%%%%%%%%%%%%%%%%%%%%%%%%%%%%%%%%%%%%%%%%%%%%%%%%%%%%%%%%
\section{Derivation of the inside-to-after gluon propagator}
\label{App:inside-to-after-gluon-prop} 
In this Appendix, we present a detailed derivation of the inside-to-after gluon propagator at Eikonal order, i.e. we will consider the gluon propagator for $x^+>L^+/2$ and $-L^+/2<y^+<L^+/2$. For this purpose, we start with the gluon propagator in pure $\A^-$ background at Eikonal order with $z^-$ dependence given in Eq. \eqref{GEikF} and consider the case with $-L^+/2<y^+<L^+/2$. Since the point $y^+$ is inside the medium, we can approximate the phase factor $e^{iy^+\check{k}^-}$ by $1$, as all other contributions coming from the phase brings an extra suppression factor, thus making the expression NEik or beyond. Moreover, since $x^+>L^+/2$ and $-L^+/2<y^+<L^+/2$, we have necessarily $x^+>y^+$, so that we can drop the instantaneous contribution and the $y^+>x^+$ contribution in Eq. \eqref{GEikF}. Finally, as in the derivation performed inside-to-inside quark propagator, we would like to integrate our expression over $\k$. Thus, we rewrite the factors of $\k$ in the expression as derivates acting on the phase. All in all, the medium contribution to the inside-to-after gluon propagator in a pure dynamical $\A^-$ background can be written as  
\begin{align}
\label{eq:IA_gluon_1}
 G_{F}^{\mu\nu}(x,y) \big|_{ \text{Pure}\,  \A^-}^{\text{IA}} &= 
\int \! \frac{d^{3}\uq}{(2\pi)^{3}} \ 
 \ e^{-ix \cdot \check{q}}  \ \theta(q^{+}) \
\int \! \frac{d^{3}\uk}{(2\pi)^{3}}  \
e^{iy^-k^+} \ e^{-i\y\cdot \k}  \ \theta(k^{+}) 
\ \frac{1}{(q^{+}\!+\!k^{+})} 
\ \int \! d^{2}\z\ e^{i \k\cdot \z}  
 \int \! dz^{-} \  e^{-ik^{+}z^{-}} \ 
 \nn \\ 
 & \hspace{-1cm}
 \times
 \Big[  - g^{\mu}_{i} g^{i \nu} + \frac{g^{\mu}_{i}  \eta^{\nu}}{k^{+}}
 \Big(-i \frac{\overleftarrow{\partial}}{\partial \z^{i}}\Big) 
 + \frac{\eta^{\mu} g^{\nu}_{i} \q^{i}}{q^{+}} 
 + \frac{\eta^{\mu}\eta^{\nu}}{q^{+}k^{+}}\q^{i}\Big(-i \frac{\overleftarrow{\partial}}{\partial \z^{i}}\Big)  
 \Big]
 \ e^{-i\q\cdot \z} \ e^{iq^{+} z^{-}} \
U_{A}(x^{+},y^{+},\z,z^{-})         
\end{align}
After performing integration by parts for the derivatives in $\z$, one can rewrite Eq. \eqref{eq:IA_gluon_1} as 
\begin{align}
\label{eq:IA_gluon_2}
 G_{F}^{\mu\nu}(x,y) \big|_{\text{Pure}\,  \A^-}^{\text{IA}} &= 
\int \! \frac{d^{3}\uq}{(2\pi)^{3}} \ 
e^{-ix \cdot \check{q}}  \ \theta(q^{+}) \
\!\!\int \! \frac{d^{3}\uk}{(2\pi)^{3}}  \
e^{iy^-k^+} \ e^{-i\y\cdot\k}  \ \theta(k^{+}) 
\ \int \! d^{2}\z\ e^{i \z\cdot(\k- \q)}
\int \! dz^{-} \  e^{-i(k^{+} -q^{+})z^{-}} \! \frac{1}{(q^{+}\!+\!k^{+})} 
\nn   \\
& \hspace{-1.5cm}
\times
\Big[  - g^{\mu}_{i} g^{i \nu}
 + \frac{g^{\mu}_{i}  \eta^{\nu}}{k^{+}}\Big(i \frac{\overrightarrow{\partial}}{\partial \z^{i}}\Big) 
 + \frac{g^{\mu}_{i}  \q^{i} \eta^{\nu}}{k^{+}} 
 + \frac{\eta^{\mu} g^{\nu}_{i} \q^{i}}{q^{+}} 
 + \frac{\eta^{\mu}\eta^{\nu}}{q^{+}k^{+}}\q^{i}\Big(i \frac{\overrightarrow{\partial}}{\partial \z^{i}}\Big) 
 + \frac{\eta^{\mu}\eta^{\nu}}{q^{+}k^{+}}\q^{i} \q^{i}\Big]
 \ U_{A}(x^{+},y^{+},\z,z^{-})         
\end{align}
Since the only $\k$ dependence in Eq. \eqref{eq:IA_gluon_2} is in the phase factors, the integration over  $\k$ can be performed trivially yielding to 
\begin{align}
 G_{F}^{\mu\nu}(x,y) \big|_{\text{Pure}\,  \A^-}^{\text{IA}} &= 
\int \! \frac{d^{3}\uq}{(2\pi)^{3}} \ e^{-ix \cdot \check{q}}  \theta(q^{+}) 
\!\! \int \! \frac{dk^{+}}{(2\pi)} \ e^{iy^-k^+}  \theta(k^{+}) 
\!\! \int \! d^{2}\z\ e^{-i \z\cdot\q} \  \delta^{2}(\z-\y)
\int \! dz^{-} \  e^{-i(k^{+} -q^{+})z^{-}}  \frac{1}{(q^{+}\!+\!k^{+})}  \
\nn \\
& \hspace{-1.5cm}
\times
 \Big[  - g^{\mu}_{i} g^{i \nu}
 + \frac{g^{\mu}_{i}  \eta^{\nu}}{k^{+}}\Big(i \frac{\overrightarrow{\partial}}{\partial \z^{i}}\Big) + \frac{g^{\mu}_{i}  \q^{i} \eta^{\nu}}{k^{+}} + \frac{\eta^{\mu} g^{\nu}_{i} \q^{i}}{q^{+}} + \frac{\eta^{\mu}\eta^{\nu}}{q^{+}k^{+}}\q^{i}\Big(i \frac{\overrightarrow{\partial}}{\partial \z^{i}}\Big) + \frac{\eta^{\mu}\eta^{\nu}}{q^{+}k^{+}}\q^{i} \q^{i}\Big]\
U_{A}(x^{+},y^{+},\z,z^{-})         
 \end{align}
For a highly boosted target, the $z^-$ dependence is slow in the Wilson line due to Lorentz time dilation. If one expands the Wilson line around a position $z^-=y^-$, the zeroth term gives contribution at Eikonal order while the first term contributes at NEik order. Since we are interested in deriving the medium contribution to inside-to-after gluon propagator at Eikonal order, one should only keep the zeroth order term in the expansion. In that case, the only $z^-$ dependence appears at the phase which can be integrated trivially and it gives a delta function in $k^+-q^+$. Performing the integrations over $\z$ and $k^+$, one obtains 
\begin{align}
\label{eq:IA_pure_A_minus_gluon}
 G_{F}^{\mu\nu}(x,y) \big|_{\text{Eik},\, \text{Pure}\, \A^-}^{\text{IA}} &= 
\int \! \frac{d^{3}\uq}{(2\pi)^{3}} \ 
 \ e^{-ix \cdot \check{q}}  \ \frac{\theta(q^{+})}{2q^{+}} 
 \ e^{iy^-q^+}   \ e^{-i \q\cdot \y} 
 \nn \\
 & \hspace{-1.5cm}
 \times
 \Big[  - g^{\mu}_{i} g^{i \nu}
 + \frac{i g^{\mu}_{i}  \eta^{\nu}}{q^{+}} \frac{\overrightarrow{\partial}}{\partial \y^{i}} 
 + \frac{g^{\mu}_{i}  \q^{i} \eta^{\nu}}{q^{+}} 
 + \frac{\eta^{\mu} g^{\nu}_{i} \q^{i}}{q^{+}} 
 + \frac{i \eta^{\mu}\eta^{\nu}}{q^{+}q^{+}}\q^{i} \frac{\overrightarrow{\partial}}{\partial \y^{i}} 
 + \frac{\eta^{\mu}\eta^{\nu}}{q^{+}q^{+}}\q^{i} \q^{i} \Big]
 \ U_{A}(x^{+},y^{+},\y, y^{-})        
 \end{align}
Since the point $y^+$ is inside the medium, one can also get a contribution from the insertion of $\A_{\perp}$ field on the $y^+$ side that would come with the instantaneous part of the gluon propagator (see the discussion for the power counting at the beginning of \ref{subsec:A_perp-to_II_quark}). Thus, the medium contribution to the inside-to-after gluon propagator that stems from the insertion of the transverse background field at eikonal order reads 
\begin{align}       
\delta G_{F}^{\mu\nu}(x,y) \big|_{\text{Eik},\,  \A_{\perp}}^{\rm IA} = 
\int d^{4}w \ G_{F}^{\mu\mu'}(x,w) \big|_{\A^-,\, \text{Eik}}^{\rm IA}
 \ \operatorname{X}^{3g}_{\mu'\nu'}(w) 
 \ G_{F}^{\nu'\nu}(w,y) \big|_{\text{Inst.}\, ,\A^-}
\end{align}
Here, we have the instantaneous gluon propagator from point $y$ to point $w$ and at point $w$ the effective single $\A_{\perp}$ insertion factor,  $\operatorname{X}^{3g}_{\mu'\nu'}(\underline{w})$ that is given in Eq. \eqref{three gluon vertex insertion}, is inserted. Then from point $w$ to $x$ we have the pure $\A^-$ contribution to the inside-to-after gluon propagator at eikonal order which is given in Eq. \eqref{eq:IA_pure_A_minus_gluon}. Putting everything together and again gradient expanding the Wilson line around $w^-=y^-$ and keeping only the zeroth order term in the expansion, we get 
\begin{align}       
\delta G_{F}^{\mu\nu}(x,y) \big|_{\text{Eik},\,  \A_{\perp}}^{\rm IA} &= 
\int d^{4}w  \  \int \! \frac{d^{3}\uq}{(2\pi)^{3}} \ \theta(q^{+})
 \ \frac{ e^{-ix \cdot \check{q}} }{2q^+}  
 \  \int \! \frac{dk^{+}}{2\pi} \ e^{iw^-q^+}\    \ e^{-i\q\cdot \w} \
U_{A}(x^{+},w^{+},\w, y^{-})  \big[ gf^{a'b'c}\big] 
\nn \\
& 
\times
 \Big[ ig^{\mu j} \eta^{\nu} - \frac{i \eta^{\mu} \q^{j} \eta^{\nu}}{q^{+}}  \Big] 
\ \Big[  2\overleftarrow{\frac{d}{dw^{-}}}\A_{c}^{j}(w)+  \A_{c}^{j}(w) \ \overrightarrow{\frac{d}{dw^{-}}}\Big]  
i \delta^{2}(\w - \y)
\ \delta(w^{+}- y^{+})  \ \frac{ e^{-i(w^{-}-y^{-})k^{+}}} {k^{+}k^{+}} 
\end{align}
Acting with the derivatives on the phases one gets 
\begin{align}       
\delta G_{F}^{\mu\nu}(x,y) \big|_{\text{Eik},\,  \A_{\perp}}^{\rm IA} &= 
 \int \! \frac{dk^{+}}{2\pi}  \  \int \! \frac{d^{3}\uq}{(2\pi)^{3}} \ \theta(q^{+})
 \ \frac{ e^{-ix \cdot \check{q}} }{2q^+}  \ \int \! dw^{-} \ e^{i(q^+ -k^+)w^-}  \int \! dw^{+} \ \delta(w^{+}- y^{+}) 
 \int \! d^{2}\w \ \delta^{2}(\w - \y) \ e^{-i\q\cdot \w}  
 \nn \\
 &
 \times
\big[ gf^{a'b'c}\big] \
 \Big[ -g^{\mu j} \eta^{\nu} + \frac{ \eta^{\mu} \q^{j} \eta^{\nu}}{q^{+}}  \Big] 
 \Big[  (2q^+ - k^+) i \A_{c}^{j}(w)\Big]   
 \ \frac{ e^{iy^{-}k^{+}}} {k^{+}k^{+}} \ U_{A}(x^{+},w^{+},\w, y^{-}) 
\end{align}
Gradient expanding $\A_{c}^{j}(w)$ as well around  $w^-=y^-$, the integrations over $w^-$, $w^+$ and $\w$ can be performed trivially and it yields to 
 \begin{align}
 \label{eq:IA_gluon_A_perp}
\delta G_{F}^{\mu\nu}(x,y) \big|_{\text{Eik},\,  \A_{\perp}}^{\rm IA} &= 
g\int \! \frac{d^{3}\uq}{(2\pi)^{3}} \ \theta(q^{+})
 \ \frac{ e^{-ix \cdot \check{q}} }{2q^+} \ e^{-i\q\cdot\y}  
 \  \frac{ e^{iy^{-}q^{+}}} {q^{+}} \ U_{A}(x^{+},y^{+},\y, y^{-}) 
 \big[T \cdot \A^{j}(y)\big]
 \Big[ g^{\mu j} \eta^{\nu} - \frac{ \eta^{\mu} \q^{j} \eta^{\nu}}{q^{+}}  \Big]
\end{align}
In order to get the inside-to-after gluon propagator at eikonal accuracy one adds the pure $\A^-$ contribution given in Eq. \eqref{eq:IA_pure_A_minus_gluon} and the transverse background field contribution given in Eq. \eqref{eq:IA_gluon_A_perp}. The final result is given in \eqref{eq:inside-after_gluon} which in the static target field approximation reads Eq. \eqref{IA G_Eik}.   

\bibliography{mybib_New}
\end{document}